\documentclass[sigconf]{acmart}
\usepackage{xcolor}
\definecolor{mypurple}{HTML}{b479d2}
\definecolor{mygreen}{HTML}{b3d055}
\definecolor{myteal}{HTML}{8ccdbf}
\definecolor{myyellow}{HTML}{f9db95}
\definecolor{myred}{HTML}{ef8477}
\AtBeginDocument{%
  }

\copyrightyear{\copyright{} Xuetong WANG, Ching Christie Pang, Pan HUI, 2025. This is the author's version of the work. It is posted here for your personal use. Not for redistribution. The definitive Version of Record was published in ACM Digital Library}
\acmYear{2025}
\setcopyright{none}
\acmConference[UIST '25]{The 38th Annual ACM Symposium on User Interface Software and Technology}{September 28-October 1, 2025}{Busan, Republic of Korea}
\acmDOI{10.1145/3746059.3747617}
\acmISBN{979-8-4007-2037-6/2025/09}

\def\markup{0}   % 1: mark-up ; 0: clean version 
\if\markup1 %mark-up version
    \def\finalmarkup{1} %0:major revision 1:final revision
    \if\finalmarkup1
        \usepackage[normalem]{ulem}
        \newcommand{\todo}[1]{}
        \definecolor{finalhighlight}{rgb}{0,0,0.75}
        \newcommand{\finalrv}[1]{{\leavevmode\color{finalhighlight}#1}}
        \newcommand{\soutMajor}[1]{}
        \newcommand{\soutFinal}[1]{{\color{red}\sout{#1}}}
    \else
        \usepackage[normalem]{ulem}
        \definecolor{MajorHighlight}{rgb}{0,0,0.75}
        
        \definecolor{todo}{rgb}{0.75,0,0}
        \newcommand{\todo}[1]{}
        \newcommand{\finalrv}[1]{#1}
        \newcommand{\soutMajor}[1]{{\color{pink}\sout{#1}}}
        \newcommand{\soutFinal}[1]{}
    \fi
\else
    
    \newcommand{\todo}[1]{}
    \newcommand{\finalrv}[1]{#1}
    \newcommand{\soutMajor}[1]{}
    \newcommand{\soutFinal}[1]{}
\fi

\begin{document}

\title{Talking Spell: A Wearable System Enabling Real-Time Anthropomorphic Voice Interaction with Everyday Objects}
\renewcommand{\shorttitle}{Talking Spell}

\author{Xuetong WANG}
\email{xwangdd@connect.ust.hk}
\affiliation{%
  \institution{The Hong Kong University of Science and Technology}
  \city{Hong Kong}
  \country{China}
}
\orcid{0000-0002-9625-2012}

\author{Ching Christie Pang}
\email{ccpangaa@connect.ust.hk}
\affiliation{%
  \institution{The Hong Kong University of Science and Technology}
  \city{Hong Kong}
  \country{China}
}
\orcid{0000-0003-4704-2403}

\author{Pan Hui}
\authornote{Corresponding author}
\authornote{Pan Hui is also affiliated with Hong Kong University of Science and Technology, and University of Helsinki, Finland}
\affiliation{%
  \institution{The Hong Kong University of Science and Technology (Guangzhou)}
  \city{Guangzhou}
  \country{China}}
\email{panhui@ust.hk}
\orcid{0000-0001-6026-1083}

\renewcommand{\shortauthors}{Xuetong Wang et al.}

\begin{abstract}
  %% Background: VAs --> AI companionship 
  % Virtual assistants (VAs) have become ubiquitous in daily life, manifesting in smartphones and smart devices. This widespread adoption has ignited interest in AI companions designed to enhance user experiences and foster emotional connections. However, these companions are often embedded into specific objects — such as glasses, home assistants, or dolls — forcing users to forge emotional bonds with unfamiliar items. This can lead to feelings of detachment and impersonality, highlighting the need for more relatable and engaging solutions.
  Virtual assistants (VAs) have become ubiquitous in daily life, integrated into smartphones and smart devices, sparking interest in AI companions that enhance user experiences and foster emotional connections. However, existing companions are often embedded in specific objects—such as glasses, home assistants, or dolls—requiring users to form emotional bonds with unfamiliar items, which can lead to reduced engagement and feelings of detachment. To address this, we introduce Talking Spell\footnote{https://github.com/poiuytxw/TalkingSpell}, a wearable system that empowers users to imbue any everyday object with speech and anthropomorphic personas through a user-centric radiative network.
  % In response, we introduce Talking Spell, an innovative system that imbues objects with speech. Through a user-centric radiative network structure, user can cast the spell on any object with our "scope" and "wand", thereby enabling a transformative form of human-object interactions. 
Leveraging advanced computer vision (e.g., YOLOv11\cite{yolo} for object detection), large vision-language models (e.g., QWEN-VL\cite{qwen-vl} for persona generation), speech-to-text and text-to-speech technologies, Talking Spell guides users through three stages of emotional connection: acquaintance, familiarization, and bonding.
  % Our approach combines advanced computer vision algorithms and large-scale language models (LLMs) to guide users in establishing three stages of emotional connection: 1) acquaintance, 2) familiarization, and 3) bonding.
  %% Contribution and Findings
  %% We present findings from a user study that guided our implementation, along with experiments that validate the effectiveness of our final system. 
  We validated our system through a user study involving 12 participants, utilizing Talking Spell to explore four interaction intentions: entertainment, companionship, utility, and creativity. The results demonstrate its effectiveness in fostering meaningful interactions and emotional significance with everyday objects. Our findings indicate that Talking Spell creates engaging and personalized experiences, as demonstrated through various devices, ranging from accessories to essential wearables.
\end{abstract}


\begin{CCSXML}
<ccs2012>
<concept>
<concept_id>10003120.10003123.10011760</concept_id>
<concept_desc>Human-centered computing~Systems and tools for interaction design</concept_desc>
<concept_significance>500</concept_significance>
</concept>
<concept>
<concept_id>10003120.10003138.10003140</concept_id>
<concept_desc>Human-centered computing~Ubiquitous and mobile computing systems and tools</concept_desc>
<concept_significance>500</concept_significance>
</concept>
</ccs2012>
\end{CCSXML}

\ccsdesc[500]{Human-centered computing~Systems and tools for interaction design}
\ccsdesc[500]{Human-centered computing~Ubiquitous and mobile computing systems and tools}

\begin{comment}
\begin{CCSXML}
<ccs2012>
 <concept>
  <concept_id>00000000.0000000.0000000</concept_id>
  <concept_desc>Do Not Use This Code, Generate the Correct Terms for Your Paper</concept_desc>
  <concept_significance>500</concept_significance>
 </concept>
 <concept>
  <concept_id>00000000.00000000.00000000</concept_id>
  <concept_desc>Do Not Use This Code, Generate the Correct Terms for Your Paper</concept_desc>
  <concept_significance>300</concept_significance>
 </concept>
 <concept>
  <concept_id>00000000.00000000.00000000</concept_id>
  <concept_desc>Do Not Use This Code, Generate the Correct Terms for Your Paper</concept_desc>
  <concept_significance>100</concept_significance>
 </concept>
 <concept>
  <concept_id>00000000.00000000.00000000</concept_id>
  <concept_desc>Do Not Use This Code, Generate the Correct Terms for Your Paper</concept_desc>
  <concept_significance>100</concept_significance>
 </concept>
</ccs2012>
\end{CCSXML}

\ccsdesc[500]{Do Not Use This Code~Generate the Correct Terms for Your Paper}
\ccsdesc[300]{Do Not Use This Code~Generate the Correct Terms for Your Paper}
\ccsdesc{Do Not Use This Code~Generate the Correct Terms for Your Paper}
\ccsdesc[100]{Do Not Use This Code~Generate the Correct Terms for Your Paper}
\end{comment}
\keywords{Embodied and Explorable Interaction; AI Companionship; Wearable; On-body Devices; Ubiquitous Computing; Human-Object Interaction; Large Language Models (LLMs)}


\maketitle
\begin{figure}[h!]
  \includegraphics[width=\linewidth]{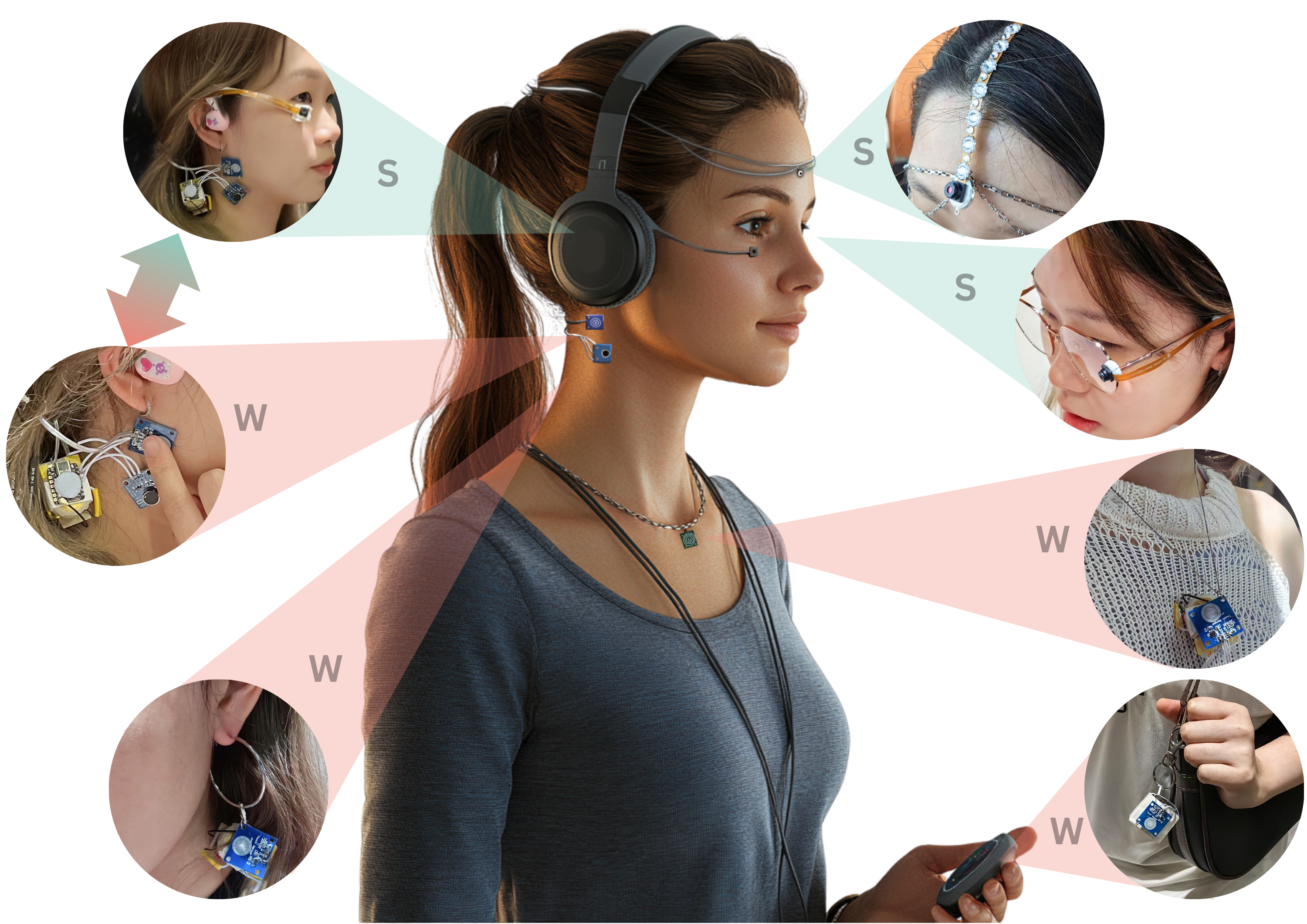}
  \caption{Talking Spell consists the scope (camera for detecting objects) and wand (sensor to allow spell casting and vibration motor to inform useful "spell". Example scope devices we created include (colored in green, clockwise from top-left): head chain, glasses, and bone conduction headphones; example wand devices we created include (colored in red, clockwise from top-left): a necklace with sensor pendant, bag charm, earrings, and a part of bone conduction headphones. Note that the two components (top-right) form an all-in-one device in the bone conduction headphones. }
  \Description{}
  \label{fig:teaser}
\end{figure}
\section{Introduction}
%% Background 
In recent years, the advent and omnipresence of artificial intelligence (AI) has significantly enhanced user experiences and improved human-computer interaction (HCI) in virtual assistants (VAs) \cite{Alkatheiri2022}, particularly through text and audio interfaces \cite{Gilles2022}. VAs have emerged as one of the mainstream interaction paradigms, exemplified by applications in various formats, ranging from smartphone applications and home assistants to wearable devices like glasses and smartwatches. Examples include voice-oriented assistant platforms such as Alexa and Siri, alongside text-oriented platforms like ChatGPT and DeepSeek \cite{Maedche2019}. Consequently, extensive research efforts have focused on enhancing personalized and customized experiences, making these VAs more accessible and user-friendly \cite{Cowan2017, Yang2019}.

%% Problem / Research Gap 
Despite these advancements, current VAs primarily function as isolated entities — devices or applications that users engage with directly \cite{Maedche2019}. For instance, Siri operates through iPhones, while Alexa is integrated into a range of smart home devices. These VAs typically serve as centralized hubs \cite{Ammari2019, Lenhart2023}, responding to user queries and commands but lacking a comprehensive understanding of their contextual environment \cite{Castillo2021, Ammari2019}. While capable of managing tasks and providing information, existing applications fall short of delivering the nuanced interaction potential that more personalized and context-aware systems could offer. Users are restricted to interacting with a singular AI entity - be it smart home devices, AI dolls, or wearables - thereby missing opportunities for dynamic and personalized experiences. This limitation underscores the need for a more flexible and creative approach to AI interaction that transcends the boundaries of traditional VAs.

%% Motivation (Reason for Anthropomorphism)
In this context, anthropomorphism explains our concept. It refers to the tendency to attribute human-like characteristics to non-human entities \cite{Epley2007}. Research indicates that anthropomorphizing objects can significantly influence individuals' psychological and emotional connections with them \cite{Bowlby1969, Mikulincer2005}. By endowing non-human objects with human-like traits, anthropomorphism alters users' relationships with these entities, thereby shifting their emotional and cognitive responses \cite{Wan2021}. This suggests that leveraging anthropomorphism in the design of AI companions could enhance user engagement and satisfaction, motivating our proposal for a more integrated and context-aware approach to AI interaction. Anthropomorphism frequently appears in creative works and movies. As a consequence, we drew inspiration from such depictions, particularly \textit{the Talking Spell in Harry Potter}, a charm that imbues objects with the ability to speak.

%% User Scenario
% Consider a scenario in which a user, while in their kitchen, looks at a potted plant and contemplates its care. They might ask their AI assistant, “What is this plant?” The assistant provides valuable information about the plant's needs, such as watering requirements and sunlight preferences. However, the user is merely communicating with the assistant, which acts as a middleman, rather than engaging directly with the plant itself. This interaction feels impersonal and detached, as the user misses the opportunity for a more engaging experience—one where they could converse directly with the plant, asking questions or sharing thoughts about its growth. What if users could communicate directly with the plant, asking it questions and receiving responses in real-time, creating a more personalized and engaging interaction? 
\finalrv{Consider a scenario when someone feel sad and wish to confide but avoid human support due to social or privacy concerns,he then turn to an AI assistant for comfort and advice. However, such comfort, delivered through screens or unfamiliar hardware, often feels impersonal and detached, lacking the warmth of interactive engagement. What if users are able to communicate directly with a familiar, frequently hugged stuffed toy while sharing their troubles could foster a deeper emotional connection and enhanced support through tactile interaction?}

%% Our Proposal and Tech (so need to add the part on actual model used. Ref on the Introduction of Augmented Physics https://dl.acm.org/doi/pdf/10.1145/3654777.3676392)
In this paper, we propose Talking Spell, a novel approach to imbue objects with speech. Our innovative system leverages AI to create a unique interaction paradigm, enabling users to "cast spells" on detected objects, thus facilitating personalized communication with these items. Talking Spell includes two hardware components (Figure \ref{fig:teaser}): \textbf{scope} for object detection and \textbf{wand} for enabling the communication with a sensor to trigger the spell casting (voice input) and a vibration motor to hint and inform successful triggers (see Section \ref{sec:hardware}). 
% Our system employs a combination of a camera for visual input, sensor for... and audio output through headphones, embedding the application into everyday worn essentials and accessories such as necklace, headbands, and other wearables (Figure 1), thereby enhancing accessibility.

%% User Workflow and Design
% The user experience with Talking Spell begins when the user wears the application. The integrated camera detects objects within the environment, and users can "cast" a spell by recording the name of the object for a duration of ten seconds. This process allows the system to store the object's identity in its memory. Once the object is recognized in subsequent interactions, users can communicate with it, establishing a unique connection. This functionality opens avenues for practical applications, such as intelligent note-taking and reminders, as well as imaginative scenarios where users can interact with objects like plants or toys, enhancing the storytelling experience.

%% Contribution 1: Existing HOI focus on the computer vision technology, but how to utilize the technology and make it to a playful, actual interaction remains unsolved --> existing relevant works on anthromorphization works on application with display or virtual aids in interaction. 
The idea of anthropomorphism and AI companionship is not new \cite{Osawa2006, Osawa2012, Osawa2009}, but this paper contributes in three key ways. First, we introduce a novel human-object interaction (HOI) communication pipeline. While many studies have concentrated on HOI detection through computer vision methodologies \cite{Gkioxari2018, Chao2018}, they often neglect the exploration of actual application that facilitate HOI communication. Existing works on anthropomorphism and virtual companionship like \textit{displaying robot} \cite{Osawa2006} and \textit{Robovie} \cite{Osawa2012} mainly focus on controlling experiments involving single objects, specifically an iris board and a humanoid printer, respectively.  However, such approaches do not suffice for free form of conversation to all everyday objects, nor is it adequate to meet the technological limitations imposed by advances in AI and computer vision development. Therefore, we develop a pipeline to \textbf{acquaintance} \textit{(detect and 'meet' the object)}, \textbf{familiarization} \textit{(anthropomorphize the object by imbuing with speech - casting the talking spell)}, and \textbf{bonding} \textit{(recall conversation and develop bonding)}. To the best of our knowledge, our work is the first to explore and demonstrate this user-centric human-object communication generation and new paradigm for AI companionship. This is the first approach in anthropomorphizing everyday object with the help of AI and novel device design. 

%% Contribution 2: Design Space of Human object interaction tools
Second, we expand the design space of human-object interaction tools and anthropomorphism with our findings from a preliminary user study of Talking Spell. Our findings, derived from a comprehensive preliminary usability study, reveal that Talking Spell not only opens up innovative forms of interaction but also enhances creativity in conjunction with emerging AI technologies. We identify four key benefits that emerge from our research: (1) Talking Spell offers augmented interaction and accessibility, providing a seamless and inclusive user experience; (2) it allows personalized experiences through customized personas and voices, which strengthen emotional connections; (3) the playful and whimsical nature of the interactions encourages creativity by inspiring novel exploration; (4) it demonstrates robust contextual awareness, informing future designs of practical and engaging AI creative tools.

%% Contribution 3: Talking Spell as a multifunctional tool with deformable device
Third, we explore the potential applications of Talking Spell in both practical and fantastical contexts, emphasizing its versatility as both an intelligent assistant and a playful companion. We conclude that Talking Spell is instrumental with four use-case scenarios from \textbf{recreation} to \textbf{practical}: (a) \textbf{\textit{entertainment}}, (b) \textbf{\textit{creativity}}, (c) \textbf{\textit{companion}}, and (d) \textbf{\textit{utility}} (see Figure \ref{fig:scenarios}). These interaction intents not only enhance user engagement but also illustrate the transformative potential of AI companions across diverse settings. For example, in education, it can serve as a personalized tutor, while in creative fields, it acts as an inspiring brainstorming partner.

Finally, our main contributions are as follows:
\begin{enumerate}
\item Talking Spell, a tool for anthropomorphizing daily objects and imbuing object with speech.
\item A set of interaction possibilities that expand the design space for human-object interaction.
\item Insights to explore AI companions as versatile applications in four interaction intents from our user study (N=12).
\end{enumerate}

\section{Related Work}
\subsection{Virtual Assistants}
AI assistants play a crucial role in enhancing human experiences but often create barriers that inhibit direct user-object interactions. Although these systems are designed to facilitate human tasks, many rely on a third-person interface that complicates interactions, leading to a fragmented user experience. AI acts as a bridge yet isolates users from the objects they interact with \cite{Su2024}. This limitation curtails the potential of AI to enrich user experiences due to structural constraints in its processing paradigms.

Despite the intention to streamline communication, AI systems frequently complicate user interactions. While AI-mediated conversations can enhance empathy, they disrupt natural dialogue flow, resulting in less authentic connections \cite{Shama2023}. This intermediary role can diminish user confidence in engaging with technology and their surroundings \cite{Su2024}. Additionally, tools designed for visually impaired users often require mediated processing of intentions and environmental contexts, which can hinder situational awareness necessary for intuitive technology use \cite{Naayini2025}. The increasing reliance on AI as intermediaries highlights the need for systems that empower users to engage directly with objects, enhancing their agency.

AI systems, while effective in improving operational efficiency, often fail to foster intuitive interactions between users and their environments \cite{Jarrahi2018}. These systems typically prioritize functional tasks over emotional engagement, missing opportunities for richer interactions \cite{Duan2024}. The transactional nature of AI interactions diminishes genuine connections, lacking emotional depth \cite{Glikson2020}. To address these limitations, AI technology must evolve to balance functionality with emotional resonance, supporting relational user experiences \cite{Su2024}. By integrating emotional intelligence, AI can better understand and respond to nuanced human emotions, fostering deeper dialogues and connections \cite{Mossbridge2024}. Promoting genuine interactions is essential for refining human-object relationships and addressing the shortcomings of current AI systems.

\subsection{\finalrv{Designing AI Agent Interaction Interface}}
\finalrv{When interaction interfaces involve both physical and digital elements, the concept of a natural user interface (NUI) is frequently referenced, though prior research highlights the complexity of defining its "naturalness." Norman et al. \cite{norman2010natural} argue that NUIs are not inherently natural but are useful. O'Hara et al. \cite{NUI2013Ohara} expand on this, interpreting naturalness as intuitiveness, ease of use, and learnability rather than inherent naturalness. Vatavu\cite{nonnatural2025Vatavu} proposes a non-natural interaction design, intentionally diverging from users' intuitive expectations and physical-world naturalness to achieve efficient interaction while maintaining high usability and effectiveness. These studies question the terminology of "naturalness" but emphasize the usability of interfaces that bridge physical and digital realms. Consequently, in this paper, we prioritize system usability, effectiveness, and intuitiveness over "naturalness" as evaluation criteria.

Additionally, some researchers have focused on the transferability of agents in digital-physical interfaces. Ogawa et al. introduced the ITACO system \cite{ITACO}, a user-centered radial network of transferable agents that provides continuous contextual support within an environment. Building on this, Ravi Tejwani et al. \cite{tejwani2020migratabledefense,tejwani2020migratable} explored how agent transferability influences user perception. These studies anchor interactions around users, unifying context across different media (e.g., wearable devices or robots) to facilitate seamless interaction. Our work addresses a gap in prior research by incorporating the context of the media hosting the agents, enhancing user experience through more personalized and context-aware agent adjustments.}

\subsection{AI as Companion \& Emotional Connection}
Artificial Intelligence (AI) companions play an important role in fostering emotional connections \cite{Purington2017}, yet these interactions often remain superficial due to the inherent limitations of AI systems. While they can provide emotional support, AI companions typically offer generic experiences rather than personalized ones. Scholars suggest that deeper, authentic connections could enhance emotional engagement\cite{Roshanaei2024}. Recent research indicates that young adults find AI interactions helpful for stress relief and social pressure avoidance, despite the lack of genuine connection \cite{Zhang2025}. This highlights the complex nature of AI's role in emotional landscapes, presenting both opportunities and challenges.

AI companions often create an illusion of emotional support, primarily designed to satisfy immediate emotional needs without true depth. For instance, AI may simulate empathetic responses, giving users a false sense of understanding \cite{Curry2023}. Such interactions can prevent users from forming meaningful relationships with real individuals, potentially leading to emotional detachment \finalrv{\cite{rostami2023,george2023,efthymiou2025}}. While the accessibility and quick gratification of AI companionship are beneficial, they raise concerns about the authenticity of emotional connections. The inability of AI to fully understand human emotions results in interactions that may feel hollow, prompting reflection on our reliance on artificial entities, which lack the nuanced empathy found in human relationships \cite{Curry2023, Zhang2025}.

Additionally, the Media Equation Theory posits that people tend to treat computers and other media as if they were human, responding to them with social and emotional reactions \cite{Reeves1996}. This theory suggests that users may project their emotional needs onto AI companions, expecting them to fulfill roles that they are not designed to meet. As a result, AI companions struggle to establish genuine emotional bonds. Emotional theory suggests that emotional relationships require reciprocal interactions and mutual understanding, which current AI capabilities often fail to support \cite{Curry2023, Roshanaei2024}. Users may develop heightened expectations, leading to disappointment when AI companions do not meet their emotional needs \cite{Zhan2019}. To foster meaningful interactions, AI must better understand and respond to authentic human emotions \cite{Kirk2025}. As such, the discussion over anthropomorphism and object attachment is potentially vital in AI companionship.

\subsection{Anthropomorphism and Human-Object Interaction: Imbue Objects with Speech}
Anthropomorphism, the attribution of human-like traits, mental states, and behaviors to non-human entities, is a widely recognized method for fostering relationships between humans and AI \cite{Kühne2022, Salles2020}. This approach often manifests through naming and customizing appearances, both of which enhance user engagement. \finalrv{Naming, a culturally significant practice, has long been used to assign identity and personality to individuals, pets, and objects \cite{Horne1996, Mcconnell2005}, thereby strengthening socio-emotional bonds \cite{Stoner2018}.} In human-object interactions, naming items like cars or computers is common and influences first impressions, attitudes toward technology, and even purchasing intentions \cite{Boersma2019, Purington2017, Liu2021}. Furthermore, self-assigned names can deepen psychological ownership and reflect user identity, amplifying the object’s perceived value \cite{Stoner2018, Wrigley2021}.

Beyond naming, anthropomorphic appearances have proven effective in social robotics and AI interactions. Studies on robots \cite{Hegel2008} and chatbots \cite{Chong2021} demonstrate that human-like designs elicit positive evaluations and enrich human-robot interactions \cite{Jakub2015}. However, anthropomorphism carries risks, such as cognitive biases favoring machines in decision-making \cite{Proudfoot2011} or increased vulnerability in hybrid teamwork, as seen in military contexts where it may expose personnel to harm \cite{Carpenter2013}. These findings highlight the dual nature of anthropomorphism as both a facilitator and a potential liability in human-AI dynamics.

Existing applications, such as Displaying Robot \cite{Osawa2006} and Robovie \cite{Osawa2012}, explore anthropomorphism through controlled experiments with specific objects (e.g., an iris board and a humanoid printer). Scholars also proposed an anthropomorphization method using attachable human-like features such as eyes and arms to the targeted agent, and asked users to use non-verbal cues to interact with the agent \cite{Osawa2009}. While insightful, these studies are limited to single-object interactions and lack the flexibility to support free-form conversations with diverse everyday objects. Moreover, they do not fully address the technological advancements in AI and computer vision that enable broader applications. Despite the historical role of naming and anthropomorphism in shaping companionship, their combined impact on AI-driven object interactions remains underexplored. This study seeks to address this gap by examining how Talking Spell leverages these elements to enhance user-object relationships across multiple intents.

\section{Talking Spell: System Design}
\subsection{Overview}

In this section, we introduce Talking Spell, a machine learning-integrated system that endows objects with speech capabilities through the utilization of large language models (LLMs). This system empowers users to actively engage with their personal belongings, enabling both object recognition and verbal dialogue. Talking Spell addresses the constraints of conventional AI assistants, which are frequently tethered to specific hardware platforms, as well as the impersonal nature of existing communication networks. Moreover, it overcomes the limitations of AI companionship products that typically foster emotional connections solely with unfamiliar entities, establishing instead a user-centric framework for interaction.

Additionally, Talking Spell functions as a wearable AI assistant device. In contrast to smartphones or other non-portable AI assistants, its compact hardware design allows seamless integration into users’ everyday wearable accessories, rendering it unobtrusive. Through a three-phase process — Acquaintance, Familiarization, and Bonding — users can progressively deepen their relationships with familiar objects, leveraging pre-existing emotional attachments. This system introduces a novel paradigm for human-object interaction. This paper primarily explores four interaction intents — entertainment, utility, creativity, and companionship — investigating the influence of Talking Spell on the connections between individuals and objects within these contexts.
\subsection{Hardware Design}
\label{sec:hardware}

In this section, we present the hardware design of Talking Spell, which comprises two distinct components: the scope and the wand. The scope operates as an observational module, akin to an "eye," responsible for capturing images and executing object detection tasks. The wand, in contrast, functions as an activation mechanism, initiating the dialogue system powered by a large language model (LLM). In the subsequent section, we provide a comprehensive overview of these components, while the final section elaborates on the considerations and implementation of our portable design.
\subsubsection{Scope} \label{sec:scope}

\begin{figure}
    \centering
    \includegraphics[width=\linewidth]{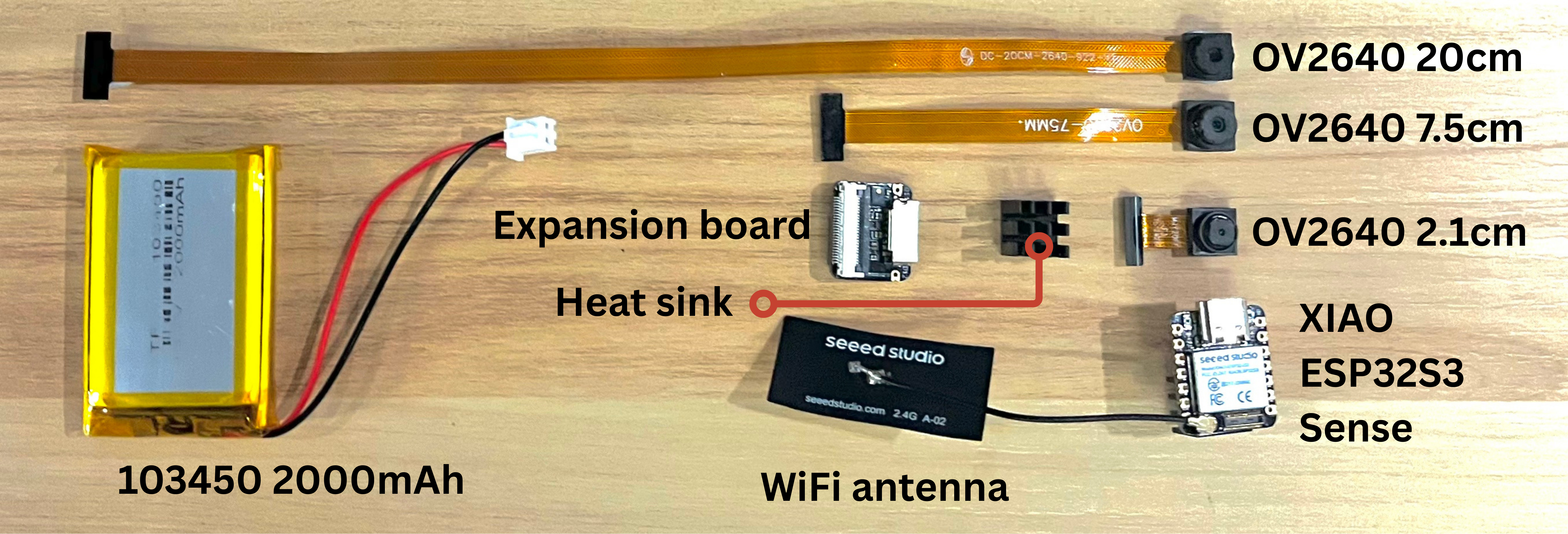}
    \caption{The exploded view diagram of the scope’s components.}
    \label{fig:scope-design}
\end{figure}

\autoref{fig:scope-design} illustrates the exploded view of the components utilized in the design of the scope. The scope employs the XIAO ESP32S3 Sense as its primary processing board, integrated with a B2B connector integrated expansion board connecting the camera OV2640. The scope facilitates video streaming via a web server to a designated IP address by establishing a connection to a computer-generated hotspot. A local computer can retrieve real-time image frames by accessing this IP address. To maintain image quality, the scope should be positioned at eye level, ensuring that the captured images closely align with the visual perspective of the user. Additionally, to accommodate various wearable needs, we selected three sizes of the OV2640 camera module: 2.1cm, 7.5cm, and 20cm.
\subsubsection{Wand}  \label{sec:wand}

\begin{figure}
    \centering
    \includegraphics[width=\linewidth]{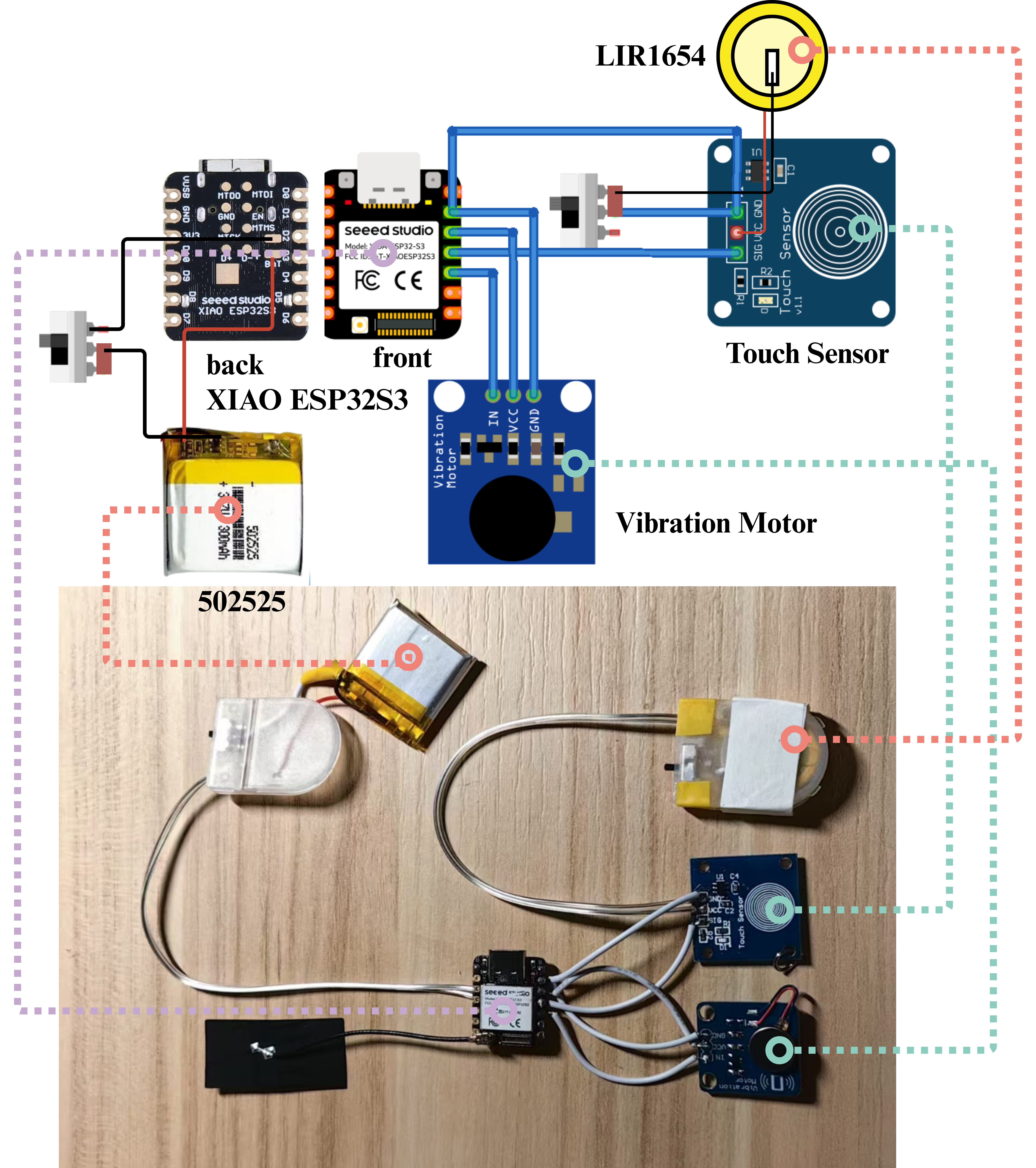}
    \caption{Wand Design. Top: Circuit schematic diagram. Bottom: Exploded view diagram of components.}
    \label{fig:wand-design}
\end{figure}

\autoref{fig:wand-design} presents the design of the wand, with the upper section depicting the circuit schematic diagram and the lower section illustrating an exploded view of the wand’s components, captured as a real-world assembly diagram. The wand employs the XIAO ESP32S3 as its primary processing board. Its design integrates this main board with two toggle switches, a touch sensor module (TTP233), a vibration motor module, and two 3.7V batteries (LIR1654, 502525 3.7v 300mah Li-Ion Polymer Battery), which independently power the components. Upon detecting a signal from the touch sensor, the main board transmits this information via Bluetooth to the computer, initiating a dialogue request. Subsequently, it receives a recording initiation signal from the computer, activating the vibration motor module. This sequence facilitates a user interaction workflow wherein pressing the touch sensor commences recording, vibration delivers haptic feedback, and releasing the sensor terminates the recording process.
\subsubsection{Portable Design}

Our design fully embodies the pursuit of portability. To ensure that the hardware of Talking Spell seamlessly integrates into users' daily lives, we made careful trade-offs in size. We chose the XIAO ESP32S3 Sense as the visual module, which measures only 21mm x 18mm, making it smaller than other ESP-CAM modules. Instead of powering through the USB-C port, we utilized the external removable 103450 Lithium-ion polymer rechargeable Battery supply on the back. We also reduced the image size to 320x320 pixels to ensure image transmission quality via Wi-Fi (see \autoref{fig:scope-design}).

In the design of the wand, we initially considered using Wi-Fi for signal transmission. However, after testing, we found that the batteries that met size requirements could not provide sufficient current to activate the Wi-Fi module. As a result, we opted for Bluetooth for low-power transmission. Given the small size and capacity of the chosen batteries, they could provide only a limited discharge current. To prevent functional loss due to current sharing among modules, we implemented separate power supplies for the touch module and the main board to ensure the proper operation of each module.

\textcolor{black}{After multiple iterations, we ultimately ended up with a design size of 35mm x 25mm x 25mm.}
\subsection{Usage Flow and Implementation}

In this section, we initially conceptualize the three distinct stages of interaction: Acquaintance, Familiarization, and Bonding. Subsequently, we present the comprehensive user workflow, followed by a detailed discussion of the implementation specifics for each stage.

To conceptualize the three stages, we draw upon the analogy of establishing a friendship between two individuals:

\textit{When encountering an unfamiliar individual, the initial phase entails observing them and forming a preliminary impression (\textbf{Acquaintance}). As interactions progress and deepen, even after a period of separation, the individual can be recognized upon a subsequent encounter, with this recognition linked to the prior impression (\textbf{Familiarization}). Throughout this progression, sustained dialogue serves to continuously reinforce the relational bond between the two parties (\textbf{Bonding})}.

Within the framework of Talking Spell, upon first receiving the device, the user equips the scope and wand and initiates the process by selecting a nearby object to establish an "acquaintance" with it. Following this initial interaction, Talking Spell generates an "impression" of the object, constructing an anthropomorphic persona (Acquaintance). Even after a period during which the object is absent from the user’s or scope’s field of view, the scope retains the capability to recognize the object upon its reappearance and accurately retrieve the previously established anthropomorphic persona (Familiarization). By activating the wand, the user can imbue the recognized object with speech capabilities, enabling verbal communication. This personified representation influences the selection of the voice tone and governs the stylistic characteristics of the object’s responses during the interaction.

\subsubsection{\textbf{Acquaintance}: Data Collection and Impression creation}

\begin{figure}
    \centering
    \includegraphics[width=\linewidth]{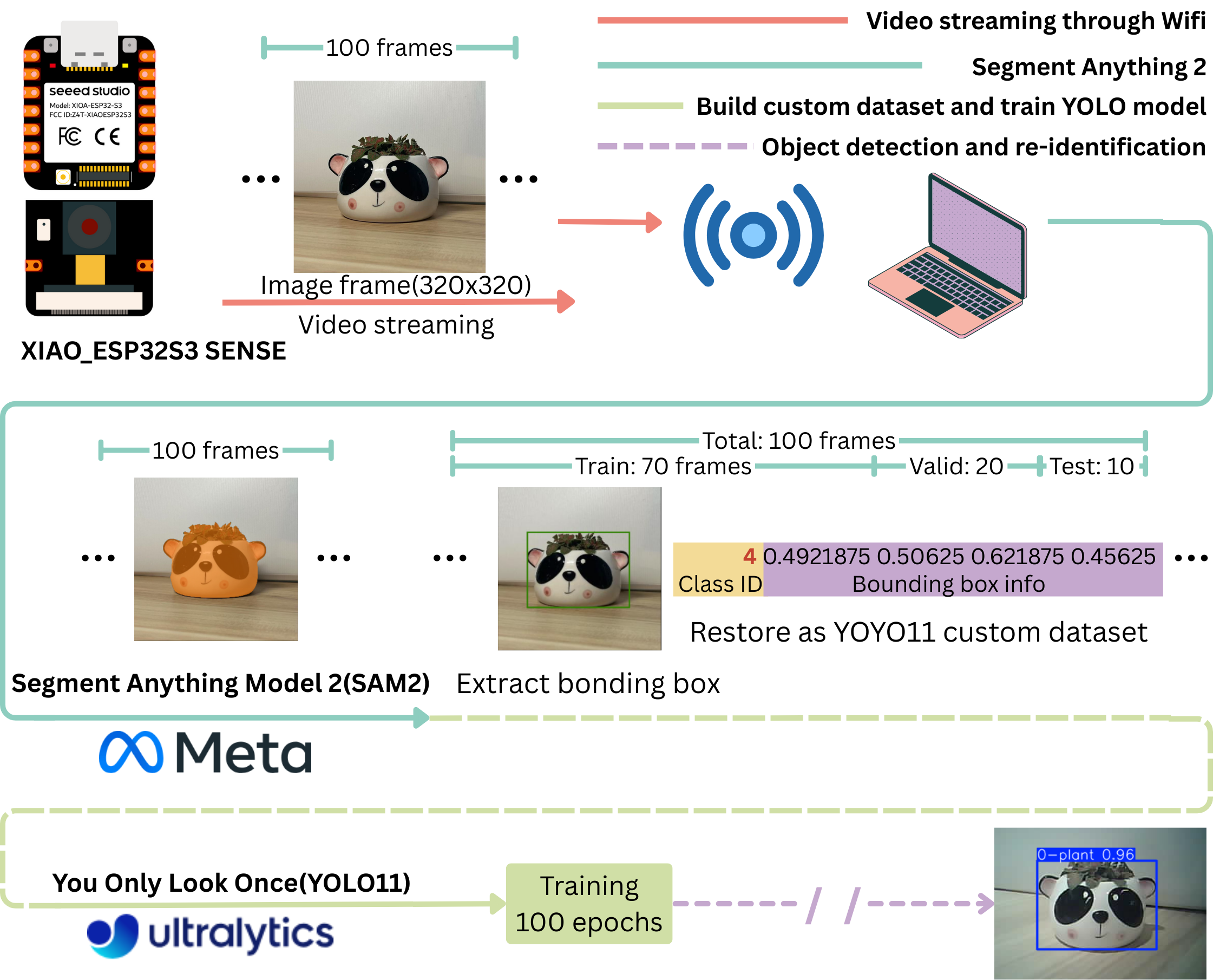}
    \caption{Acquaintance workflow for Talk Spell. The dashed section refers to the connection with the familiarization stage.}
    \label{fig:acquaintance}
\end{figure}

As outlined in the preceding section, the Wi-Fi module integrated into the system captures images and transmits them to a designated IP address via a Wi-Fi network (see \autoref{fig:acquaintance}). A local computer retrieves these image frames by accessing the specified IP address. When a user selects an object to imbue with speech capabilities, the computer stores 100 image frames starting from receiving the capture signal. These frames are processed using the segmentation algorithm from the Segment Anything 2 model \cite{SAM2} to extract the primary content (mask) of the object. By computing the bounding box for the extracted mask, the resulting data is formatted to be compatible with the training requirements of the YOLOv11 model.

\begin{figure}
    \centering
    \includegraphics[width=\linewidth]{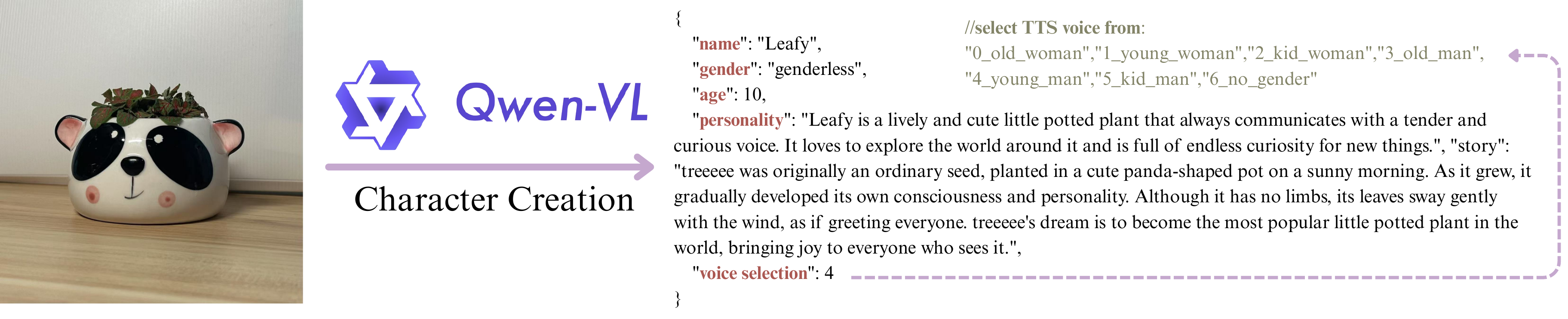}
    \caption{Create anthropomorphic persona using large visual language Model}
    \label{fig:qwen}
\end{figure}

Concurrently, a captured video frame is processed by QWEN-VL \cite{qwen-vl}, a large-scale vision-language model accessed via the Aliyun QWEN-VL API. Through a predefined prompt, an anthropomorphic persona of the object is generated and stored in a JSON file. This representation encapsulates attributes such as the object’s name, gender, age, personality, background story, and selected voice characteristics. The large vision-language model is tasked with selecting an appropriate voice based on these attributes. In this study, the available voice options include elderly female, young female, child female, elderly male, young male, child male, and gender-neutral voices (see \autoref{fig:qwen}).

\subsubsection{\textbf{Familiarization}: Re-identification and Impression retrieval}

After preparing the dataset in a format compatible with YOLOv11 training, we  adopted a train-validation-test split ratio of 7:2:1 and forward them for training\cite{yolo}. This training process leveraged the services provided by Ultralytics\footnote{https://github.com/ultralytics/ultralytics} for 100 epochs with patience=25 for early stopping.%每次遇到和新的object建立连接时，我们都以先前的training model作为pretrained model。 
When we establish a connection with a new object, we use the previously trained model as a pretrained model to support the process.
The resulting model empowers the system to re-identify objects upon their reappearance within the camera view. Furthermore, based on the detected object class, the system retrieves the corresponding JSON file generated during the acquaintance phase, enabling context-aware processing.

\subsubsection{\textbf{Bonding}: LLMs and TTS supported dialogue system}

\begin{figure} [h]
    \centering
    \includegraphics[width=\linewidth]{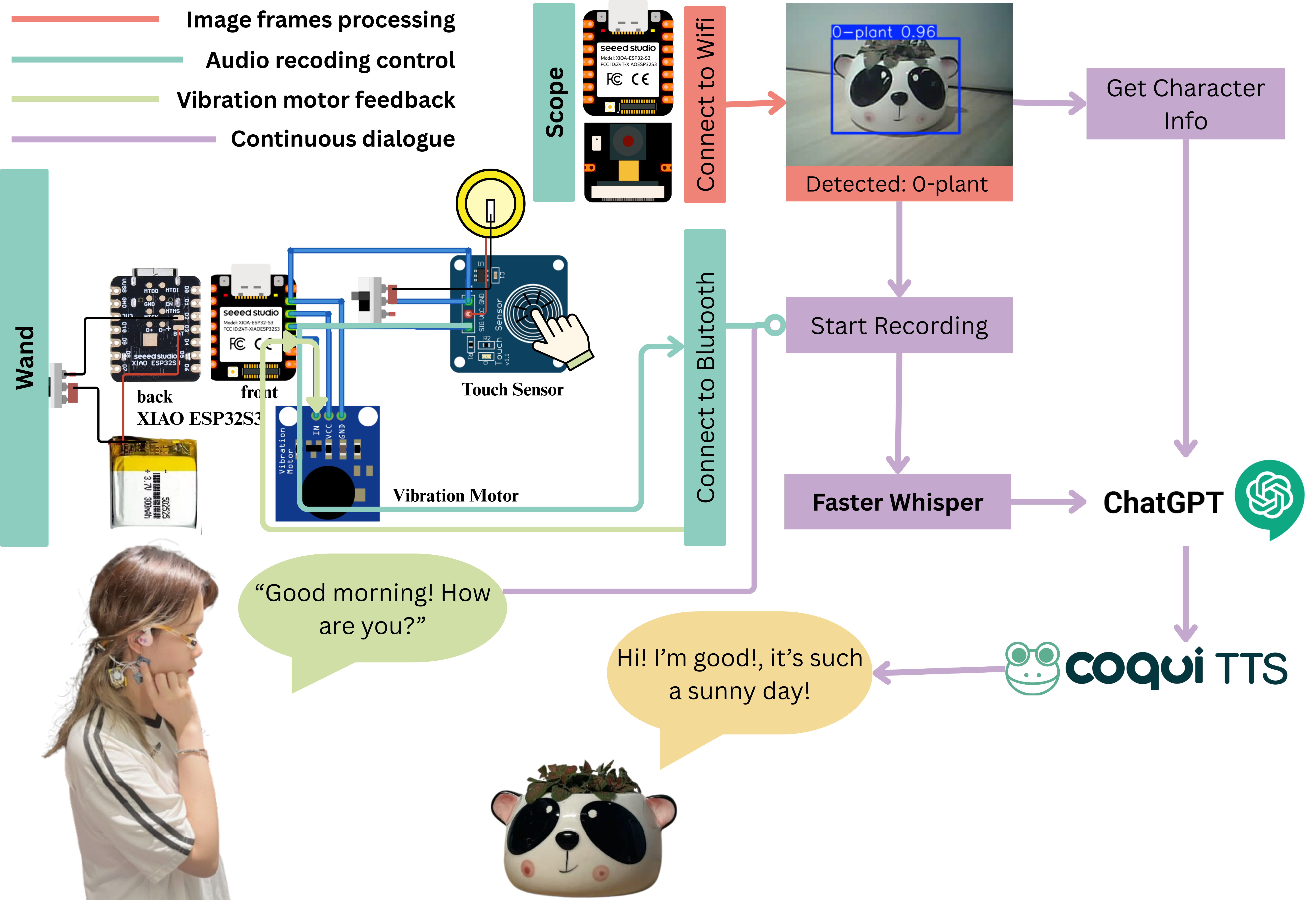}
    \caption{Usage flow of the bonding stage.}
    \label{fig:bonding}
\end{figure}

Following the training phase, the system is capable of re-identifying known objects with improved accuracy. The algorithms executed on the computational platform further enhance this functionality by integrating speech capabilities, enabling users to actively initiate dialogues and iteratively strengthen the bonding process. Figure \ref{fig:bonding} illustrates the workflow of a single bonding cycle. During this process, the system continuously monitors for the detection of recognized objects and dynamically adapts its persona based on the object’s category, thereby facilitating a contextually responsive interaction.

By interacting with the touch sensor on the wand, a recording request signal is transmitted via Bluetooth to the computer. If an object is detected, the recording process initiates, and a success signal is sent back to the wand, simultaneously activating the vibration motor to provide haptic feedback. The recording ceases once the user releases the touch sensor, and the audio is saved as an MP3 file. This MP3 file is then processed using the Faster Whisper algorithm\footnote{https://github.com/SYSTRAN/faster-whisper?tab=readme-ov-file} for speech-to-text conversion. The resulting text is combined with the persona to generate a formatted JSON document, where the persona serves as the system prompt. Subsequently, the Azure ChatGPT Turbo-3.5 API is invoked to process the json input, and the generated response is converted to speech using the Coqui TTS framework\footnote{https://github.com/coqui-ai/tts}, employing the voice timbre selected within the persona. This audio output is delivered through the user’s headphones.
To ensure efficient audio transmission speed, we instruct the GPT model to insert markers at specific positions—such as the end of a sentence or natural breathing pauses—during response generation. Prior to text-to-speech (TTS) conversion, these markers are detected to segment lengthy responses into smaller parts, enabling parallel TTS processing. This approach ensures the fluency of the dialogue.
Both the user’s voice input and ChatGPT’s responses are appended to the JSON file, forming a chat history that serves as input for subsequent interactions, thereby endowing the object with short-term memory. 

\begin{figure}[h]
    \centering
    \includegraphics[width=\linewidth]{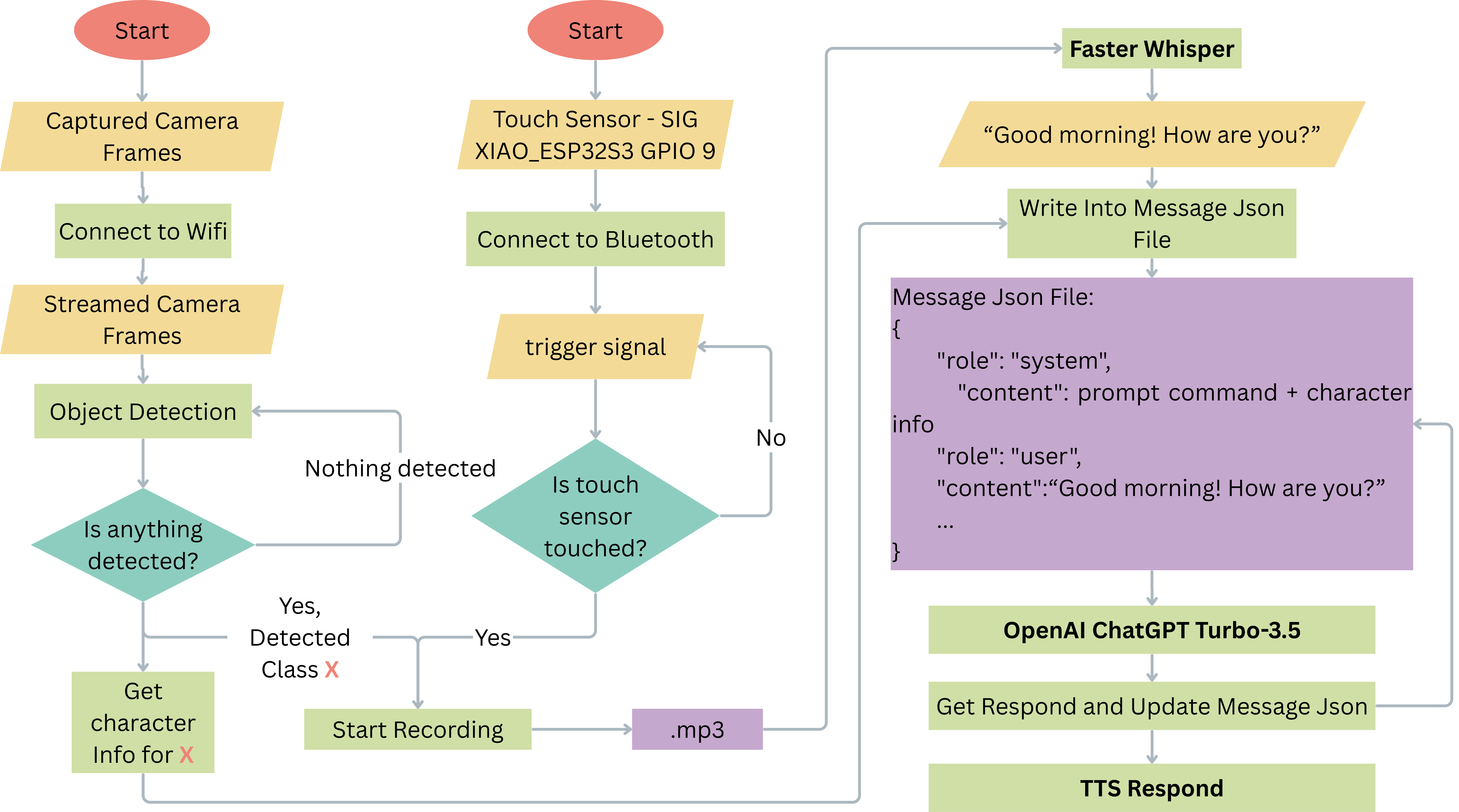}
    \caption{Flowchart of the bonding stage. Oval:start/end; \textcolor{mygreen}{Green} rectangle: processing; \textcolor{mypurple}{Purple} rectangle: generated document; Diamond:decision; Parallelogram: input/output}
    \label{fig:flowchart}
\end{figure}

 In this study, the system retains a maximum of 10 conversation records between the user and the object. When the dialogue exceeds 5 cycles, a "last-in, first-pop" principle is applied to manage the memory buffer. \autoref{fig:flowchart} illustrates the workflow of the Bonding stage, encompassing the processing steps, decision-making processes, involved files, and transmitted signals.

\subsubsection{Support for establishing connections with multiple objects and multi-lingual interaction}

Our system facilitates the establishment of emotional relationships with multiple objects. To this end, we implement independent classification and precise detection mechanisms for distinct objects, ensuring that each object possesses a unique persona and that memory associations between them remain separate and unambiguous. As depicted in the right panel of \autoref{fig:scenarios}, we utilize Talking Spell as a central anchor to construct a user-centric, radial communication network encompassing multiple objects. Users can seamlessly transition from interacting with one object to another, while our system supports multilingual interactions in both Chinese and English.

\subsection{Use-Case Scenarios }\label{sec:scenario}
\begin{figure}[h!]
  \includegraphics[width=\linewidth]
    {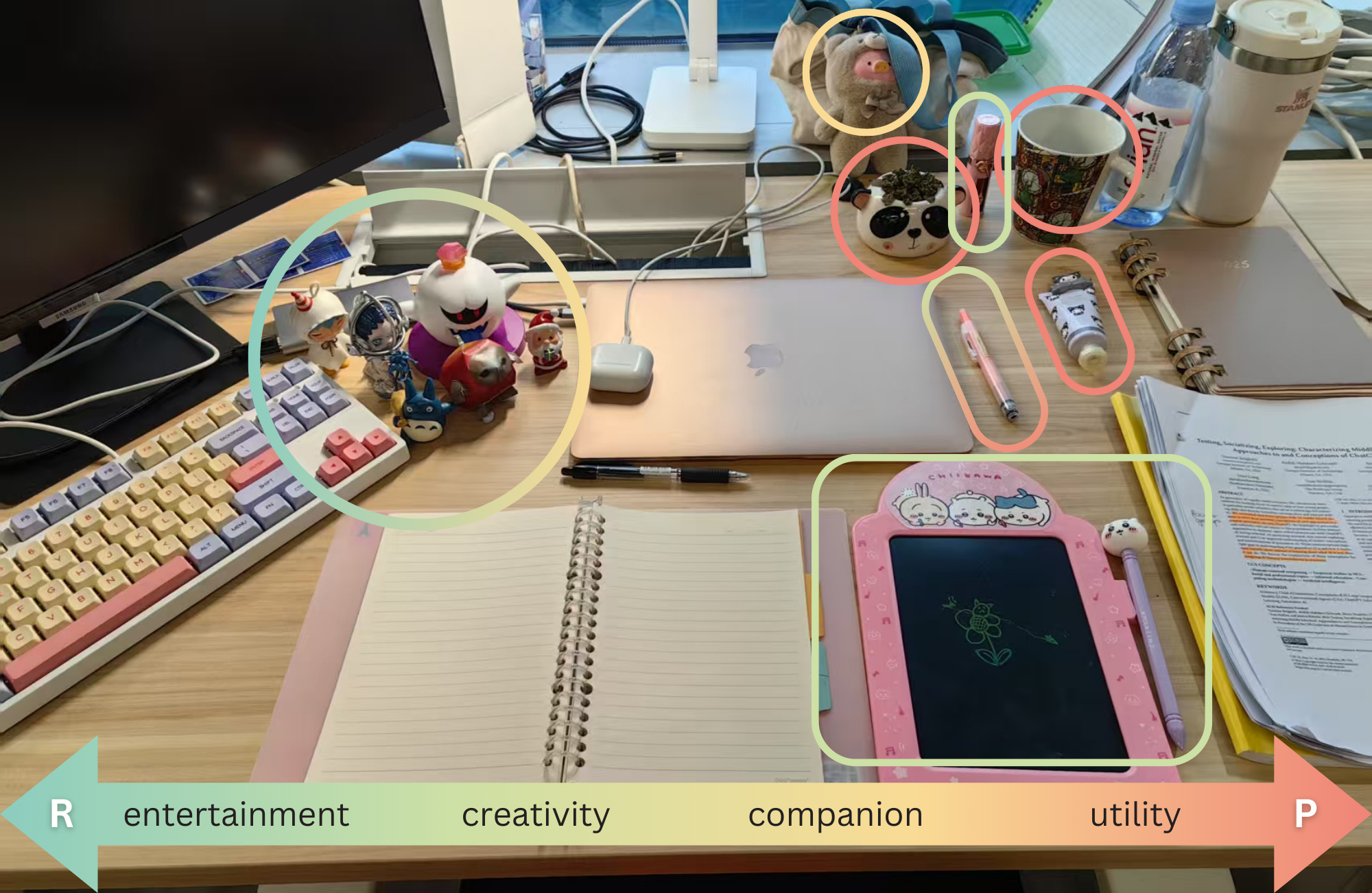}\hfill
      \caption{Talking Spell encompasses four primary interaction intents, spanning recreational (R) to practical (P) applications: \textcolor{myteal}{entertainment}, \textcolor{mygreen}{creativity}, \textcolor{myyellow}{companion}, and \textcolor{myred}{utility}. }
  \label{fig:objects}
\end{figure}

\begin{figure}[h]
    \includegraphics[width=\linewidth]
    {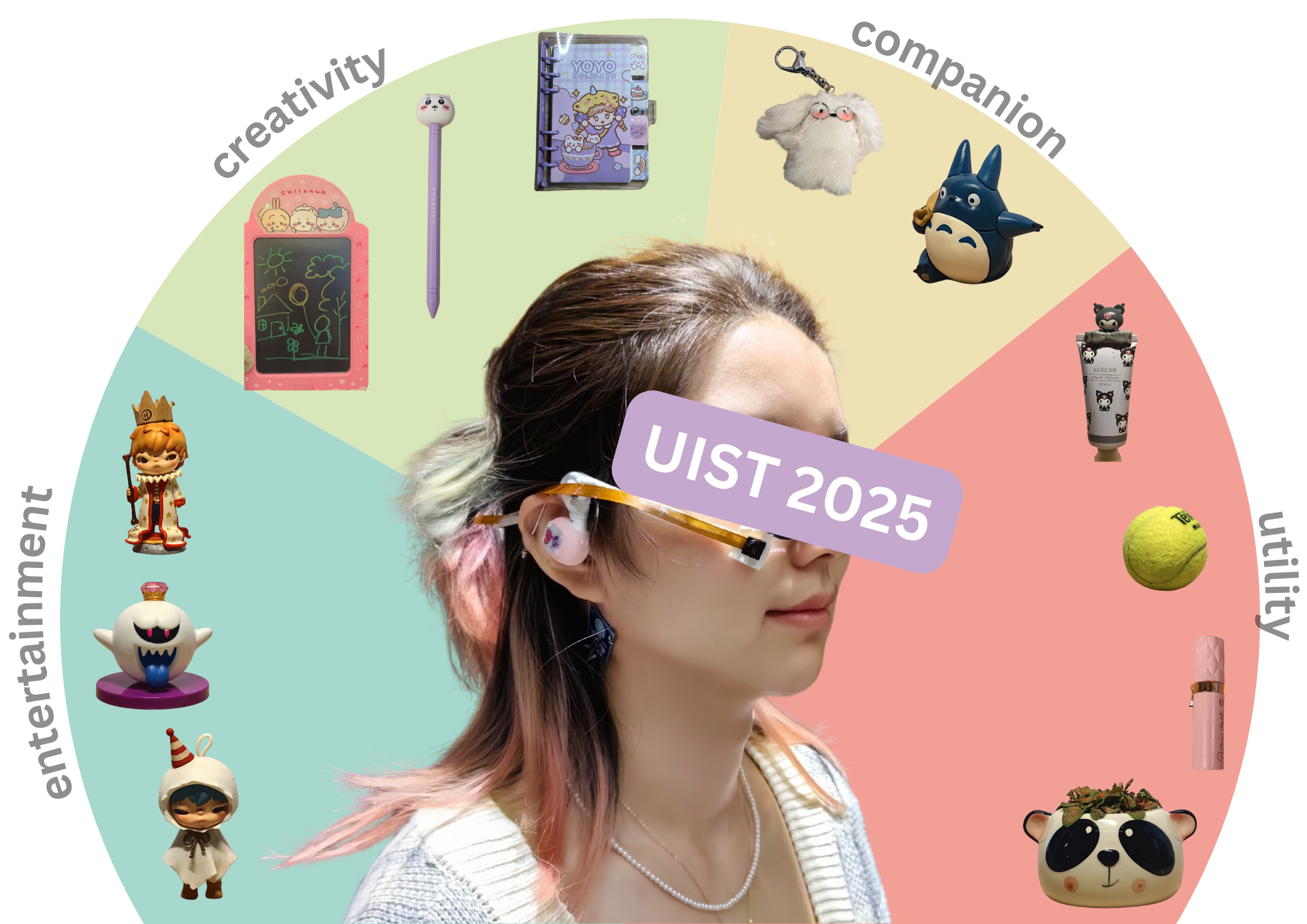}
  \caption{Talking Spell follows the user-centric radiative network structure, positioning the user as the centroid in the interaction flow, where user interacts with the objects directly without the interfere from any VAs. }
  \label{fig:scenarios}
\end{figure}

To effectively demonstrate the versatility of our design, we present four distinct use-case scenarios corresponding to the four main interactions intents identified — namely, \textbf{entertainment}, \textbf{creativity}, \textbf{companion}, and \textbf{utility} — as illustrated in Figure \ref{fig:objects}, showing how users can interact with various objects on a typical office desk via Talking Spell. These scenarios underscore the adaptability and broad applicability of our approach. This section describes the four intents from recreational to practical application, highlighting our user-centric radiative network structure (Figure \ref{fig:scenarios}), where user is centroid in the interaction paradigm.  

\begin{figure}[h!]
    \centering
    \includegraphics[width=\linewidth]{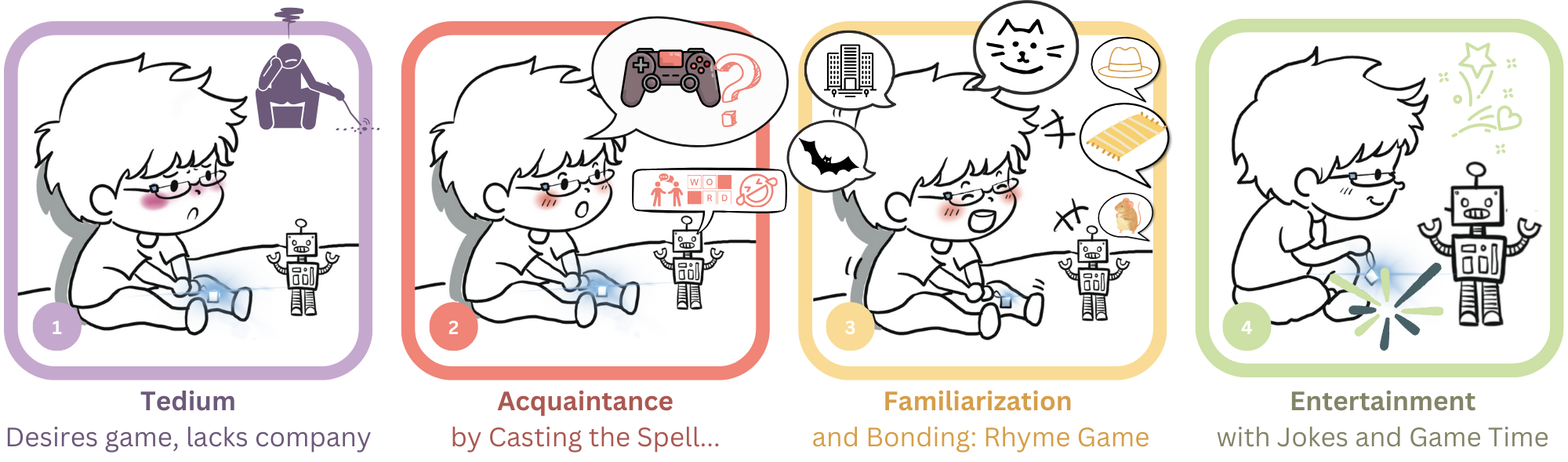}
    \caption{Stages in Entertainment with Talking Spell: from tedium to companionship through engaging games.}
    \label{fig:entertainment}
\end{figure}

\textbf{\textit{Entertainment}.}
The entertainment interaction intent within the Talking Spell refers to the capacity to deliver amusement and joy, thus enriching user engagement through dynamic conversational exchanges. This scenario is particularly valuable in addressing moments of boredom or isolation, as depicted in Figure \ref{fig:entertainment}. It depicts the four stages of emotional connection in entertainment, which a boring youngster can play Rhyme Challenge with his robot via Talking Spell.

Imagine being stuck in a room alone, without company or the ability to play multiplayer games. In such a context, traditional entertainment options, such as mobile games, may not be feasible or satisfying, especially when alone and without external resources. Talking Spell addresses this limitation by transforming object into active participant in recreational activities. Objects gain the ability to facilitate games and light-hearted interactions tailored to a solitary user. 

Examples include classic activities such as \textit{Solitaire}, \textit{Spelling Word Challenges}, \textit{the Alphabet Insult Game}, \textit{Twenty Questions}, or a \textit{Whimsical Twist on  Paper, Scissors, Stones} like \textit{Rock, Paper, and Whatever}. Beyond structured games, the entertainment intent extends to spontaneous and playful exchanges between the user and objects, particularly in the storytelling and role play scenarios. This may involve the object sharing light-hearted anecdotes tied to its own "experiences," offering fun facts or trivia about similar objects, or recounting favorite routines associated with its use. For example, a mundane bric-a-brac can humorously narrate its "adventures" or provide intriguing historical tidbits about its kind, enhancing the user’s amusement. 

\textbf{\textit{Creativity (co-work and co-design)}.}
The creativity interaction intent within Talking Spell emphasizes the collaborative nature of co-design and co-work between users and anthropomorphized objects, fostering an innovative environment for artistic and intellectual exploration. Unlike conventional AI virtual assistants that operate as detached entities, Talking Spell imbues physical objects with agency, enabling direct and engaging interactions that enhance creative expression. 

\begin{figure}[h]
    \centering
    \includegraphics[width=\linewidth]{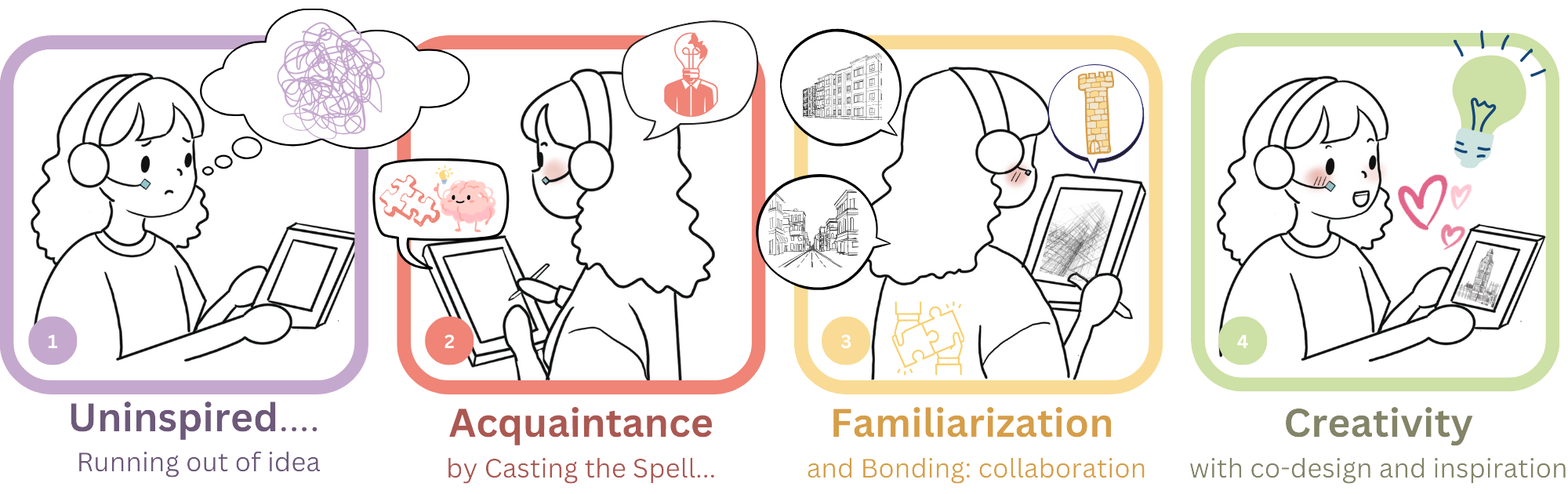}
    \caption{Stages in Creativity with Talking Spell:  from creative stagnation to inspiration through co-work and co-design.}
    \label{fig:creativity}
\end{figure}

This application scenario is exemplified in Figure \ref{fig:creativity}, which depicts an uninspired artist who collaborates with her drawing pad to co-design a novel artwork through Talking Spell. The protagonist engages with the drawing pad not merely as a passive tool but as a co-creator equipped with AI-driven capabilities tailored for artistic endeavors. This anthropomorphic method contrasts sharply with traditional AI approaches, where interactions often lack physical immediacy or contextual relevance. Talking Spell allows the drawing pad to be a dynamic collaborator, offering a more immersive and intuitive experience.

The creativity intent encompasses a broad spectrum of activities that leverage the object’s animated presence to stimulate imagination and productivity. These include artistic or creative projects directly involving the object (e.g., \textit{co-drawing or sculpting}), \textit{crafting stories} inspired by the object's characteristics, or exploring musical influences and connections tied to their identity when it comes to musical instruments. Additionally, Talking Spell can enable the object to generate creative prompts and ideas, provide comments, opinions, and constructive feedback, and assist in design simulations or prototype development. Practical applications extend to creating mood boards, supporting content creation, and even facilitating virtual workshops, where the object serves as both a tool and a collaborator. As such, Talking Spell redefines the creative process, positioning everyday objects as active contributors rather than inert instruments. This offers a novel paradigm for creativity in a user-centric context.

\textbf{\textit{Companion}. }
The companion interaction intent is to transform objects into sources of emotional support, companionship, and meaningful interaction, thereby addressing the user’s psychological and social needs. This intent is particularly significant in scenarios where users seek solace or connection. 

\begin{figure}[h]
    \centering
    \includegraphics[width=\linewidth]{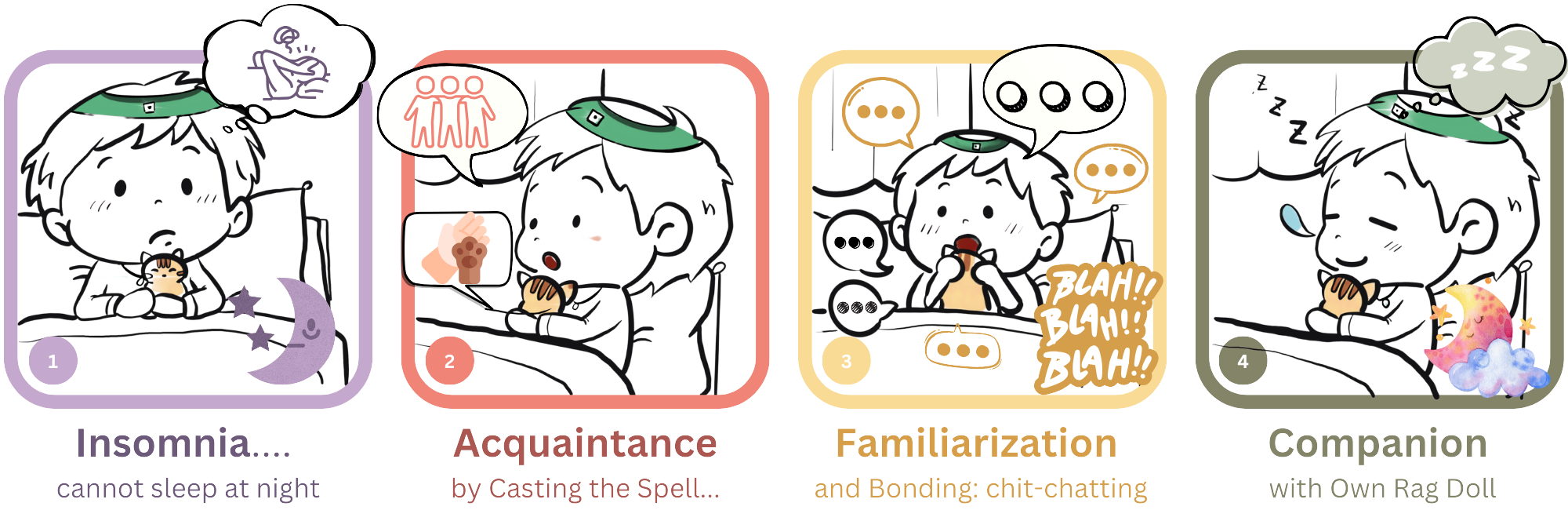}
    \caption{ Stages in Companion with Talking Spell: from isolated to accompanied. }
    \label{fig:companion}
\end{figure}

One such use case scenario, as illustrated in Figure \ref{fig:companion}, features an insomnia-afflicted protagonist who finds comfort in late-night conversations with a rag doll, a cherished object that has accompanied since birth. The pre-existing emotional bond between the user and the doll, rooted in years of personal history, enables the object to engage in a more profound, supportive dialogue.

This anthropomorphic approach distinguishes Talking Spell from traditional AI virtual assistants, which typically lack the physical presence and sentimental resonance of a tangible object. In our depicted scenario, the rag doll transcends its role as a passive keepsake, becoming an active conversational partner that alleviates the protagonist's sleeplessness through deep and emotionally resonant exchanges. 

The activities facilitated under the companion intent encompass a wide range of interactions that reinforce the object’s role as a mental and emotional ally. These include sharing memories associated with the object, recounting how it was acquired, and exploring its emotional significance to the user's milestones. Conversations may also relate to questions, advice, or even secrets the object might "hold" or offer in its animated state. Furthermore, the object can embody familial significance—representing traditions, superstitions, or unique characteristics that distinguish it—enhancing its status as a companion with a rich narrative identity. This highlights how Talking fosters psychological well-being, leveraging the power of anthropomorphism to bridge the gap between the inanimate and the deeply personal.

\textbf{\textit{Utility}. }
The utility interaction intent in Talking Spell focuses on transforming everyday objects into useful tools that improve practicality and support daily tasks. By endowing objects with functions like note-taking, recording, progress tracking, and reminders, objects can act as active assistants instead of just passive items. Talking Spell enhances object functionality to better meet user needs.

\begin{figure}[h]
    \centering
    \includegraphics[width=\linewidth]{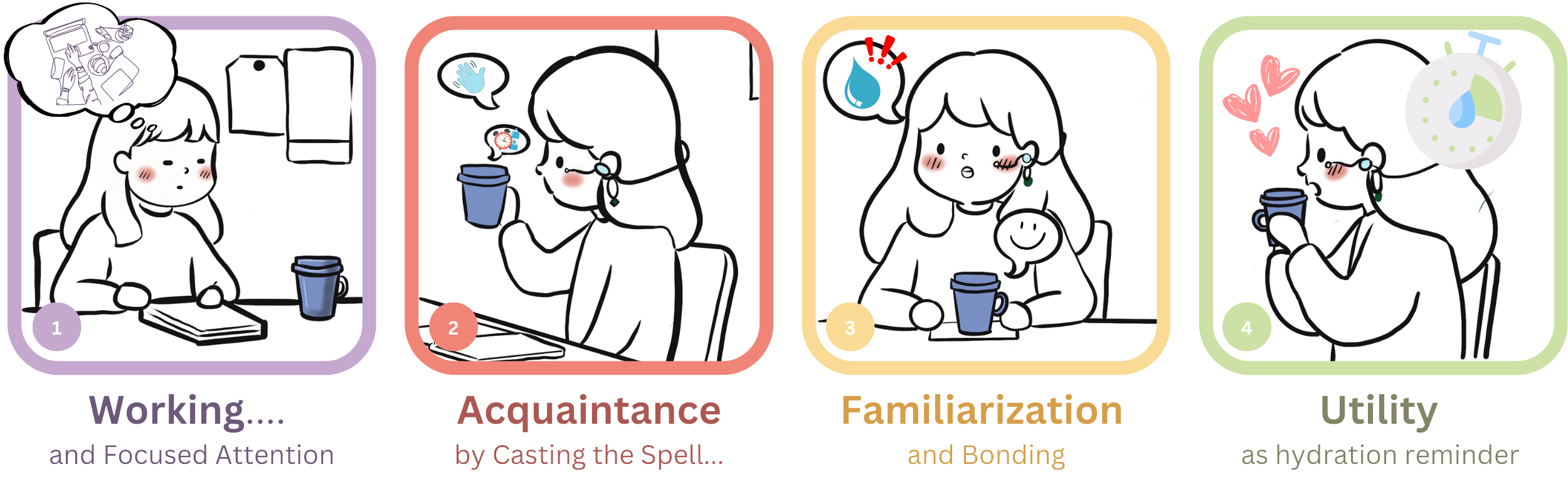}
    \caption{Stages in Utility with Talk Spell: from a decoration to functional assistant.}
    \label{fig:utility}
\end{figure}

One exemplary use-case scenario, depicted in Figure \ref{fig:utility}, features a protagonist deeply engrossed in a focused work session, surrounded by a busy and demanding environment. In this context, Talking Spell imbued her cup with speech and turned it to a hydration reminder that prompts her to pause and drink, thereby supporting her well-being amidst a task-heavy schedule. 

Examples of utility include a plant that reminds users to relax, a water bottle that tracks hydration, and a notebook that takes notes. Each object is designed to utilize its natural features for practical benefits. For instance, a plant could monitor stress levels and suggest breaks, while a notebook might record ideas or track project milestones. These scenarios highlight how Talking Spell can easily fit into daily routines, providing reminders and updates that improve productivity and organization. Unlike traditional tools that need manual input or external devices, this anthropomorphic approach integrates these functions directly into the objects, creating a more intuitive user experience.

This intent amplifies the practical value of everyday items and balances efficiency with a personalized, object-centric design. Such transformations highlight the innovative potential of Talking Spell in optimizing both individual performance and routine management.

\subsection{Technical Evaluation}

\begin{figure}
    \centering
    \includegraphics[width=\linewidth]{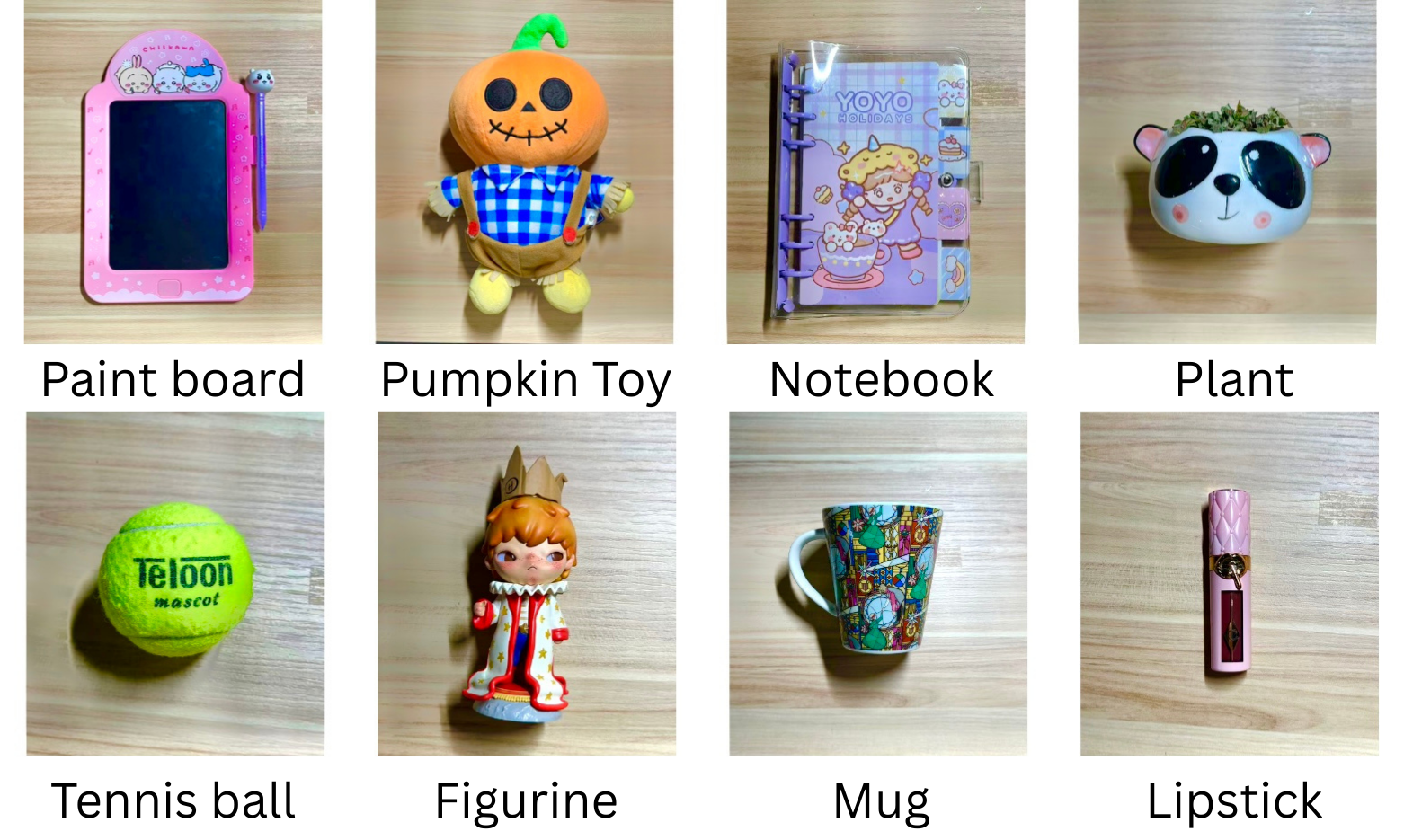}
    \caption{Objects involved in the technical evaluation session.}
    \label{fig:eval_obj}
\end{figure}

In this section, we conduct a technical evaluation of our system by dividing it into three components: visual processing, persona generation, and dialogue system. To ensure the generalizability of the results, we select eight distinct objects for testing, as shown in \autoref{fig:eval_obj}. All local processing for the tests is performed using an NVIDIA GeForce RTX 3080 Ti Laptop GPU with 16384 MiB of memory.
\subsubsection{Method}

\begin{figure}
    \centering
    \includegraphics[width=\linewidth]{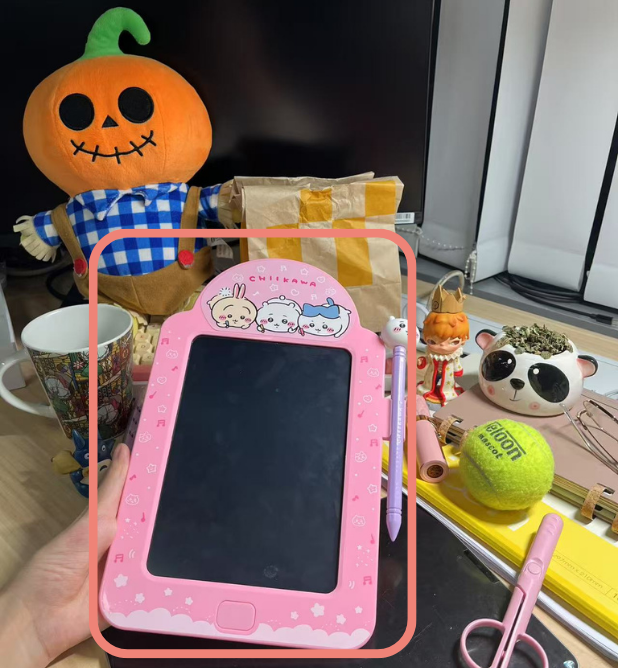}
    \caption{Example Environments for Evaluation Experiments. Segmentation/Detection Subject: Drawing Board.}
    \label{fig:env}
\end{figure}

\textbf{Vision Processing. }We follow the standard Acquaintance - Familiarization workflow to sequentially process the selected objects through data collection, segmentation, data integration, and model iteration. During this process, we record and analyze any potentially significant data, including the time taken for image processing and quantitative results from model iterations. To align with real-world application scenarios, we conduct data extraction and detection in environments with significant "noise" (see \autoref{fig:env}), and the user study is performed under similar conditions to ensure ecological validity. After completing all model iterations, we evaluate the detection accuracy across various categories in real-world testing scenarios.

\textbf{Persona Generation}
To evaluate the quality of the anthropomorphic personas generated by our algorithm, we input the same prompts and images into other large-scale models, namely GPT-3.5 and Grok-3, to generate personas for the same objects. We compare the generated results in terms of voice tone selection and age. Researchers establish a scoring mechanism focusing on the generation of personality and background story, with scores ranging from 1 to 7 (1: does not meet expectation at all, 7: most meets expectation). The evaluation rubric centers on \textit{which persona suits your expectation the most when imagining talking to the objects}. We recruited three participants (1 female, 2 male) and provided them with a randomized table to score the personas independently.

\textbf{Dialogue System}

In this section, we evaluate only the usability and response speed of the dialogue system. We randomly select four objects and conduct multiple rounds of dialogue to test their performance.

\subsubsection{Results}
\textbf{Vision Processing. }
Our system performs the image capture and segmentation stages without errors, with each object taking 17 seconds from the segmentation task to integration into a formatted database. We simulate real-world usage scenarios by conducting training each time a new category is introduced. The training parameters are presented in \autoref{tab:train}, where a plus sign indicates that each training session builds upon the results of the previous one. Four of the categories trigger early stopping around 60 epochs. The training time increases with the number of categories, but the overall duration remains short (less than 10 minutes). The average processing time per image is 3.04 ms.

\begin{table}[h]
    \centering
    \caption{Statistical result from interactive training}
    \label{tab:train}
    \begin{tabular}{@{}lcccccccc@{}}
        \toprule
        Category           & Training Epochs & Time (hours) & Speed (ms) \\ \midrule
        Paint board        & 61               & 0.032        & 2.5        \\
        +Pumpkin Toy        & 62               & 0.046        & 3.3        \\
        +Notebook           & 100              & 0.088        & 2.8        \\
        +Plant              & 64               & 0.063        & 3.2        \\
        +Tennis ball        & 58               & 0.071        & 3.7        \\
        +Figurine           & 54               & 0.077        & 3.0        \\
        +Mug                & 83               & 0.133        & 3.3        \\
        +Lipstick           & 92               & 0.165        & 2.5        \\ \bottomrule
    \end{tabular}
\end{table}

\autoref{tab:val} presents the performance of the model, encompassing eight categories, on the validation set. The results demonstrate a bounding box precision exceeding 99\%, a mean average precision (mAP) of 0.995, and outstanding performance across various IoU thresholds.
\begin{table}[h]
    \centering
    \caption{Yolo11n validation summary (8 classes in total)}
    \label{tab:val}
    \resizebox{0.48\textwidth}{!}{
    \begin{tabular}{@{}lccccccc@{}}
        \toprule
        Class       & Images & Instances & Box (P) & R   & mAP50 & mAP50-95\\ \midrule
        all         & 160    & 160       & 0.997   & 1   & 0.995 & 0.969 \\
        0-board     & 20     & 20        & 0.999   & 1   & 0.995 & 0.914 \\
        1-pumpkin   & 20     & 20        & 0.997   & 1   & 0.995 & 0.995 \\
        2-notebook  & 20     & 20        & 0.998   & 1   & 0.995 & 0.915 \\
        3-plant     & 20     & 20        & 0.997   & 1   & 0.995 & 0.995 \\
        4-tennis    & 20     & 20        & 0.997   & 1   & 0.995 & 0.995 \\
        5-figurine  & 20     & 20        & 0.997   & 1   & 0.995 & 0.995 \\
        6-mug       & 20     & 20        & 0.997   & 1   & 0.995 & 0.954 \\
        7-lipstick  & 20     & 20        & 0.996   & 1   & 0.995 & 0.990 \\ \bottomrule
    \end{tabular}}
\end{table}

We further validate our model’s performance through empirical experiments. \autoref{tab:real} presents the results of randomly sampling 200 consecutive frames for each object and analyzing their detection outcomes. Five categories achieve 100\% continuous detection, while the remaining categories all exceed 90\%. The table also reports the average image processing time and mean confidence level, demonstrating that our model can perform detection tasks rapidly, stably, and accurately.
\begin{table}[h]
    \centering
    \caption{Object detection real-world testing performance (200 frames in total)}
    \label{tab:real}
    \resizebox{0.48\textwidth}{!}{
    \begin{tabular}{@{}p{1.5cm}p{1cm}p{1.5cm}p{1cm}p{1.4cm}p{1cm}@{}}
        \toprule
        \textbf{Metrics} & \textbf{Accuracy (\%)} & \textbf{Average Time (ms)} & \textbf{SD (ms)} & \textbf{Average Confidence (>0.75)} & \textbf{SD} \\ \midrule
        Paint board & 91.5 & 11.3265 & 4.1147 & 0.9718 & 0.0075 \\
        Pumpkin & 98.5 & 10.6345 & 2.6201 & 0.9572 & 0.0096 \\
        Notebook & 100 & 11.1565 & 3.4735 & 0.9206 & 0.0146 \\
        Plant & 100 & 11.2635 & 2.8081 & 0.9232 & 0.0260 \\
        Tennis ball & 100 & 11.3800 & 3.4605 & 0.9659 & 0.0032 \\
        Figurine & 100 & 11.2575 & 3.9481 & 0.9269 & 0.0177 \\
        Mug & 100 & 11.874 & 3.3021 & 0.9231 & 0.0137 \\
        Lipstick & 98.5 & 11.5045 & 3.7931 & 0.9220 & 0.0316 \\ \bottomrule
    \end{tabular}}
\end{table}

Through the experiments described above, we demonstrate the exceptional capability of our technical pipeline in visual processing.

\textbf{Persona Creation.}
 By comparing the personas generated by three large-scale models using the same prompts and images, we find that the QWEN-VL model consistently produces age and voice tone selections that align with at least one of the other models across all categories. Participant ratings indicate that QWEN-VL (Mean = 6.2917, SD = 1.0826) outperforms GPT-3.5 (Mean = 3.2917, SD = 1.2676) and Grok-3 (Mean = 3.8333, SD = 1.6854) in terms of how well the generated stories and personalities meet user expectations. This outcome may stem from various factors, such as differences in prompt characteristics and training corpora. Nevertheless, these results provide preliminary evidence of our pipeline’s sufficient capability in persona generation, confirming the suitability of the selected large-scale model within our framework.

\textbf{Dialogue system.}
We conducted a total of 47 dialogue rounds with four objects, with a cumulative input duration of 136.14 seconds (Mean = 3.0941, SD = 1.083). By employing separators, the GPT-generated response text is segmented into 95 short sentences. The average TTS processing time per short sentence is 1.7732 seconds (SD = 0.7781), with an average real-time factor of 0.6297 (SD = 0.1649).

\section{User Study} \label{user-study}

We evaluated our system through a user study involving 12 participants (\textcolor{black}{6 female, 6 male}), aged between \textcolor{black}{18 and 27}. The purpose of the user study was to assess the system’s usability and to gather user experiences and feedback regarding the application of Talking Spell across various scenarios. Additionally, we investigated user preferences concerning the wearability of Talking Spell as a wearable device.

\begin{figure}[h]
    \centering
    \includegraphics[width=\linewidth]{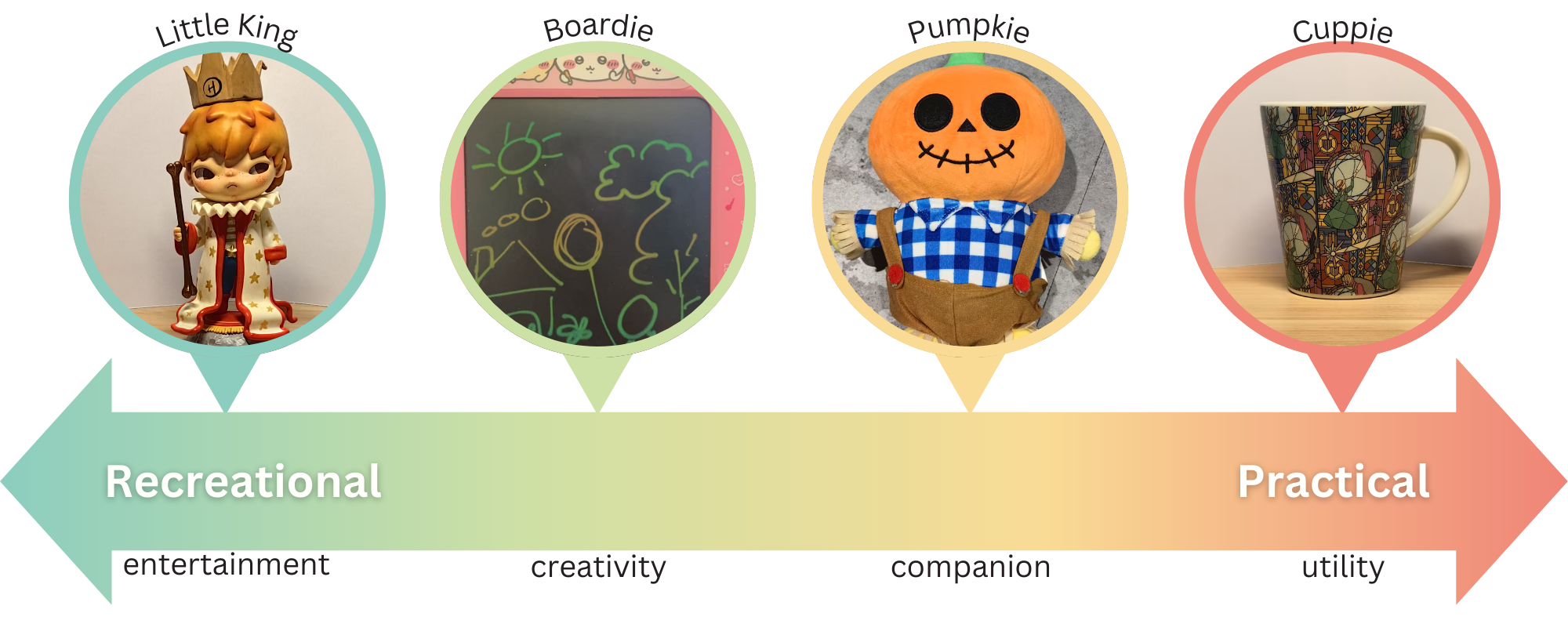}
    \caption{Talking Spell emphasizes the accessibility and versatility in its application and features of anthropomorphism. To demonstrate we design four default use case scenarios according to the four types of objects interaction from recreational to practicality. These four representative objects are used in a section of the user study that is covered in section \ref{user-study}: User Study. }
    \label{fig:use-case}
\end{figure}
\subsection{Method}
We recruited 12 participants (P1-P12), most of whom had prior experience with AI voice assistants or AI companionship products (12 out of 12). The study comprised three sessions: a brainstorming session, a dialogue with a predefined object in four intents, and a connection-building exercise with own personal items. The entire experiment lasted 60 minutes. Upon completion, participants were required to complete a follow-up survey and a brief interview.
\subsection{Study condition}

In \autoref{sec:scenario}, we proposed four interaction intents for using Talking Spell: entertainment, creativity, companionship, and utility. To enhance user understanding, we predefined four objects corresponding to these interaction intents (see \autoref{fig:use-case}). These objects have already undergone the Acquaintance and Familiarization stages to establish usable personas, though they lack prior chat histories.

\subsection{Procedure}
\subsubsection{Brainstorm}

We first obtained consent from the participants to participate in the experiment and then introduced them to the study’s objectives. We presented the predefined objects to the participants, along with their corresponding interaction intents. Participants were invited to envision four cases, imagining aspects such as the themes of dialogues with these objects, potential outcomes, and the background stories and voice tones associated with the objects. We instructed participants to document these reflections in writing on paper. We also invited participants to talk aloud in their later experiences, alongside the conversation to objects.

\subsubsection{Dialogue with a predefined object}

At this stage, users directly enter the Bonding stage, engaging in dialogue with predefined objects. We first demonstrate to users how to operate Talking Spell. Subsequently, users are equipped with the scope and instructed to hold the wand, interacting with each of the four predefined objects.

\subsubsection{Construct connection with personal items}

During participant recruitment, we required participants to bring a familiar personal item. Throughout the experiment, we utilized these participant-provided items to conduct the three stages: Acquaintance, Familiarization, and Bonding. To facilitate a stronger connection between participants and their items, we allowed participants to autonomously assign names and presented them with a persona generated by a large vision language model, which they could adjust according to their preferences.

\subsubsection{Follow-up Survey and Interview}

Following the completion of the three experimental sessions, participants were asked to complete a questionnaire. This questionnaire included their ranked experiences with Talking Spell across the four interaction intents, along with the reasons for their rankings, as well as the reasons for selecting specific personal items during the connection-building stage. The System Usability Scale (SUS) \cite{SUS} was employed to evaluate the system’s usability. Additionally, we designed a custom questionnaire to assess our system across seven dimensions: usability, personalization, creativity, emotional connection, practicality, trust, and overall satisfaction. We also inquired about participants’ likelihood to recommend our system to others, as part of the overall satisfaction assessment.

Following the survey, we conducted interviews with participants, focusing on their exploration of wearable options for Talking Spell. We provided participants with three wearing methods for the scope and four for the wand, allowing them to select their preferred options and explain their reasoning. Any ideas beyond the designs we offered were documented. We also followed up the questions related to the seven assessment dimensions, asking for the reasoning and explanation behind the choices. Additionally, we inquired about participants’ expectations for future design enhancements and upgrades to Talking Spell, as well as any shortcomings they perceived in the current experience.

\subsection{Results}
This section presents the outcomes of our preliminary user study and the follow-up interviews.

\subsubsection{User Evaluation}
%% SUS
We evaluated the system usability score (SUS), overall engagement, and the system’s usefulness. Our system achieved
an overall SUS score of 79.09, with a standard deviation (SD) of 10.97 and A graded. We gauged Talking Spell by the Net Promotor Score (NPS) into the bargain. NPS was calculated as 41.67\% (N = 12, with 7 Promoters and 2 Detractors), reflecting a positive likelihood of recommendation. When asked, ``\textit{On a scale of 0 to 10, how likely are you to recommend the Talking Spell to a friend or colleague?}'' Participants provided a mean score of 8.67 (median = 10, SD = 1.92), further affirming the favorable reception of the system.

\begin{table}[h]
    \centering
    \caption{Frequency Distribution of Rankings for Talking Spell Interaction Intents}
    \label{tab:rank}
    \begin{tabular}{@{}lcccc@{}}
        \toprule
        {Scenario}                          & \textbf{Rank 1} & \textbf{Rank 2} & \textbf{Rank 3} & \textbf{Rank 4} \\ \midrule
        Companionship                     & 6                & 3                & 2                & 1  \\
        Utility/Task-oriented             & 3                & 5                & 3                & 1  \\
        Recreation/Entertainment           & 5                & 4                & 2                & 1  \\
        Creativity                         & 3                & 4                & 4                & 1 \\ \bottomrule
    \end{tabular}
\end{table}

Preferences for the four interaction intents—companionship, utility/task-oriented, recreation/entertainment, and creativity—were assessed through a frequency distribution of rankings, as shown in Table \ref{tab:rank}. Companionship was most frequently ranked first (6 participants), followed by recreation/entertainment (5 ranked 1st), highlighting their appeal. Utility/task-oriented and creativity showed varied but balanced rankings, with 5 and 4 participants ranking them 2nd, respectively, indicating consistent but less dominant interest. This distribution underscores the system’s versatility, with companionship and entertainment emerging as particularly valued features. In conclusion, anthropomorphism is generally more favored in recreational contexts than in practical ones. This aligns with studies suggesting that anthropomorphism in technology can enhance the sense of companionship in emotional and mental supports.

Participant comments offered insights into the system’s strengths and areas for enhancement. P5 praised its engagement, stating, ``\textit{I love it so much as it is engaging.}'' Conversely, P3 observed that object responses, while displaying general traits, felt ``\textit{relatively generic},'' hindering deeper bonds, and suggested objects initiate conversations occasionally to improve the experience. P2 proposed, ``\textit{Users could provide personalized object backstories to enhance immersion and closeness},'' emphasizing customization. P11 noted, ``\textit{Sometimes it doesn’t hear me clearly; it could repeat and ask again when unsure},'' pointing to speech recognition issues. P12 added, ``\textit{It’d be better if we could record our own voices},'' suggesting greater personalization.

%% NPS = 41.67% (N=12, 7 Promoters and 2 Destractors)
%% Overall
%% On a scale of 0 to 10, how likely are you to recommend Talking Spell to a friend or colleague?
%% (Mean: 8.67, Median: 10, SD:	1.92)
%% Rank

\subsubsection{Participants' Feedback}
\label{sec:participants-feedback}

\begin{figure}[h]
  \includegraphics[width=\linewidth]{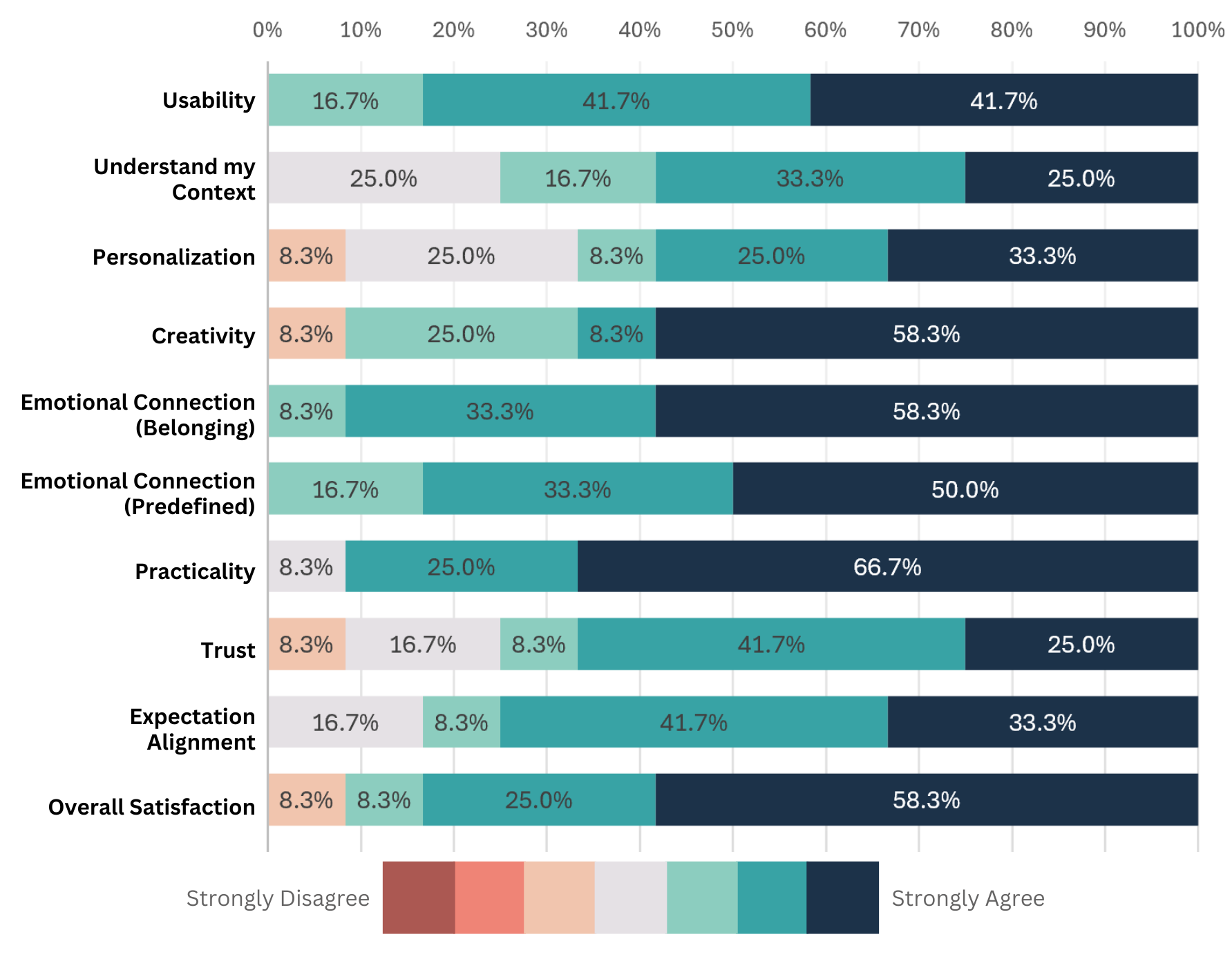}
  \caption{Participant (N=12) feedback on the seven assessment dimension to evaluate the user experiences and perception.}
  \label{fig:participant-feedback}
\end{figure}

Participants generally expressed that our system enhances engagement and emotional connection while offering high practicality due to its versatility in adapting to various real-world scenarios, especially when compared to typical virtual assistants, which are often perceived as additional entities. They provided invaluable feedback on Talking Spell and its future implementation. Below, we summarize their insights:

\textbf{\textit{Enhancing Engagement and Practicality through Versatile Design.}}
Talking Spell significantly enhances user engagement and emotional connection, distinguishing it from traditional virtual assistants. However, its practicality elicited mixed feedback, with 66.7\% of users adopting a neutral stance on this aspect (see Figure \ref{fig:participant-feedback}). Participants emphasized the system’s adaptability to diverse real-world scenarios as a key strength, facilitating its seamless integration into daily routines. For example, P6 described the experience as \textit{transformatively interesting}, highlighting how it delivered \textit{warm yet hilarious} replies even in formal settings. She recounted asking a cup, ``\textit{Can I pour Coke in you and drink it?}'' to which cuppie playfully responded with a refusal and claimed that coke as unhealthy drinks. This example underscores Talking Spell’s engaging appeal and adaptability. Nevertheless, the neutral feedback on practicality suggests an opportunity to refine its versatile design by incorporating features that more effectively address users’ everyday needs. This insight can guide future development toward enhancing the practical applications of anthropomorphism in the system.

%% Perception over the seven assess dimensions

\textbf{\textit{Elevating Interaction through Unexpected Voice Importance, Rather than Other Sensory Modality. }}
Voice interaction emerged as an unexpectedly critical element in participants’ experiences with Talking Spell, challenging initial assumptions that verbal exchanges alone might lack depth. Participants (P4, P5, P6, P7) reported that the system’s voice-based communication fostered meaningful and dynamic interactions with objects, proving surprisingly satisfying. P4, who originally expected a boring and not too attached user experiences, highly rated the system as such interaction \textit{leaves more imaginative spaces} for him. This underscores the need for robust voice functionality as a foundational feature, enhancing user satisfaction and deepening engagement within the system.

\begin{figure}[h]
  \includegraphics[width=\columnwidth]
    {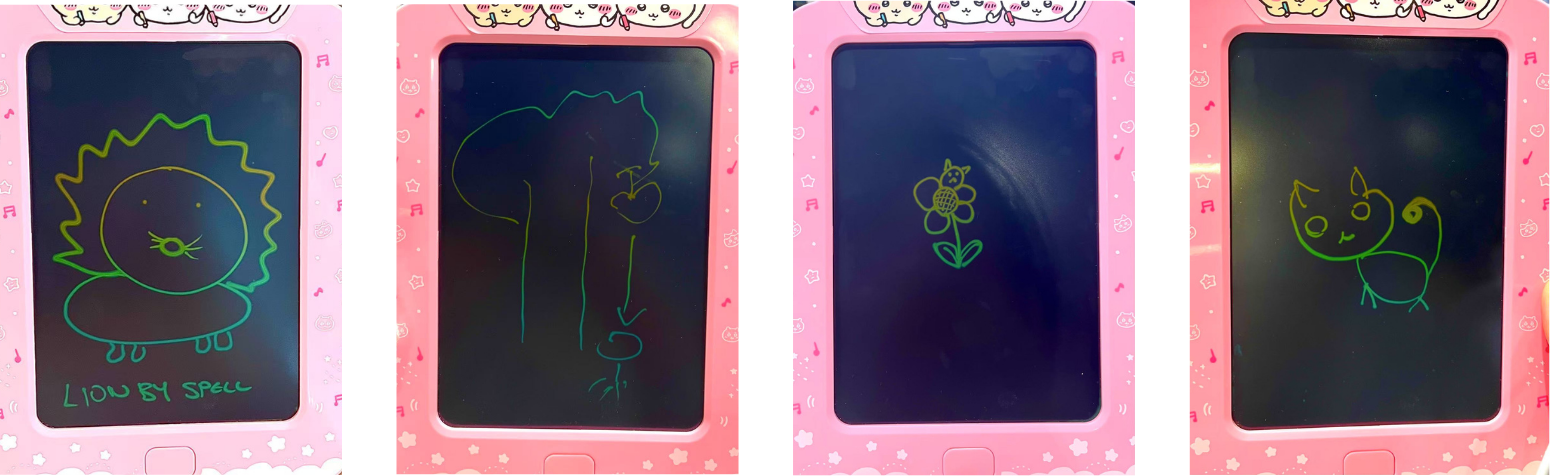}\hfill
  \caption{Sketches by the participants with the help from Boardie.}
  \label{fig:sketch}
\end{figure}

\textbf{\textit{Encouraging Creativity through Co-Design Collaboration. }}
Talking Spell inspires creativity through its co-design and collaborative features, even among participants who did not consider themselves artistic, not being art lovers and described themselves as lacking confidence in drawing. The system’s supportive feedback, such as compliments during drawing tasks evidenced by collected sketches (Figure \ref{fig:sketch}), boosted engagement and extended the appeal of virtual collaboration. This capability not only enriched the creative process but also expanded the imaginative potential of user-object partnerships, positioning Talking Spell as a valuable tool for fantasy-driven cooperation.

\textbf{\textit{Facilitating Emotional Connection through Personal Belongings. }}
Talking Spell excels at fostering emotional connections, particularly when participants interacted with personal belongings rather than predefined objects. For example, participants like P2, P7, and P10 highlighted the comfort derived from objects related to own hobbies (such as \textit{tennis ball }from P3), dolls and plush toys, noting how personal significance amplified their attachment. 

\finalrv{P10 expressed, ``\textit{When I spoke to my old stuffed bear, it felt like I was talking to a lifelong friend who understood me.}'' This sentiment underscores the profound sense of companionship that anthropomorphized objects can evoke through Talking Spell. Similarly, P1 reflected, ``\textit{Each time I interacted with my plant, it was as if it became more than just a decoration; it became a part of my daily routine that I genuinely cared for.}'' These insights highlight the transformative potential of Talking Spell in fostering emotional bonds, as users reported feelings of comfort, nostalgia, and connection. While empirical evidence is still needed, participants (P1, P2, P7, P10) noted that their attachment to these items was significantly enhanced through their interactions with the system.}

An interesting observation and example from P7 with Cuppie highlights that the anthropomorphic design often trumped practical utility. Animated objects trigger associations, such as a cup (Cuppie) prompting drinking without explicit cues just by its existence. While warm, imaginative personalities with human-like flaws further strengthened emotional bonds and encouraged sustained interaction.

\textbf{\textit{Needs for Proactive Object Interaction. }}
Participants (P3, P11) expressed a clear need for objects within Talking Spell to exhibit greater proactivity, moving beyond merely responding to user prompts. They desired objects that could initiate conversations or actions independently, enhancing the sense of companionship and reducing the effort required to sustain interactions. Such proactive behavior could make the system feel more lifelike and dynamic, deepening emotional connections and aligning with expectations for a more reciprocal relationship with animated objects.

\textbf{\textit{Call for Multi-agent Interaction. }}
Feedback also revealed a demand for multi-agent interaction, where multiple objects could communicate not only with the user but also with each other. A participant (P7) envisioned richer scenarios, such as group conversations or collaborative tasks among objects, which would create a more immersive and interactive environment. By enabling social dynamics among animated entities, Talking Spell could enhance both its realism and utility, broadening its applications for practical and entertainment purposes.

\textbf{\textit{Need for More Voice Options to Enhance User Experience.}}
A notable critique from eight participants focused on the need for enhanced vocal interaction, pointing to issues like \textit{mechanical tones, slow speech, unclear delivery}, and \textit{repetitive voice options across objects}. Users observed redundancy in voices between predefined items and personal belongings, such as an anime character pendant by P6, suggesting that greater customization and variety could address these limitations. Improving voice quality and diversity is essential to elevate the system’s \soutFinal{naturalness}\finalrv{intuitiveness}, appeal, and overall user satisfaction.

\section{Open Source Application}
To enable other researchers to replicate and extend our work, we have made our circuit designs available at: \textbf{not available now due to the review anonymous consideration}.

\section{Example Portable Applications}
To the best of our knowledge, Talking Spell is the first system that enables human communication with objects through advanced technology, using a hybrid system and hardware design. It transforms the idea of anthropomorphism into practical applications. Our main design focus is on system accessibility and facilitating seamless interaction between users and any objects - be it a bric-a-brac or a plant. 

Prior research on VA\cite{GazePointAR,GesPrompt,BodyTouch,10.1145/3659625} has led to many on-body and off-body widgets. Therefore, we did not prioritize application development. Instead, we believe Talking Spell can enhance engagement and practicality through creative anthropomorphism. Our focus is on usability while remaining open to novel body placements. We have created several example devices to demonstrate the versatility of our approach, which are covered in this section (also see Figure \ref{fig:teaser}).

Talking Spell consists of a scope for spell target (see section \ref{sec:scope}), and wand for spell casting (see section \ref{sec:wand}). We start by discussing the all-in-one wearable that embodies both scope and wand, then review numerous examples for both components respectively. The section concludes with envisioned example applications.

%% Talking Spell consists the scope (camera for detecting objects) and wand (sensor to allow spell casting and vibration motor to inform useful "spell". Example scope devices we created include (colored in green, clockwise from top-left): head chain, glasses, and bone conduction headphones; example wand devices we created include (colored in red, clockwise from top-left): a necklace with sensor pendant, bag charm, earrings, and a part of bone conduction headphones. Note that the two components (top-right) combine in the bone conduction headphones. 
 
\subsection{All-in-One Wearable}
\begin{figure}[h]
  \includegraphics[width=.45\columnwidth]
    {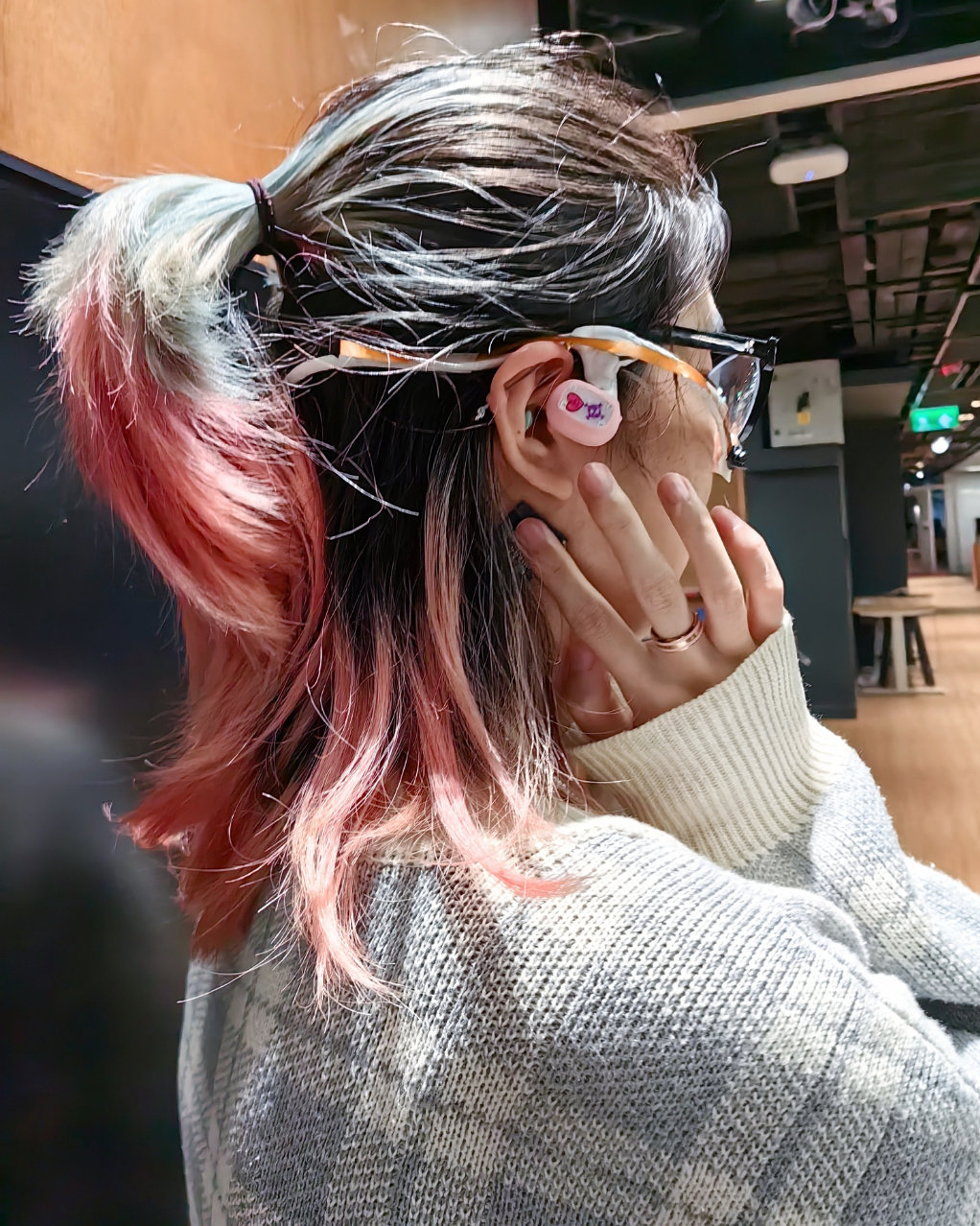}\hfill
  \includegraphics[width=.47\columnwidth]
    {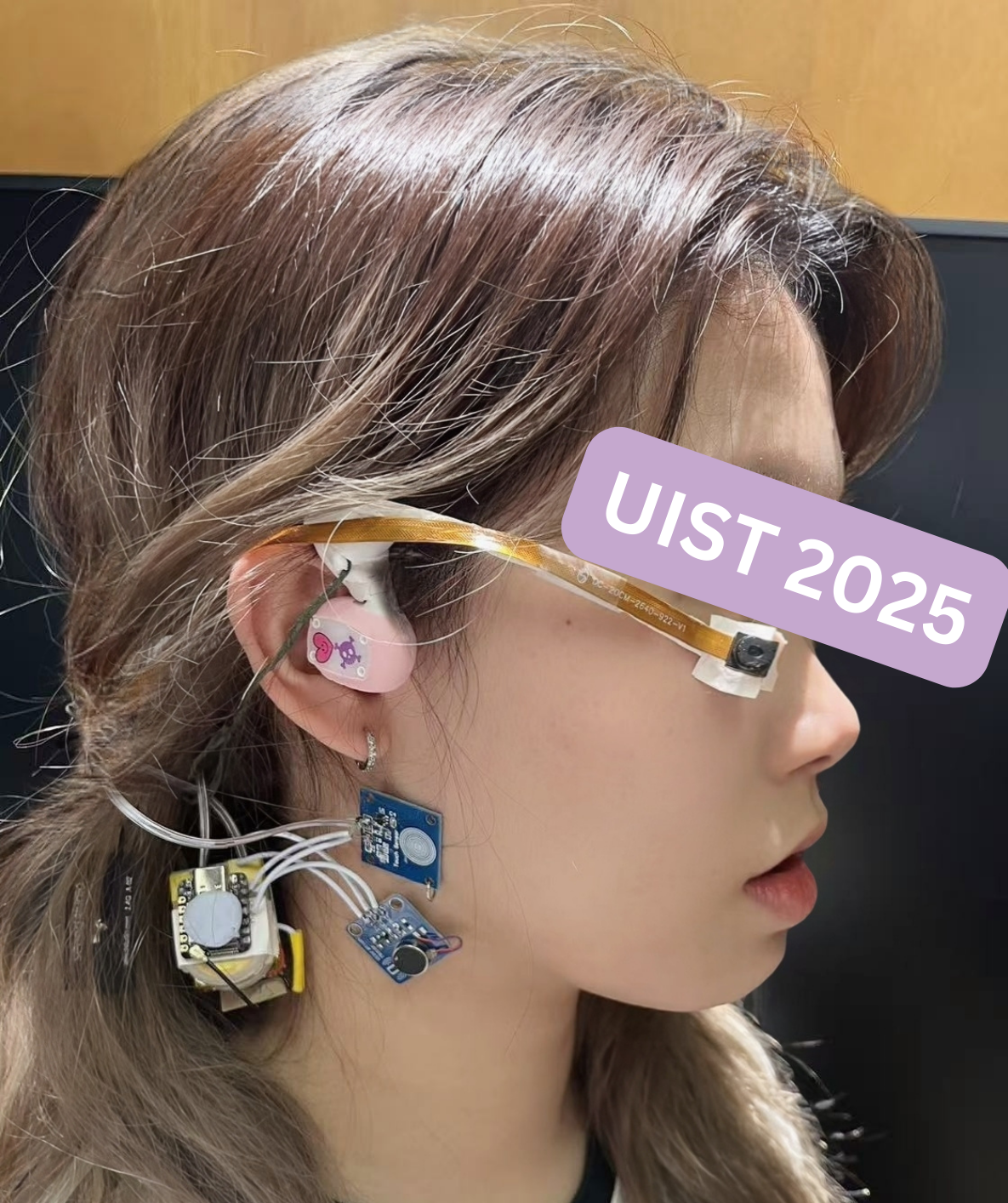}
  \caption{A bone conduction headphone that serves as the all-in-one wearable for Talking Spell: back side (\textit{left}) and side (\textit{right}).}
  \label{fig:all-in-one-user}
\end{figure}

\begin{figure}[h]
  \includegraphics[width=\linewidth]{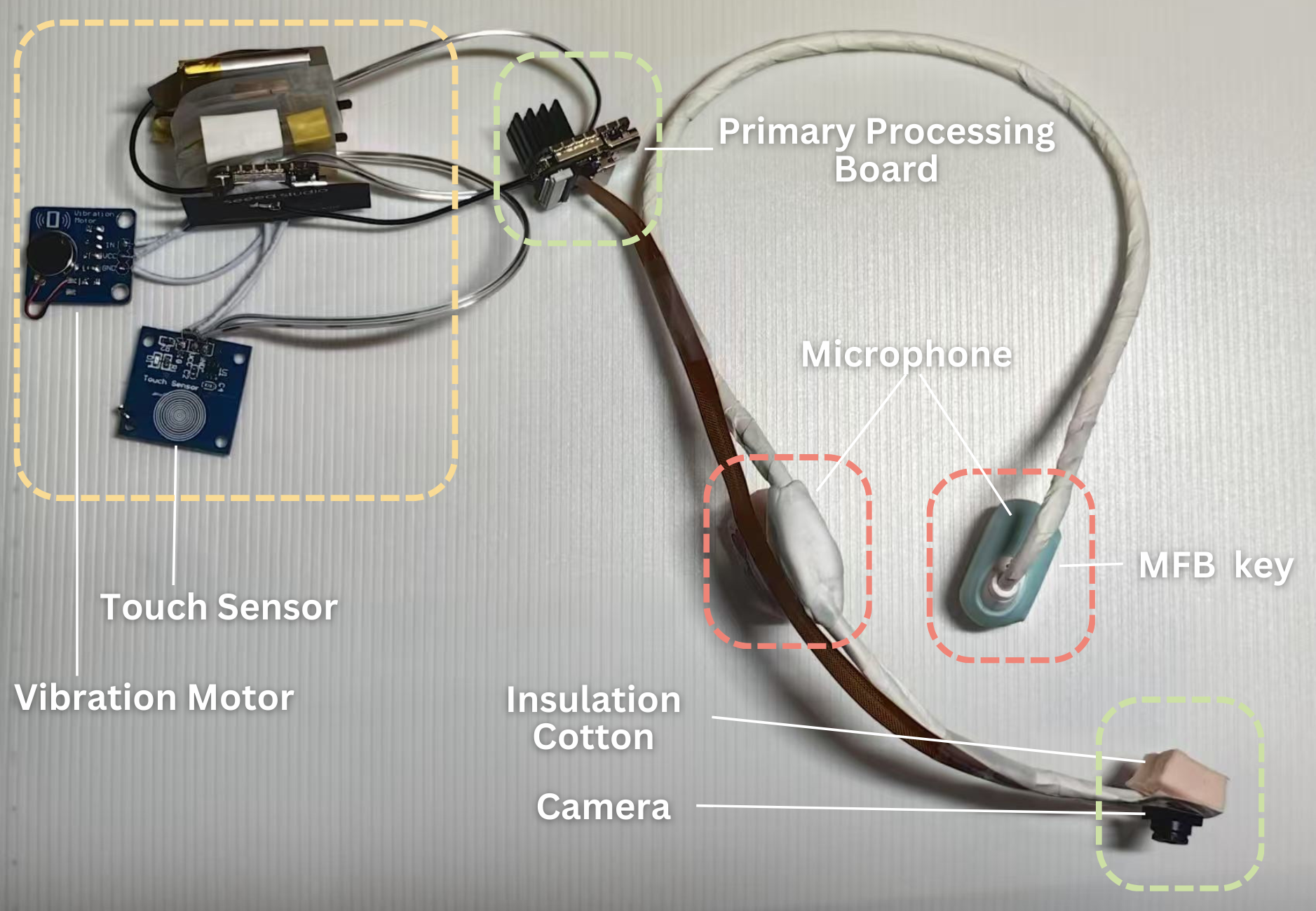}\hfill
  \includegraphics[width=\linewidth]{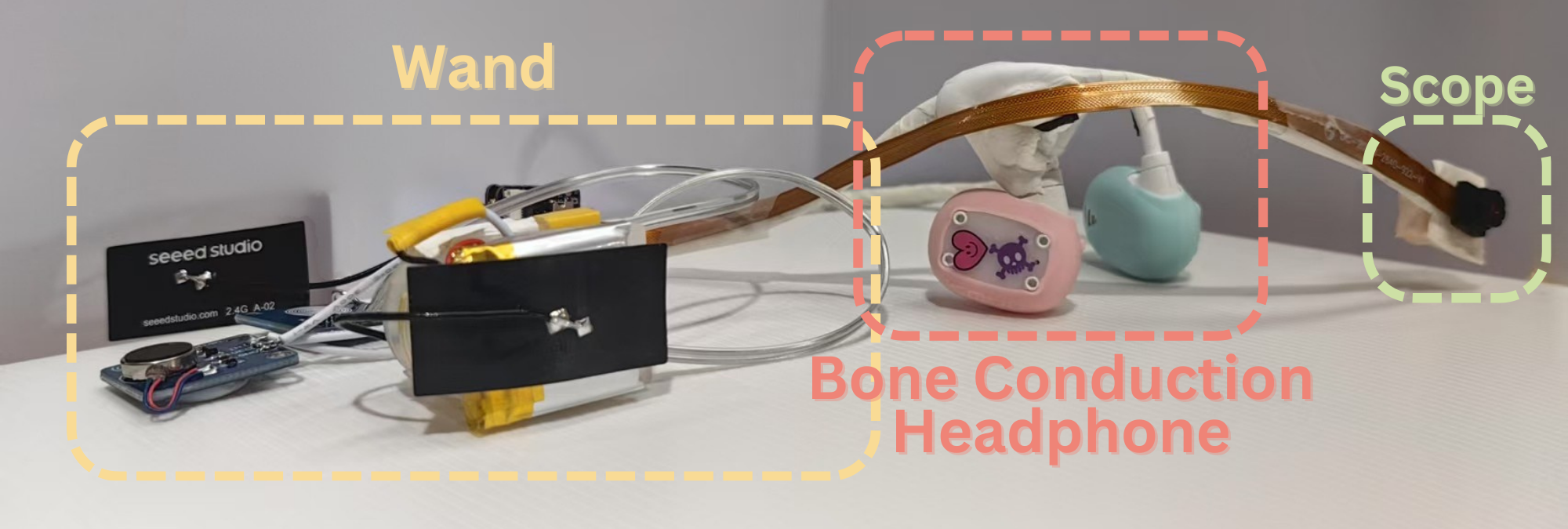}
  \caption{A bone conduction headphone as an example of Talking Spell integrated into an all-in-one wearable device.}
  \label{fig:all-in-one}
\end{figure}

Starting simple, we created an all-in-one wearable - \textbf{a bone conduction headphone} - that handles visual input, audio input and output (Figure \ref{fig:all-in-one-user}). The scope should sits where it can tract what the user sees; the wand should be placed for easy touch and vibration feedback; and the headphone / microphone manage the voice input and audio output, creating a seamless and cohesive user experience. Hence, Talking Spell is particularly suitable to appear in form of a headphone (Figure \ref{fig:all-in-one}).  

\subsection{Extensive Instances for Scope (Camera)}
\begin{figure}[h]
  \includegraphics[width=.49\columnwidth]{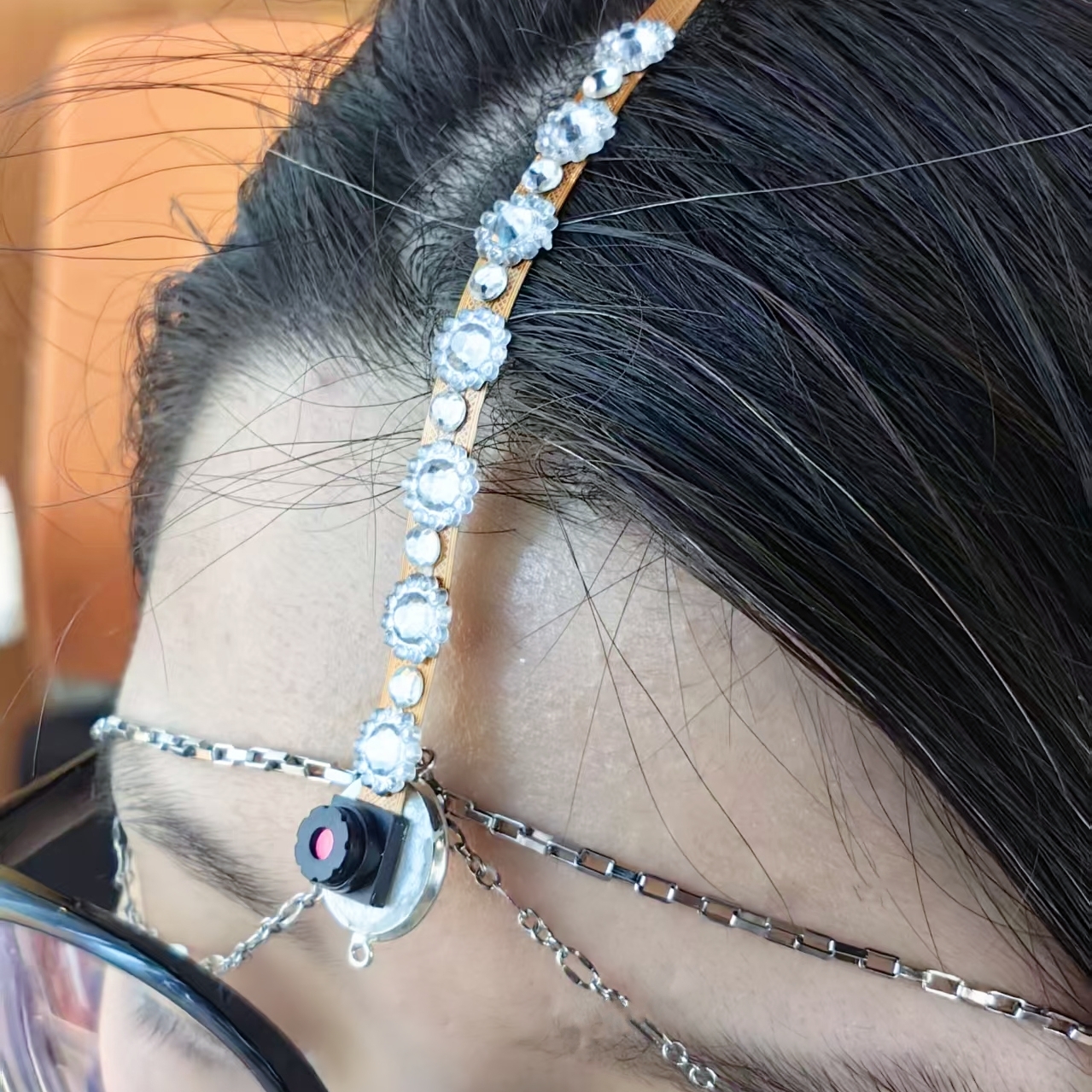}\hfill
  \includegraphics[width=.49\columnwidth]{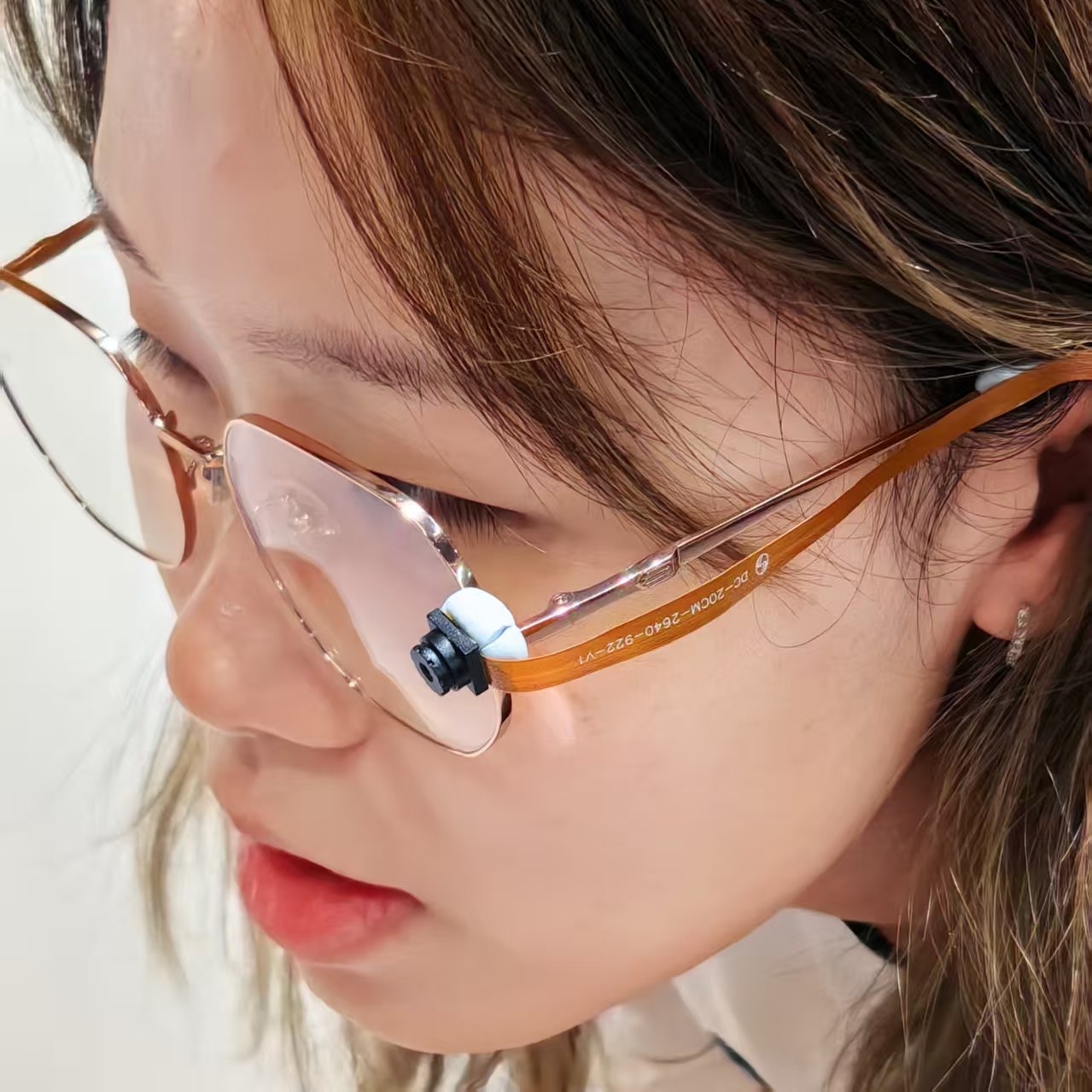}
  \\
  \includegraphics[width=.49\columnwidth]{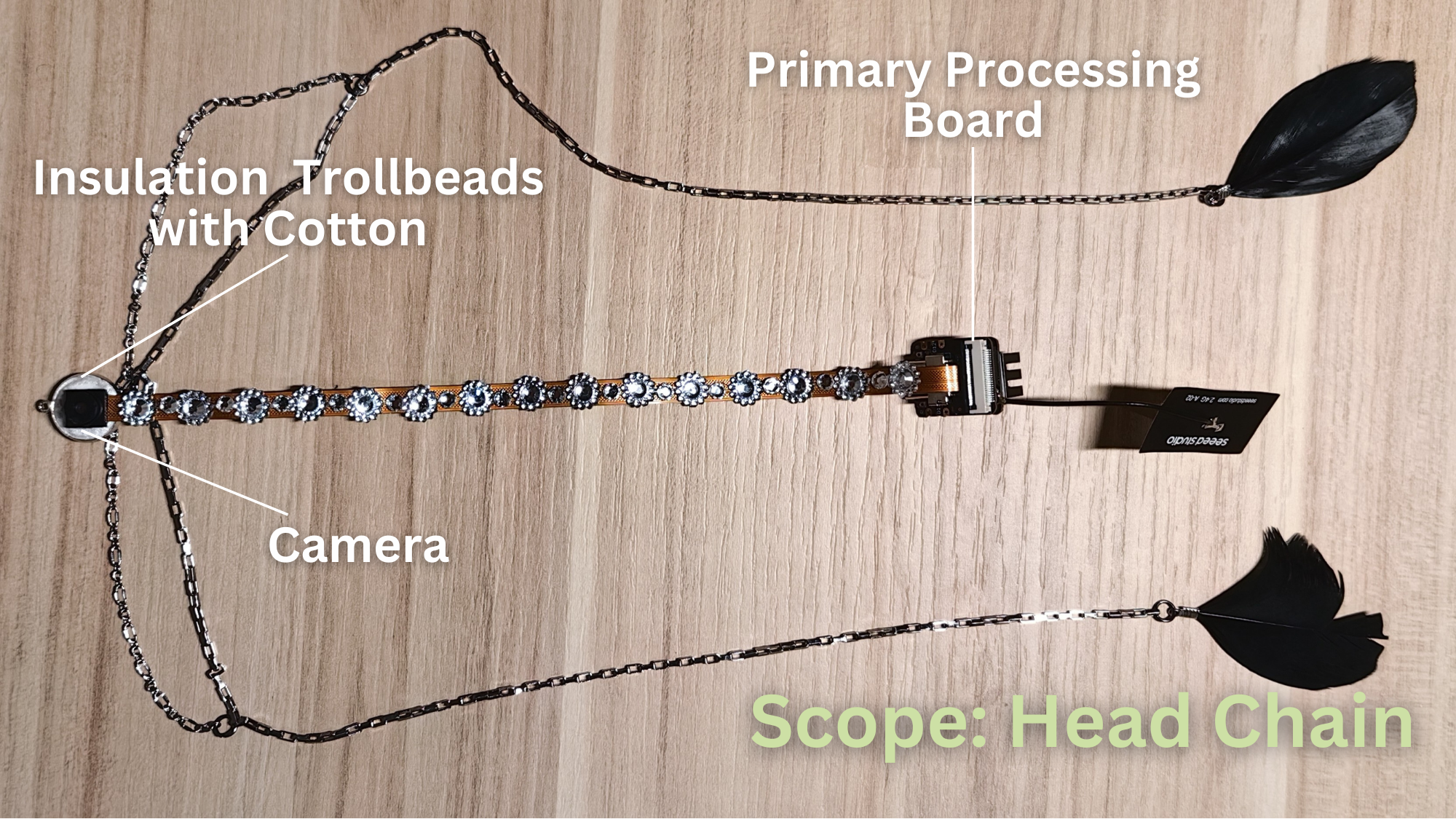}\hfill
  \includegraphics[width=.49\columnwidth]{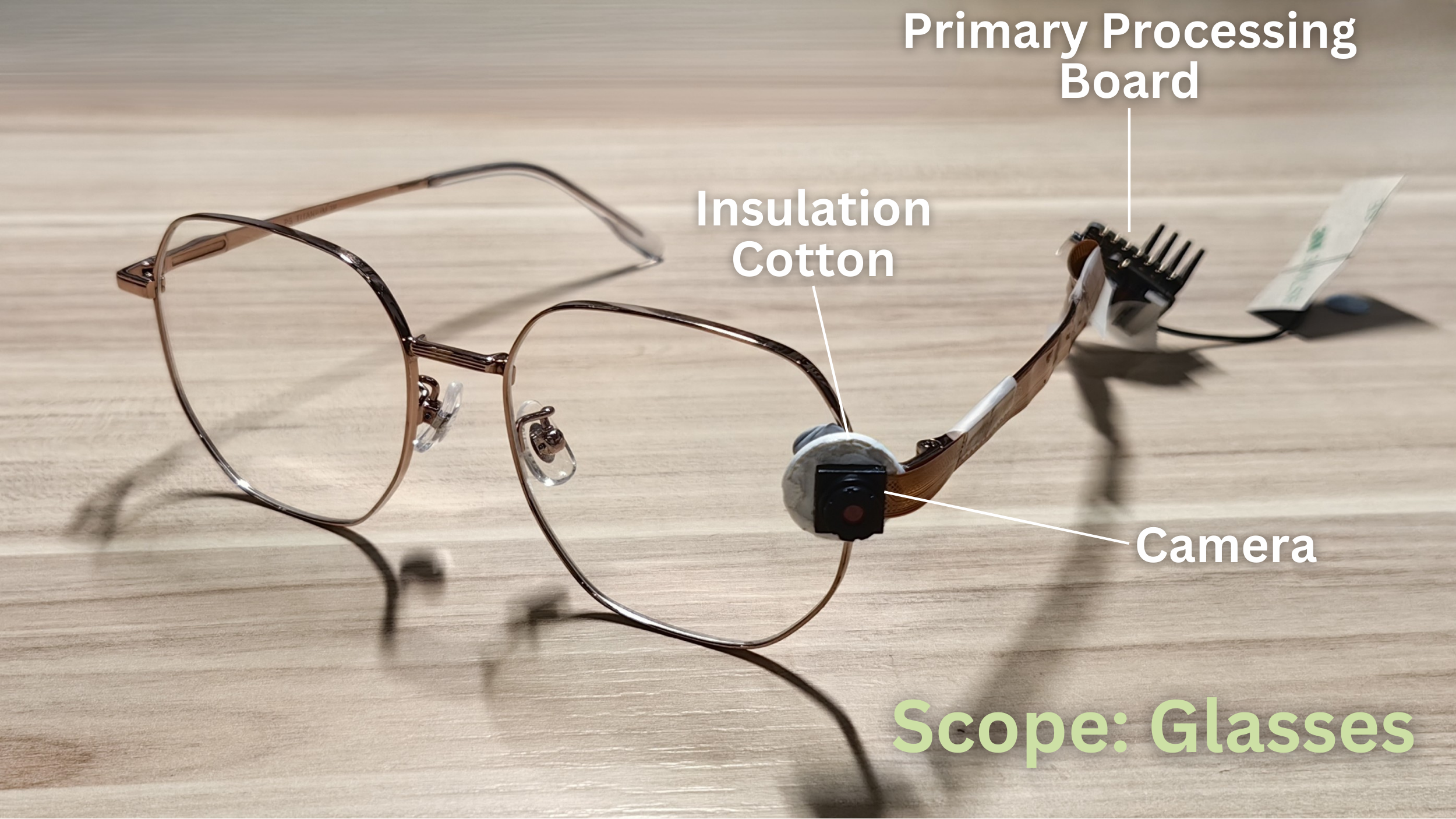}
  
  \caption{A Talking Spell scope in form of head chain and glasses, allowing users to look and detect the "spelling" targets, triggering the anthropomorphism of objects by object recognition visual input.}
  \label{fig:example-scope}
\end{figure}

As previously mentioned, the scope, specifically the camera, must be positioned to support object detection by aligning with the user’s vision, which is a key design requirement for effective interaction. To indicate, we developed two instances to demonstrate this principle: (1) \textbf{glasses} (see \textbf{\textit{top left}} in Figure \ref{fig:example-scope}), which embed the Talking Spell scope within a \textbf{wearable necessity}, leveraging their natural placement on the face for seamless object recognition, and (2) a\textbf{ head chain} (see \textbf{\textit{top right}} in Figure \ref{fig:example-scope}), which integrates the scope into \textbf{jewelries or accessories}, offering a stylish yet functional alternative for vision-based detection. Both instances contain the primary processing board, camera, and insulation layer (see \textbf{\textit{bottoms}} in Figure \ref{fig:example-scope}) with high adaptability.

\subsection{Extensive Instance for Wand (Sensor and Vibration Motor)}
\begin{figure}[h]
  \includegraphics[width=.49\columnwidth]{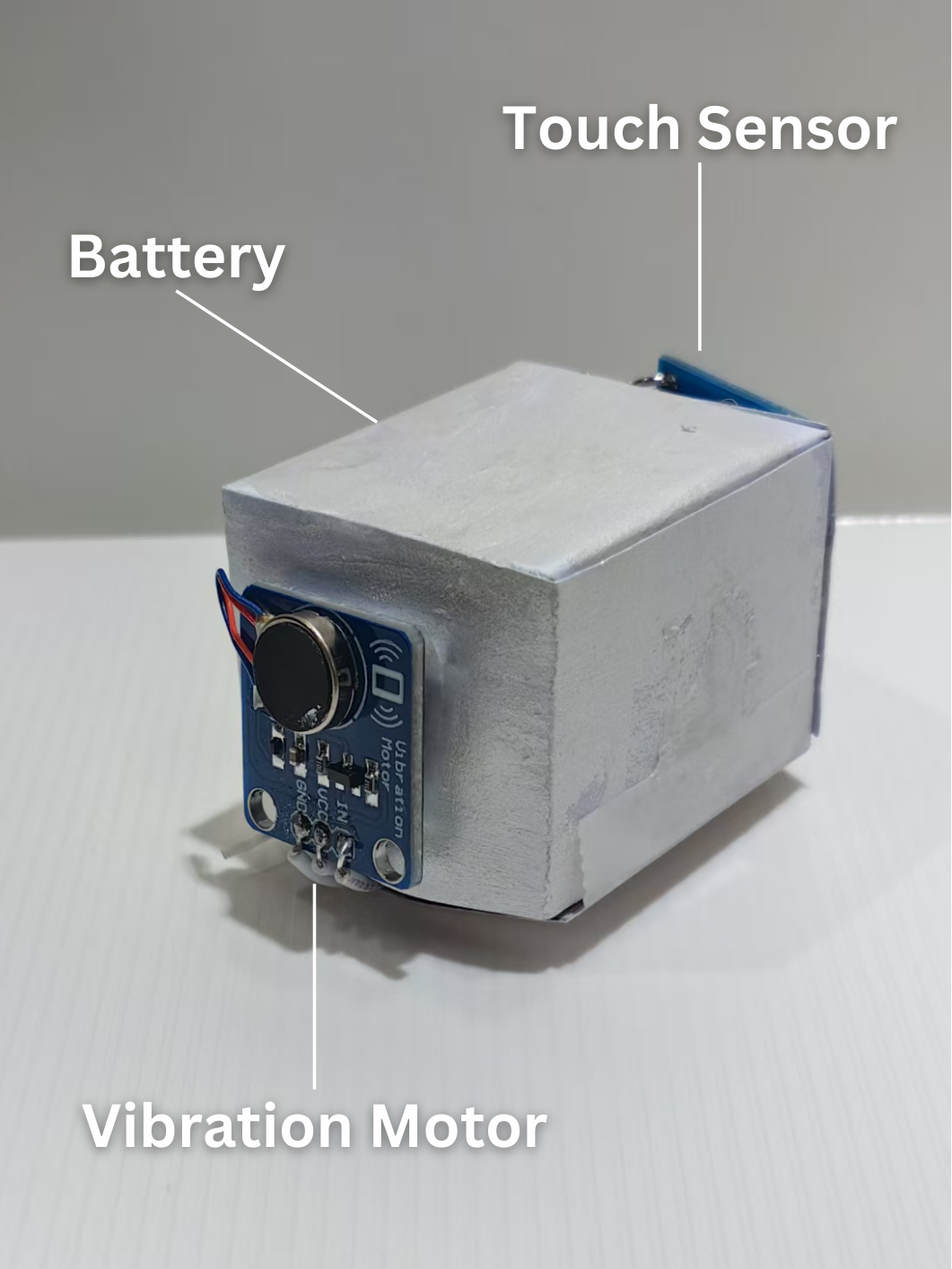}\hfill
  \includegraphics[width=.49\columnwidth]{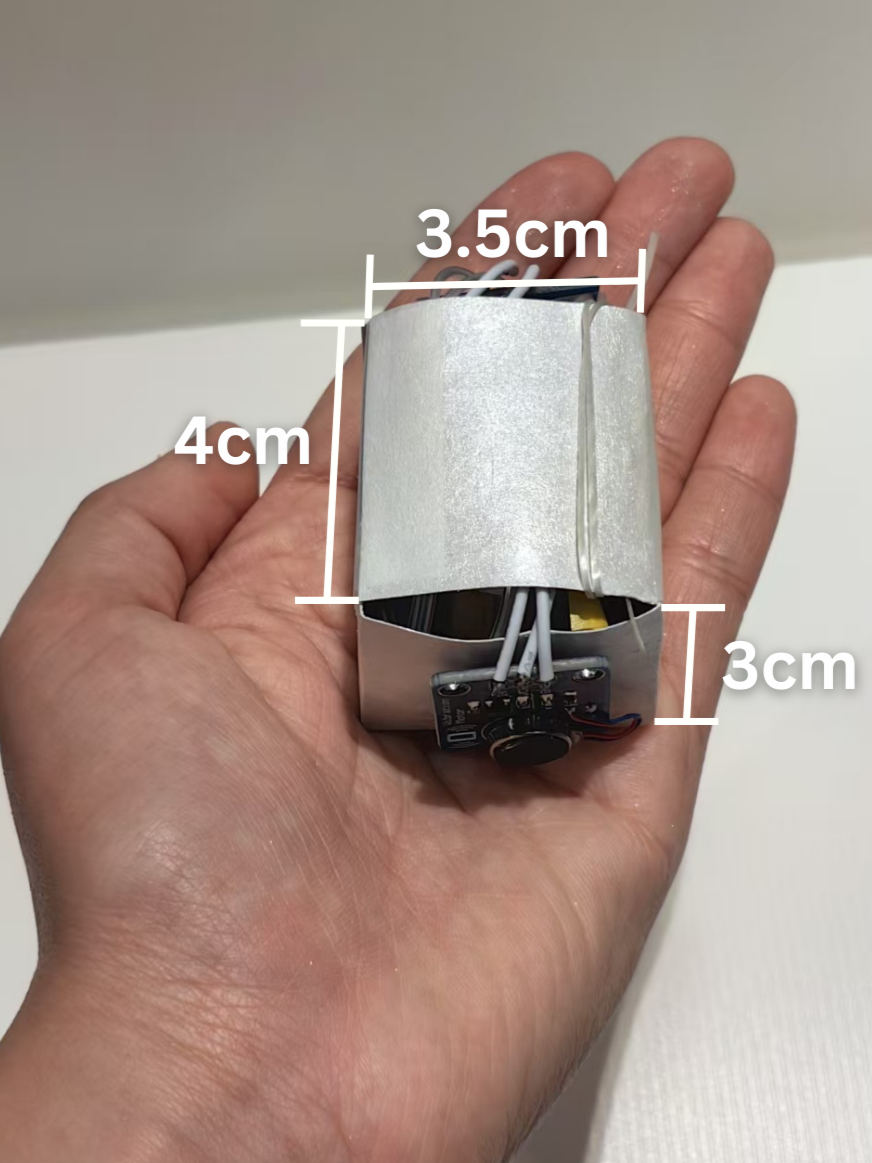}  
  \caption{A Talking Spell head chain and glasses, allowing users to look and detect the "spelling" targets, triggering the anthropomorphism of objects and initiating the conversation.}
  \label{fig:example-wand}
\end{figure}

Along with our previously introduced scope, wand is the primary mechanism for initiating and continuing conversations by activating voice input recording. It comprises three key components: a touch sensor to trigger voice input, a battery to sustain device longevity, and a vibration motor to provide tactile feedback confirming successful activation (see \textbf{\textit{left}} of Figure \ref{fig:example-wand}). These components work in tandem to ensure reliable functionality and user awareness, making the wand an essential device for seamless interaction with Talking Spell.

To optimize portability and enhance user interaction, the wand’s design adheres to specific criteria. First, it is compact, measuring 3.5 × 4 × 3 cm — roughly palm-sized — for easy carrying, as illustrated in the final design after iterative refinements (see \textbf{\textit{right}} of Figure \ref{fig:example-wand}). Second, the touch sensor is strategically positioned for effortless access, allowing users to trigger the device with minimal effort. Third, the vibration motor is engineered to fit snugly against the user’s skin, ensuring that the tactile feedback is distinctly felt and notified once the "spell" is successfully cast, reinforcing the interaction’s responsiveness.

\begin{figure}[h]
  \includegraphics[width=.33\columnwidth]{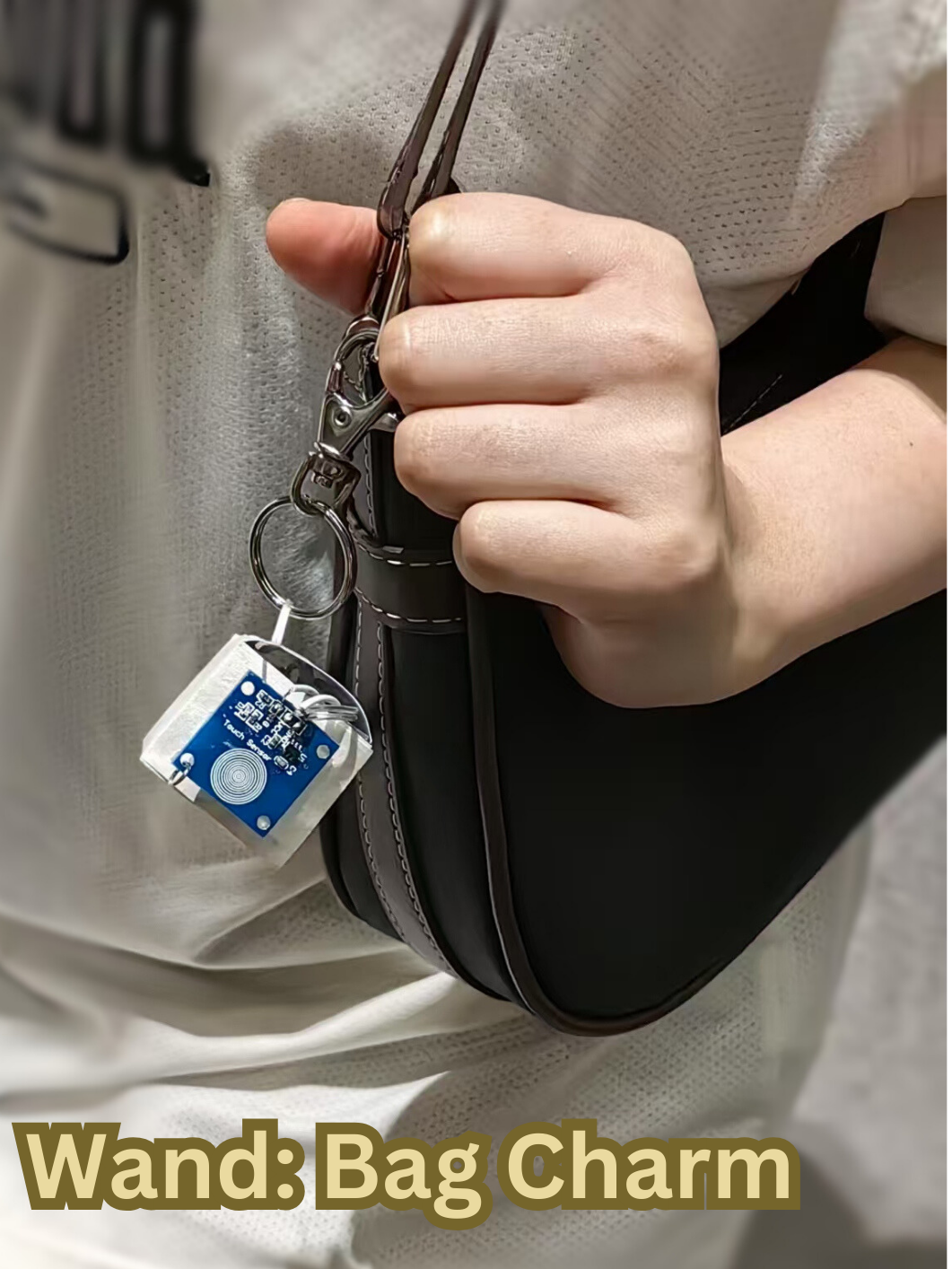}\hfill
  \includegraphics[width=.33\columnwidth]{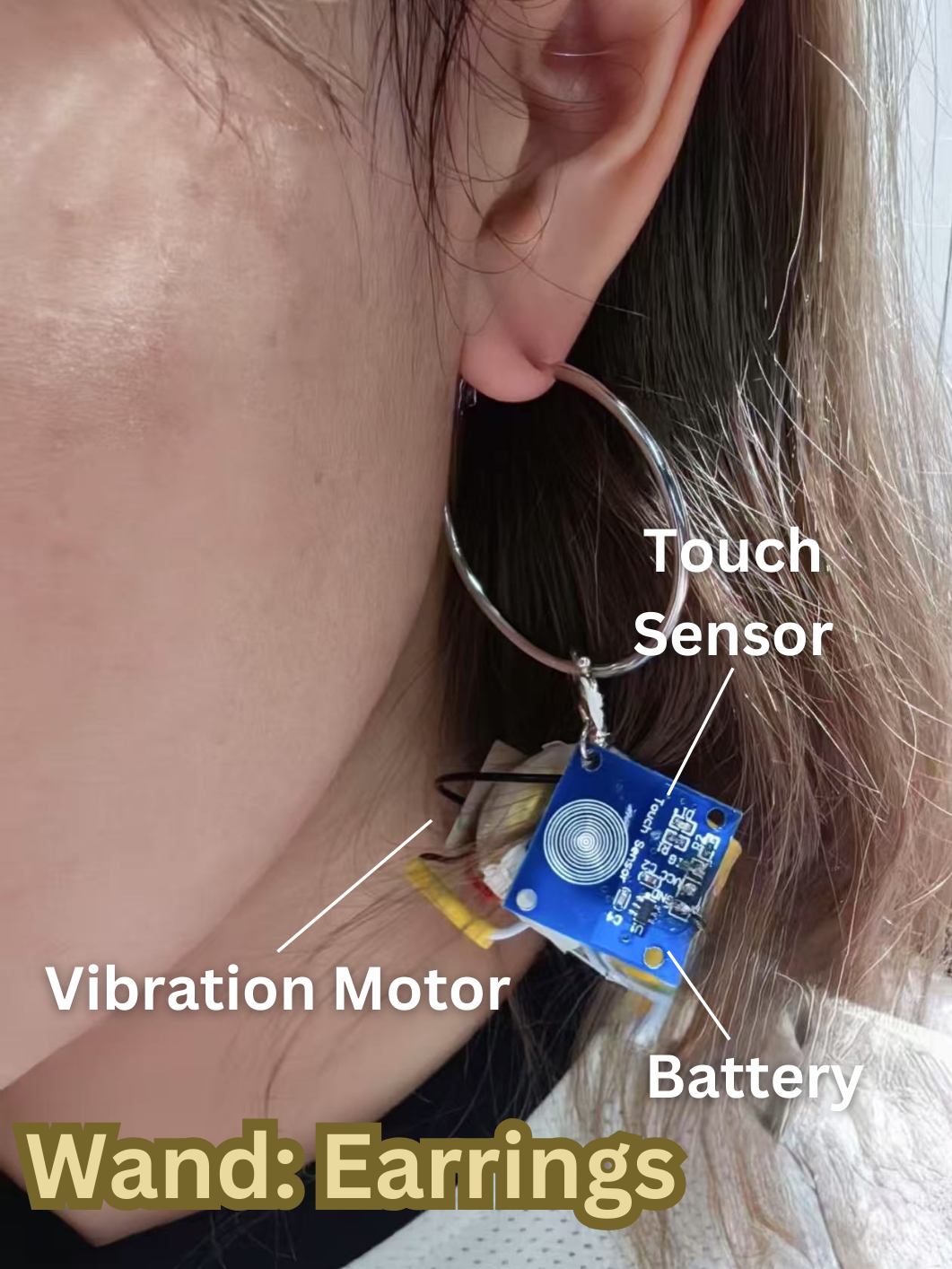}
  \includegraphics[width=.33\columnwidth]{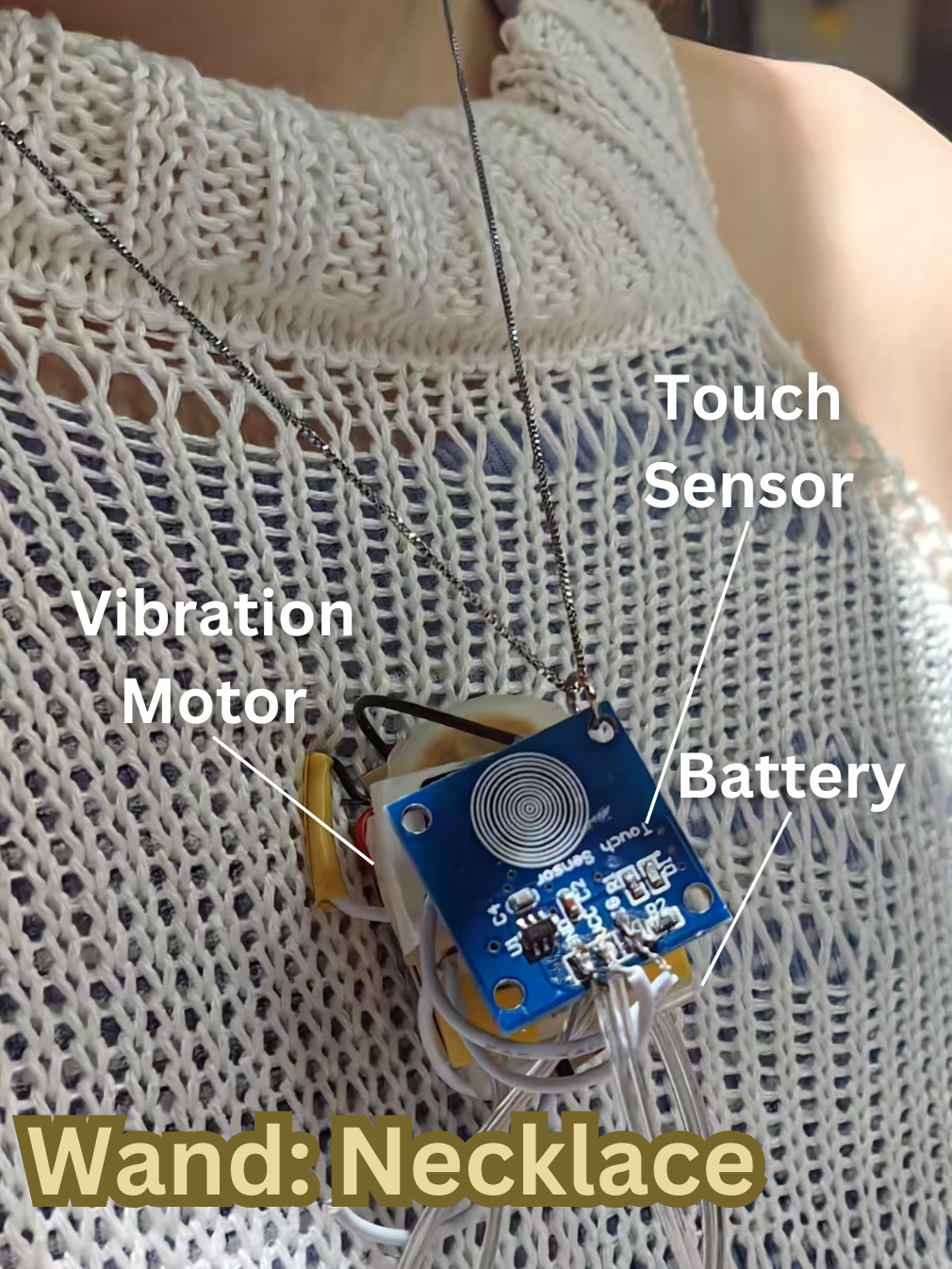}  
  \caption{A Talking Spell wand in form of charm, earrings, and necklace, allowing users to "cast the spell", initiating the conversation and recording the voice input.}
  \label{fig:wand-example-device}
\end{figure}

Based on these considerations, we instrumented three distinct design variants to demonstrate the wand’s versatility. The \textbf{charm} (see \textbf{\textit{left}} in Figure \ref{fig:wand-example-device})  functions as a lanyard that users can attach anywhere, though it requires holding to perceive the vibration feedback. Alternatively, the \textbf{earrings} (see \textbf{\textit{middle}} in Figure \ref{fig:wand-example-device}) and \textbf{necklace} (see \textbf{\textit{right}} in Figure \ref{fig:wand-example-device}) position the wand closer to the body, facilitating easier interaction and vibration detection without additional handling. These highlight our wand’s adaptability, balancing portability, accessibility, and tactile engagement to initiate conversations effectively in our Talking Spell.

\subsection{Envisioned Example Applications}

%% Future and opening
Many applications were considered during development, but are left and earmarked for future work. This offers a glimpse into Talking Spell's scalability across diverse forms and contexts. We believe that further innovation in the design of the wand and scope, as well as their integration into everyday objects, could significantly broaden the application's utility and user appeal.

%% scope and wand
For the scope, which is responsible for visual input and object detection, could be re-imagined as a headband, tie, bow tie, or nose chain, aligning with the requirement to position the camera in line with the user’s vision. Similarly, as the wand initiates conversations through touch and vibration feedback, potential designs include wearable forms such as a bracelet, belt, watch, ring, or hair clip. 

%% All in one glasses
A particularly promising avenue lies in the development of smart glasses as an all-in-one Talking Spell device. This medium fulfills all core design requirements: audio input and output via integrated microphones and speakers, visual input through a camera aligned with the user’s field of view, and tactile input to trigger the conversation and avoid unwanted recording input. As smart glasses are already a common necessity or accessory for certain individuals, they represent an ideal platform for consolidating the system’s components. This integration could streamline the user experience, eliminating the need for separate devices while capitalizing on an established wearable form factor.

%% closing
Talking Spell’s potential applications show its flexibility to meet diverse user needs, aiming for intuitive, aesthetic designs that enhance environmental interaction, though deferred to future work. We aim to pave the way for more enrichment in the interplay between users and their surroundings.

\subsection{\finalrv{Actionable Implications For Practitioners}}
\finalrv{
The insights from our user study illuminate actionable implications for practitioners aiming to utilize Talking Spell, an anthropomorphic AI system integrated with portable wearables, across diverse domains.

For daily companionship, participants expressed a strong desire for deeper emotional connections with personal belongings, indicating that anthropomorphized objects could provide comfort and meaningful interaction. For instance, as detailed in the third part of our user study, P4, a university student engaging with a childhood stuffed toy, stated, ``\textit{Interacting with my childhood toy felt like reconnecting with a part of myself.}'' This observation highlights the system’s potential to evoke nostalgia and enhance emotional well-being. Practitioners could thus explore integrating Talking Spell into everyday objects to foster greater personal engagement and emotional support in domestic settings.

In educational contexts, our findings, particularly from \autoref{sec:participants-feedback}, demonstrate that Talking Spell, when embedded in tools like a drawing pad, facilitates personalized learning and augments creative processes. Participants noted that interactions with anthropomorphized educational tools encouraged creativity through co-design collaboration. P7, reflecting on his interaction with Boardie, commented, ``\textit{It was like having a creative partner that encouraged me to explore new ideas.}'' This suggests that Talking Spell could inspire students and promote collaborative learning. Consequently, educators and developers might consider implementing Talking Spell in classrooms to create interactive learning environments that stimulate creativity and active engagement.

In the healthcare domain, our study underscores Talking Spell’s potential to address social isolation. Follow-up discussions revealed that participants (P2, P3, P9, P11) valued anthropomorphized objects as companions that offer emotional support and facilitate communication, thereby mitigating feelings of loneliness. Although our study focused on participants aged 18 to 27, our findings, together with prior research highlighting the prevalence of loneliness and the need for companionship among older adults (\cite{sorkin2002,rook1987,ten2018}, may suggest that Talking Spell could be adapted for elderly companionship products. Practitioners in healthcare and social services could explore deploying wearabke systems that enable real-time anthropomorphic voice interaction in care facilities or home environments to improve seniors’ quality of life, fostering social connections and supporting mental health through interactive companionship.

In summary, the actionable implications derived from user feedback on Talking Spell highlight its versatility across daily companionship, educational enhancement, and elderly care. By harnessing the emotional and interactive potential of anthropomorphism, practitioners can develop innovative solutions that enrich user experiences and address pressing social needs.
}

\section{Limitations \& Future Work}
In this section, we discuss the current limitations of our system and provide several design opportunities for future research.

\finalrv{\textbf{Generalizability of results.}
The user study presented in this paper, conducted with 12 participants, provides preliminary evidence of the usability and positive experiential impact of our proposed system. In future work, we plan to recruit a larger and more diverse participant pool and extend the duration of usage to enhance the generalizability and reliability of our findings.}

\textbf{Remote Deployment.} Currently, our system relies on a laptop as an information transmission hub, which also handles image processing and API calls. This setup limits the system's flexibility. In future designs, we plan to migrate information transmission to mobile devices, such as smartphones and smartwatches, and offload computational tasks to cloud servers. Additionally, our current device employs a stacked module approach to initially demonstrate its usability and potential, but it lacks sufficient integration. In the future, we aim to enhance the integration of these modules, reducing their size and weight to improve wearability and comfort.

\textbf{Object Proactive Interaction. }
Participants frequently mention active interaction during the experiments, indicating its significance as a feature. However, the mechanisms for triggering and deactivating active interaction should be explored in future work.

\textbf{Distance Measurement and Spatial Audio Integration.}
By incorporating distance detection, the system can integrate spatial information into audio feedback, thereby providing users with a more immersive experience.

\textbf{Multi-agent Interaction.}
Currently, our system supports seamless switching between different objects, but it does not yet enable simultaneous interaction with multiple objects. 
% Future work should focus on supporting interactions among multiple objects as well as interactions between objects themselves.
\finalrv{Future research should focus on supporting interactions between users and multiple objects, as well as interactions among objects themselves, while exploring the group effects arising from interactions among these anthropomorphic agents.}

\textbf{Deep online search.}
Objects in daily life inherently possess certain attributes; for instance, a Totoro figurine is associated with the movie’s voice tone and storyline. However, the vision-language model in our current system fails to provide such functionality, thus not fully aligning with users’ expectations of the object. Future work could incorporate deep online search capabilities to connect the object’s persona generation and voice tone selection with the latest information.

\textbf{Anthropomorphism in audio.}
Our user study reports that the current system exhibits certain shortcomings in audio output, as the generated speech feedback lacks anthropomorphic responses. These shortcomings may relate to emotion, intonation, and breathing, which contribute to user discomfort during interaction. In future work, adopting more anthropomorphic text-to-speech techniques could enhance user immersion.

\section{Conclusion}

We introduce Talking Spell, a portable wearable device integrated with machine learning that imbues random objects with speech capabilities and empowers them with anthropomorphic personas. Leveraging technologies such as image acquisition, segmentation, training, object detection, large vision-language models, speech-to-text, large language models (LLMs), and text-to-speech, the system enables users to engage in direct voice interactions with their possessions through three stages: acquaintance, familiarization, and bonding. Talking Spell showcases high usability and interactivity, empowering users by enhancing personalization in AI companionship, sparking creativity through co-design collaboration, and fostering emotional connections via personal belongings across diverse daily contexts. Despite its strengths, user feedback underscores the need for proactive object interactions, multi-agent capabilities, and a broader range of voice options to further boost engagement and practical utility. Future work includes optimizing hardware integration and deployment, as well as conducting in-depth exploration of AI-empowered human-object interaction networks.

\begin{comment}
\end{comment}

\begin{acks}
This work is partially supported by the National Key Research and Development Program of China under Grant No.: 2024YFC3307602, and
the Guangdong Provincial Talent Program under Grant No.:
2023JC10X009.
\end{acks}

\bibliographystyle{ACM-Reference-Format}
\bibliography{main}


\begin{thebibliography}{64}

%%% the Web address.

\ifx \showCODEN    \undefined \def \showCODEN     #1{\unskip}     \fi
\ifx \showISBNx    \undefined \def \showISBNx     #1{\unskip}     \fi
\ifx \showISBNxiii \undefined \def \showISBNxiii  #1{\unskip}     \fi
\ifx \showISSN     \undefined \def \showISSN      #1{\unskip}     \fi
\ifx \showLCCN     \undefined \def \showLCCN      #1{\unskip}     \fi
\ifx \shownote     \undefined \def \shownote      #1{#1}          \fi
\ifx \showarticletitle \undefined \def \showarticletitle #1{#1}   \fi
\ifx \showURL      \undefined \def \showURL       {\relax}        \fi
% invisible to TeX
\providecommand\bibfield[2]{#2}
\providecommand\bibinfo[2]{#2}
\providecommand\natexlab[1]{#1}
\providecommand\showeprint[2][]{arXiv:#2}

\bibitem[Alkatheiri(2022)]%
        {Alkatheiri2022}
\bibfield{author}{\bibinfo{person}{Mohammed~S. Alkatheiri}.} \bibinfo{year}{2022}\natexlab{}.
\newblock \showarticletitle{Artificial intelligence assisted improved human-computer interactions for computer systems}.
\newblock \bibinfo{journal}{\emph{Computers and Electrical Engineering}}  \bibinfo{volume}{101} (\bibinfo{year}{2022}), \bibinfo{pages}{107950}.
\newblock
\showISSN{0045-7906}
\href{https://doi.org/10.1016/j.compeleceng.2022.107950}{doi:\nolinkurl{10.1016/j.compeleceng.2022.107950}}


\bibitem[Ammari et~al\mbox{.}(2019)]%
        {Ammari2019}
\bibfield{author}{\bibinfo{person}{Tawfiq Ammari}, \bibinfo{person}{Jofish Kaye}, \bibinfo{person}{Janice~Y. Tsai}, {and} \bibinfo{person}{Frank Bentley}.} \bibinfo{year}{2019}\natexlab{}.
\newblock \showarticletitle{Music, Search, and IoT: How People (Really) Use Voice Assistants}.
\newblock \bibinfo{journal}{\emph{ACM Trans. Comput.-Hum. Interact.}} \bibinfo{volume}{26}, \bibinfo{number}{3}, Article \bibinfo{articleno}{17} (\bibinfo{date}{April} \bibinfo{year}{2019}), \bibinfo{numpages}{28}~pages.
\newblock
\showISSN{1073-0516}
\href{https://doi.org/10.1145/3311956}{doi:\nolinkurl{10.1145/3311956}}


\bibitem[Arleen~Salles(2020)]%
        {Salles2020}
\bibfield{author}{\bibinfo{person}{Michele Farisco~and Arleen~Salles, Kathinka~Evers}.} \bibinfo{year}{2020}\natexlab{}.
\newblock \showarticletitle{Anthropomorphism in AI}.
\newblock \bibinfo{journal}{\emph{AJOB Neuroscience}} \bibinfo{volume}{11}, \bibinfo{number}{2} (\bibinfo{year}{2020}), \bibinfo{pages}{88--95}.
\newblock
\href{https://doi.org/10.1080/21507740.2020.1740350}{doi:\nolinkurl{10.1080/21507740.2020.1740350}}
\showeprint{https://doi.org/10.1080/21507740.2020.1740350}
\newblock
\shownote{PMID: 32228388}.


\bibitem[Bai et~al\mbox{.}(2023)]%
        {qwen-vl}
\bibfield{author}{\bibinfo{person}{Jinze Bai}, \bibinfo{person}{Shuai Bai}, \bibinfo{person}{Shusheng Yang}, \bibinfo{person}{Shijie Wang}, \bibinfo{person}{Sinan Tan}, \bibinfo{person}{Peng Wang}, \bibinfo{person}{Junyang Lin}, \bibinfo{person}{Chang Zhou}, {and} \bibinfo{person}{Jingren Zhou}.} \bibinfo{year}{2023}\natexlab{}.
\newblock \bibinfo{title}{Qwen-VL: A Versatile Vision-Language Model for Understanding, Localization, Text Reading, and Beyond}.
\newblock
\showeprint[arxiv]{2308.12966}~[cs.CV]
\urldef\tempurl%
\url{https://arxiv.org/abs/2308.12966}
\showURL{%
\tempurl}


\bibitem[Boersma et~al\mbox{.}(2019)]%
        {Boersma2019}
\bibfield{author}{\bibinfo{person}{Reginald Boersma}, \bibinfo{person}{P~Marijn Poortvliet}, {and} \bibinfo{person}{Bart Gremmen}.} \bibinfo{year}{2019}\natexlab{}.
\newblock \showarticletitle{Naming is framing: the effects of a technological name on the interpretation of a technology}.
\newblock \bibinfo{journal}{\emph{Journal of Science Communication}} \bibinfo{volume}{18}, \bibinfo{number}{6} (\bibinfo{year}{2019}), \bibinfo{pages}{A04}.
\newblock


\bibitem[Bowlby(1969)]%
        {Bowlby1969}
\bibfield{author}{\bibinfo{person}{John Bowlby}.} \bibinfo{year}{1969}\natexlab{}.
\newblock \bibinfo{booktitle}{\emph{Attachment and loss}}.
\newblock Number~79. \bibinfo{publisher}{Random House}.
\newblock


\bibitem[Brooke et~al\mbox{.}(1996)]%
        {SUS}
\bibfield{author}{\bibinfo{person}{John Brooke} {et~al\mbox{.}}} \bibinfo{year}{1996}\natexlab{}.
\newblock \showarticletitle{SUS-A quick and dirty usability scale}.
\newblock \bibinfo{journal}{\emph{Usability evaluation in industry}} \bibinfo{volume}{189}, \bibinfo{number}{194} (\bibinfo{year}{1996}), \bibinfo{pages}{4--7}.
\newblock


\bibitem[Carpenter(2013)]%
        {Carpenter2013}
\bibfield{author}{\bibinfo{person}{Julie Carpenter}.} \bibinfo{year}{2013}\natexlab{}.
\newblock \showarticletitle{Just doesn’t look right: Exploring the impact of humanoid robot integration into explosive ordnance disposal teams}.
\newblock In \bibinfo{booktitle}{\emph{Handbook of research on technoself: Identity in a technological society}}. \bibinfo{publisher}{IGI Global}, \bibinfo{pages}{609--636}.
\newblock


\bibitem[Castillo et~al\mbox{.}(2021)]%
        {Castillo2021}
\bibfield{author}{\bibinfo{person}{Daniela Castillo}, \bibinfo{person}{Ana~Isabel Canhoto}, {and} \bibinfo{person}{Emanuel~Said and}.} \bibinfo{year}{2021}\natexlab{}.
\newblock \showarticletitle{The dark side of AI-powered service interactions: exploring the process of co-destruction from the customer perspective}.
\newblock \bibinfo{journal}{\emph{The Service Industries Journal}} \bibinfo{volume}{41}, \bibinfo{number}{13-14} (\bibinfo{year}{2021}), \bibinfo{pages}{900--925}.
\newblock
\href{https://doi.org/10.1080/02642069.2020.1787993}{doi:\nolinkurl{10.1080/02642069.2020.1787993}}
\showeprint{https://doi.org/10.1080/02642069.2020.1787993}


\bibitem[Chao et~al\mbox{.}(2018)]%
        {Chao2018}
\bibfield{author}{\bibinfo{person}{Yu-Wei Chao}, \bibinfo{person}{Yunfan Liu}, \bibinfo{person}{Xieyang Liu}, \bibinfo{person}{Huayi Zeng}, {and} \bibinfo{person}{Jia Deng}.} \bibinfo{year}{2018}\natexlab{}.
\newblock \showarticletitle{Learning to detect human-object interactions}. In \bibinfo{booktitle}{\emph{2018 ieee winter conference on applications of computer vision (wacv)}}. IEEE, \bibinfo{pages}{381--389}.
\newblock


\bibitem[Cheng and Chan(2024)]%
        {BodyTouch}
\bibfield{author}{\bibinfo{person}{Wen-Wei Cheng} {and} \bibinfo{person}{Liwei Chan}.} \bibinfo{year}{2024}\natexlab{}.
\newblock \showarticletitle{BodyTouch: Investigating Eye-Free, On-Body and Near-Body Touch Interactions with HMDs}.
\newblock \bibinfo{journal}{\emph{Proc. ACM Interact. Mob. Wearable Ubiquitous Technol.}} \bibinfo{volume}{7}, \bibinfo{number}{4}, Article \bibinfo{articleno}{152} (\bibinfo{date}{Jan.} \bibinfo{year}{2024}), \bibinfo{numpages}{22}~pages.
\newblock
\href{https://doi.org/10.1145/3631426}{doi:\nolinkurl{10.1145/3631426}}


\bibitem[Chong et~al\mbox{.}(2021)]%
        {Chong2021}
\bibfield{author}{\bibinfo{person}{Terrence Chong}, \bibinfo{person}{Ting Yu}, \bibinfo{person}{Debbie~Isobel Keeling}, {and} \bibinfo{person}{Ko de Ruyter}.} \bibinfo{year}{2021}\natexlab{}.
\newblock \showarticletitle{AI-chatbots on the services frontline addressing the challenges and opportunities of agency}.
\newblock \bibinfo{journal}{\emph{Journal of Retailing and Consumer Services}}  \bibinfo{volume}{63} (\bibinfo{year}{2021}), \bibinfo{pages}{102735}.
\newblock


\bibitem[Cowan et~al\mbox{.}(2017)]%
        {Cowan2017}
\bibfield{author}{\bibinfo{person}{Benjamin~R Cowan}, \bibinfo{person}{Nadia Pantidi}, \bibinfo{person}{David Coyle}, \bibinfo{person}{Kellie Morrissey}, \bibinfo{person}{Peter Clarke}, \bibinfo{person}{Sara Al-Shehri}, \bibinfo{person}{David Earley}, {and} \bibinfo{person}{Natasha Bandeira}.} \bibinfo{year}{2017}\natexlab{}.
\newblock \showarticletitle{" What can i help you with?" infrequent users' experiences of intelligent personal assistants}. In \bibinfo{booktitle}{\emph{Proceedings of the 19th international conference on human-computer interaction with mobile devices and services}}. \bibinfo{pages}{1--12}.
\newblock


\bibitem[Curry and Curry(2023)]%
        {Curry2023}
\bibfield{author}{\bibinfo{person}{Alba~Cercas Curry} {and} \bibinfo{person}{Amanda~Cercas Curry}.} \bibinfo{year}{2023}\natexlab{}.
\newblock \showarticletitle{Computer says “no”: The case against empathetic conversational AI}. In \bibinfo{booktitle}{\emph{Findings of the Association for Computational Linguistics: ACL 2023}}. \bibinfo{pages}{8123--8130}.
\newblock


\bibitem[Duan et~al\mbox{.}(2024)]%
        {Duan2024}
\bibfield{author}{\bibinfo{person}{Shiyu Duan}, \bibinfo{person}{Ziyi Wang}, \bibinfo{person}{Shixiao Wang}, \bibinfo{person}{Mengmeng Chen}, {and} \bibinfo{person}{Runsheng Zhang}.} \bibinfo{year}{2024}\natexlab{}.
\newblock \showarticletitle{Emotion-aware interaction design in intelligent user interface using multi-modal deep learning}. In \bibinfo{booktitle}{\emph{2024 5th International Symposium on Computer Engineering and Intelligent Communications (ISCEIC)}}. IEEE, \bibinfo{pages}{110--114}.
\newblock


\bibitem[Efthymiou et~al\mbox{.}(2025)]%
        {efthymiou2025}
\bibfield{author}{\bibinfo{person}{Fotis Efthymiou}, \bibinfo{person}{Alex Mari}, \bibinfo{person}{Ertugrul Uysal}, {and} \bibinfo{person}{Jeffrey Brooks}.} \bibinfo{year}{2025}\natexlab{}.
\newblock \showarticletitle{No Hard Feelings: The Protective Power of AI Empathy During Service Interaction Failures}.
\newblock  (\bibinfo{year}{2025}).
\newblock


\bibitem[Epley et~al\mbox{.}(2007)]%
        {Epley2007}
\bibfield{author}{\bibinfo{person}{Nicholas Epley}, \bibinfo{person}{Adam Waytz}, {and} \bibinfo{person}{John~T Cacioppo}.} \bibinfo{year}{2007}\natexlab{}.
\newblock \showarticletitle{On seeing human: a three-factor theory of anthropomorphism.}
\newblock \bibinfo{journal}{\emph{Psychological review}} \bibinfo{volume}{114}, \bibinfo{number}{4} (\bibinfo{year}{2007}), \bibinfo{pages}{864}.
\newblock


\bibitem[George et~al\mbox{.}(2023)]%
        {george2023}
\bibfield{author}{\bibinfo{person}{A~Shaji George}, \bibinfo{person}{AS~Hovan George}, \bibinfo{person}{T Baskar}, {and} \bibinfo{person}{Digvijay Pandey}.} \bibinfo{year}{2023}\natexlab{}.
\newblock \showarticletitle{The allure of artificial intimacy: Examining the appeal and ethics of using generative AI for simulated relationships}.
\newblock \bibinfo{journal}{\emph{Partners Universal International Innovation Journal}} \bibinfo{volume}{1}, \bibinfo{number}{6} (\bibinfo{year}{2023}), \bibinfo{pages}{132--147}.
\newblock


\bibitem[Gilles and Bevacqua(2022)]%
        {Gilles2022}
\bibfield{author}{\bibinfo{person}{Marl{\`e}ne Gilles} {and} \bibinfo{person}{Elisabetta Bevacqua}.} \bibinfo{year}{2022}\natexlab{}.
\newblock \showarticletitle{A review of virtual assistants’ characteristics: Recommendations for designing an optimal human--machine cooperation}.
\newblock \bibinfo{journal}{\emph{Journal of Computing and Information Science in Engineering}} \bibinfo{volume}{22}, \bibinfo{number}{5} (\bibinfo{year}{2022}), \bibinfo{pages}{050904}.
\newblock


\bibitem[Gkioxari et~al\mbox{.}(2018)]%
        {Gkioxari2018}
\bibfield{author}{\bibinfo{person}{Georgia Gkioxari}, \bibinfo{person}{Ross Girshick}, \bibinfo{person}{Piotr Doll{\'a}r}, {and} \bibinfo{person}{Kaiming He}.} \bibinfo{year}{2018}\natexlab{}.
\newblock \showarticletitle{Detecting and recognizing human-object interactions}. In \bibinfo{booktitle}{\emph{Proceedings of the IEEE conference on computer vision and pattern recognition}}. \bibinfo{pages}{8359--8367}.
\newblock


\bibitem[Glikson and Woolley(2020)]%
        {Glikson2020}
\bibfield{author}{\bibinfo{person}{Ella Glikson} {and} \bibinfo{person}{Anita~Williams Woolley}.} \bibinfo{year}{2020}\natexlab{}.
\newblock \showarticletitle{Human trust in artificial intelligence: Review of empirical research}.
\newblock \bibinfo{journal}{\emph{Academy of management annals}} \bibinfo{volume}{14}, \bibinfo{number}{2} (\bibinfo{year}{2020}), \bibinfo{pages}{627--660}.
\newblock


\bibitem[Hegel et~al\mbox{.}(2008)]%
        {Hegel2008}
\bibfield{author}{\bibinfo{person}{Frank Hegel}, \bibinfo{person}{Soren Krach}, \bibinfo{person}{Tilo Kircher}, \bibinfo{person}{Britta Wrede}, {and} \bibinfo{person}{Gerhard Sagerer}.} \bibinfo{year}{2008}\natexlab{}.
\newblock \showarticletitle{Understanding social robots: A user study on anthropomorphism}. In \bibinfo{booktitle}{\emph{RO-MAN 2008-the 17th IEEE international symposium on robot and human interactive communication}}. IEEE, \bibinfo{pages}{574--579}.
\newblock


\bibitem[Horne and Lowe(1996)]%
        {Horne1996}
\bibfield{author}{\bibinfo{person}{Pauline~J Horne} {and} \bibinfo{person}{C~Fergus Lowe}.} \bibinfo{year}{1996}\natexlab{}.
\newblock \showarticletitle{On the origins of naming and other symbolic behavior}.
\newblock \bibinfo{journal}{\emph{Journal of the Experimental Analysis of behavior}} \bibinfo{volume}{65}, \bibinfo{number}{1} (\bibinfo{year}{1996}), \bibinfo{pages}{185--241}.
\newblock


\bibitem[Hu et~al\mbox{.}(2025)]%
        {GesPrompt}
\bibfield{author}{\bibinfo{person}{Xiyun Hu}, \bibinfo{person}{Dizhi Ma}, \bibinfo{person}{Fengming He}, \bibinfo{person}{Zhengzhe Zhu}, \bibinfo{person}{Shao-Kang Hsia}, \bibinfo{person}{Chenfei Zhu}, \bibinfo{person}{Ziyi Liu}, {and} \bibinfo{person}{Karthik Ramani}.} \bibinfo{year}{2025}\natexlab{}.
\newblock \showarticletitle{GesPrompt: Leveraging Co-Speech Gestures to Augment LLM-Based Interaction in Virtual Reality}. In \bibinfo{booktitle}{\emph{Proceedings of the 2025 ACM Designing Interactive Systems Conference}} \emph{(\bibinfo{series}{DIS '25})}. \bibinfo{publisher}{Association for Computing Machinery}, \bibinfo{address}{New York, NY, USA}, \bibinfo{pages}{59–80}.
\newblock
\showISBNx{9798400714856}
\href{https://doi.org/10.1145/3715336.3735769}{doi:\nolinkurl{10.1145/3715336.3735769}}


\bibitem[Jakub et~al\mbox{.}(2015)]%
        {Jakub2015}
\bibfield{author}{\bibinfo{person}{Z{\l}otowski Jakub}, \bibinfo{person}{Diane Proudfoot}, \bibinfo{person}{Yogeeswaran Kumar}, {and} \bibinfo{person}{Bartneck Christoph}.} \bibinfo{year}{2015}\natexlab{}.
\newblock \showarticletitle{Anthropomorphism: opportunities and challenges in human--robot interaction}.
\newblock \bibinfo{journal}{\emph{International Journal of Social Robotics}} \bibinfo{volume}{7}, \bibinfo{number}{3} (\bibinfo{year}{2015}), \bibinfo{pages}{347--360}.
\newblock


\bibitem[Jarrahi(2018)]%
        {Jarrahi2018}
\bibfield{author}{\bibinfo{person}{Mohammad~Hossein Jarrahi}.} \bibinfo{year}{2018}\natexlab{}.
\newblock \showarticletitle{Artificial intelligence and the future of work: Human-AI symbiosis in organizational decision making}.
\newblock \bibinfo{journal}{\emph{Business horizons}} \bibinfo{volume}{61}, \bibinfo{number}{4} (\bibinfo{year}{2018}), \bibinfo{pages}{577--586}.
\newblock


\bibitem[Kirk et~al\mbox{.}(2025)]%
        {Kirk2025}
\bibfield{author}{\bibinfo{person}{Hannah~Rose Kirk}, \bibinfo{person}{Iason Gabriel}, \bibinfo{person}{Chris Summerfield}, \bibinfo{person}{Bertie Vidgen}, {and} \bibinfo{person}{Scott~A Hale}.} \bibinfo{year}{2025}\natexlab{}.
\newblock \showarticletitle{Why human-AI relationships need socioaffective alignment}.
\newblock \bibinfo{journal}{\emph{arXiv preprint arXiv:2502.02528}} (\bibinfo{year}{2025}).
\newblock


\bibitem[Kühne and Peter(2022)]%
        {Kühne2022}
\bibfield{author}{\bibinfo{person}{Rinaldo Kühne} {and} \bibinfo{person}{Jochen Peter}.} \bibinfo{year}{2022}\natexlab{}.
\newblock \showarticletitle{Anthropomorphism in human–robot interactions: a multidimensional conceptualization}.
\newblock \bibinfo{journal}{\emph{Communication Theory}} \bibinfo{volume}{33}, \bibinfo{number}{1} (\bibinfo{date}{Oct.} \bibinfo{year}{2022}), \bibinfo{pages}{42--52}.
\newblock
\showISSN{1468-2885}
\href{https://doi.org/10.1093/ct/qtac020}{doi:\nolinkurl{10.1093/ct/qtac020}}
\newblock
\shownote{tex.eprint: https://academic.oup.com/ct/article-pdf/33/1/42/48958537/qtac020.pdf}.


\bibitem[Lee et~al\mbox{.}(2024)]%
        {GazePointAR}
\bibfield{author}{\bibinfo{person}{Jaewook Lee}, \bibinfo{person}{Jun Wang}, \bibinfo{person}{Elizabeth Brown}, \bibinfo{person}{Liam Chu}, \bibinfo{person}{Sebastian~S. Rodriguez}, {and} \bibinfo{person}{Jon~E. Froehlich}.} \bibinfo{year}{2024}\natexlab{}.
\newblock \showarticletitle{GazePointAR: A Context-Aware Multimodal Voice Assistant for Pronoun Disambiguation in Wearable Augmented Reality}. In \bibinfo{booktitle}{\emph{Proceedings of the 2024 CHI Conference on Human Factors in Computing Systems}} (Honolulu, HI, USA) \emph{(\bibinfo{series}{CHI '24})}. \bibinfo{publisher}{Association for Computing Machinery}, \bibinfo{address}{New York, NY, USA}, Article \bibinfo{articleno}{408}, \bibinfo{numpages}{20}~pages.
\newblock
\showISBNx{9798400703300}
\href{https://doi.org/10.1145/3613904.3642230}{doi:\nolinkurl{10.1145/3613904.3642230}}


\bibitem[Lenhart et~al\mbox{.}(2023)]%
        {Lenhart2023}
\bibfield{author}{\bibinfo{person}{Anna Lenhart}, \bibinfo{person}{Sunyup Park}, \bibinfo{person}{Michael Zimmer}, {and} \bibinfo{person}{Jessica Vitak}.} \bibinfo{year}{2023}\natexlab{}.
\newblock \showarticletitle{"You Shouldn't Need to Share Your Data": Perceived Privacy Risks and Mitigation Strategies Among Privacy-Conscious Smart Home Power Users}.
\newblock \bibinfo{journal}{\emph{Proc. ACM Hum.-Comput. Interact.}} \bibinfo{volume}{7}, \bibinfo{number}{CSCW2}, Article \bibinfo{articleno}{247} (\bibinfo{date}{Oct.} \bibinfo{year}{2023}), \bibinfo{numpages}{34}~pages.
\newblock
\href{https://doi.org/10.1145/3610038}{doi:\nolinkurl{10.1145/3610038}}


\bibitem[Liu et~al\mbox{.}(2021)]%
        {Liu2021}
\bibfield{author}{\bibinfo{person}{Peng Liu}, \bibinfo{person}{Qingqing Fei}, \bibinfo{person}{Jinting Liu}, {and} \bibinfo{person}{Jianqiang Wang}.} \bibinfo{year}{2021}\natexlab{}.
\newblock \showarticletitle{Naming is framing: The framing effect of technology name on public attitude toward automated vehicles}.
\newblock \bibinfo{journal}{\emph{Public Understanding of Science}} \bibinfo{volume}{30}, \bibinfo{number}{6} (\bibinfo{year}{2021}), \bibinfo{pages}{691--707}.
\newblock


\bibitem[Maedche et~al\mbox{.}(2019)]%
        {Maedche2019}
\bibfield{author}{\bibinfo{person}{Alexander Maedche}, \bibinfo{person}{Christine Legner}, \bibinfo{person}{Alexander Benlian}, \bibinfo{person}{Benedikt Berger}, \bibinfo{person}{Henner Gimpel}, \bibinfo{person}{Thomas Hess}, \bibinfo{person}{Oliver Hinz}, \bibinfo{person}{Stefan Morana}, {and} \bibinfo{person}{Matthias S{\"o}llner}.} \bibinfo{year}{2019}\natexlab{}.
\newblock \showarticletitle{AI-based digital assistants: Opportunities, threats, and research perspectives}.
\newblock \bibinfo{journal}{\emph{Business \& Information Systems Engineering}}  \bibinfo{volume}{61} (\bibinfo{year}{2019}), \bibinfo{pages}{535--544}.
\newblock


\bibitem[Mcconnell-Ginet(2005)]%
        {Mcconnell2005}
\bibfield{author}{\bibinfo{person}{Sally Mcconnell-Ginet}.} \bibinfo{year}{2005}\natexlab{}.
\newblock \showarticletitle{"Whaf s in a Name?" Social Labeling and Gender Practices}.
\newblock \bibinfo{journal}{\emph{The Handbook of Language and Gender}} (\bibinfo{year}{2005}), \bibinfo{pages}{69}.
\newblock


\bibitem[Mikulincer and Shaver(2005)]%
        {Mikulincer2005}
\bibfield{author}{\bibinfo{person}{Mario Mikulincer} {and} \bibinfo{person}{P Shaver}.} \bibinfo{year}{2005}\natexlab{}.
\newblock \showarticletitle{Mental representations and attachment security}.
\newblock \bibinfo{journal}{\emph{Interpersonal cognition}} (\bibinfo{year}{2005}), \bibinfo{pages}{233--266}.
\newblock


\bibitem[Mossbridge(2024)]%
        {Mossbridge2024}
\bibfield{author}{\bibinfo{person}{Julia Mossbridge}.} \bibinfo{year}{2024}\natexlab{}.
\newblock \showarticletitle{Shifting the Human-AI Relationship: Toward a Dynamic Relational Learning-Partner Model}.
\newblock \bibinfo{journal}{\emph{arXiv preprint arXiv:2410.11864}} (\bibinfo{year}{2024}).
\newblock


\bibitem[Naayini et~al\mbox{.}(2025)]%
        {Naayini2025}
\bibfield{author}{\bibinfo{person}{Prudhvi Naayini}, \bibinfo{person}{Praveen~Kumar Myakala}, \bibinfo{person}{Chiranjeevi Bura}, \bibinfo{person}{Anil~Kumar Jonnalagadda}, {and} \bibinfo{person}{Srikanth Kamatala}.} \bibinfo{year}{2025}\natexlab{}.
\newblock \showarticletitle{AI-Powered Assistive Technologies for Visual Impairment}.
\newblock \bibinfo{journal}{\emph{arXiv preprint arXiv:2503.15494}} (\bibinfo{year}{2025}).
\newblock


\bibitem[Norman(2010)]%
        {norman2010natural}
\bibfield{author}{\bibinfo{person}{Donald~A Norman}.} \bibinfo{year}{2010}\natexlab{}.
\newblock \showarticletitle{Natural user interfaces are not natural}.
\newblock \bibinfo{journal}{\emph{interactions}} \bibinfo{volume}{17}, \bibinfo{number}{3} (\bibinfo{year}{2010}), \bibinfo{pages}{6--10}.
\newblock


\bibitem[Ogawa and Ono(2008)]%
        {ITACO}
\bibfield{author}{\bibinfo{person}{Kohei Ogawa} {and} \bibinfo{person}{Tetsuo Ono}.} \bibinfo{year}{2008}\natexlab{}.
\newblock \showarticletitle{ITACO: Constructing an emotional relationship between human and robot}. In \bibinfo{booktitle}{\emph{RO-MAN 2008 - The 17th IEEE International Symposium on Robot and Human Interactive Communication}}. \bibinfo{pages}{35--40}.
\newblock
\href{https://doi.org/10.1109/ROMAN.2008.4600640}{doi:\nolinkurl{10.1109/ROMAN.2008.4600640}}


\bibitem[O'hara et~al\mbox{.}(2013)]%
        {NUI2013Ohara}
\bibfield{author}{\bibinfo{person}{Kenton O'hara}, \bibinfo{person}{Richard Harper}, \bibinfo{person}{Helena Mentis}, \bibinfo{person}{Abigail Sellen}, {and} \bibinfo{person}{Alex Taylor}.} \bibinfo{year}{2013}\natexlab{}.
\newblock \showarticletitle{On the naturalness of touchless: Putting the “interaction” back into NUI}.
\newblock \bibinfo{journal}{\emph{ACM Trans. Comput.-Hum. Interact.}} \bibinfo{volume}{20}, \bibinfo{number}{1}, Article \bibinfo{articleno}{5} (\bibinfo{date}{April} \bibinfo{year}{2013}), \bibinfo{numpages}{25}~pages.
\newblock
\showISSN{1073-0516}
\href{https://doi.org/10.1145/2442106.2442111}{doi:\nolinkurl{10.1145/2442106.2442111}}


\bibitem[Osawa et~al\mbox{.}(2009)]%
        {Osawa2009}
\bibfield{author}{\bibinfo{person}{Hirotaka Osawa}, \bibinfo{person}{Kentaro Ishii}, \bibinfo{person}{Toshihiro Osumi}, \bibinfo{person}{Ren Ohmura}, {and} \bibinfo{person}{Michita Imai}.} \bibinfo{year}{2009}\natexlab{}.
\newblock \showarticletitle{Anthropomorphization of a space with implemented human-like features}. In \bibinfo{booktitle}{\emph{ACM SIGGRAPH 2009 Emerging Technologies}} (New Orleans, Louisiana) \emph{(\bibinfo{series}{SIGGRAPH '09})}. \bibinfo{publisher}{Association for Computing Machinery}, \bibinfo{address}{New York, NY, USA}, Article \bibinfo{articleno}{2}, \bibinfo{numpages}{1}~pages.
\newblock
\showISBNx{9781605588339}
\href{https://doi.org/10.1145/1597956.1597958}{doi:\nolinkurl{10.1145/1597956.1597958}}


\bibitem[Osawa et~al\mbox{.}(2012)]%
        {Osawa2012}
\bibfield{author}{\bibinfo{person}{Hirotaka Osawa}, \bibinfo{person}{Yuji Matsuda}, \bibinfo{person}{Ren Ohmura}, {and} \bibinfo{person}{Michita Imai}.} \bibinfo{year}{2012}\natexlab{}.
\newblock \showarticletitle{Embodiment of an agent by anthropomorphization of a common object}.
\newblock \bibinfo{journal}{\emph{Web Intelligence and Agent Systems}} \bibinfo{volume}{10}, \bibinfo{number}{3} (\bibinfo{year}{2012}), \bibinfo{pages}{345--358}.
\newblock


\bibitem[Osawa et~al\mbox{.}(2006)]%
        {Osawa2006}
\bibfield{author}{\bibinfo{person}{Hirotaka Osawa}, \bibinfo{person}{Jun Mukai}, {and} \bibinfo{person}{Michita Imai}.} \bibinfo{year}{2006}\natexlab{}.
\newblock \showarticletitle{Anthropomorphization of an object by displaying robot}. In \bibinfo{booktitle}{\emph{ROMAN 2006-The 15th IEEE International Symposium on Robot and Human Interactive Communication}}. IEEE, \bibinfo{pages}{763--768}.
\newblock


\bibitem[Proudfoot(2011)]%
        {Proudfoot2011}
\bibfield{author}{\bibinfo{person}{Diane Proudfoot}.} \bibinfo{year}{2011}\natexlab{}.
\newblock \showarticletitle{Anthropomorphism and AI: Turing's much misunderstood imitation game}.
\newblock \bibinfo{journal}{\emph{Artificial Intelligence}} \bibinfo{volume}{175}, \bibinfo{number}{5-6} (\bibinfo{year}{2011}), \bibinfo{pages}{950--957}.
\newblock


\bibitem[Purington et~al\mbox{.}(2017)]%
        {Purington2017}
\bibfield{author}{\bibinfo{person}{Amanda Purington}, \bibinfo{person}{Jessie~G Taft}, \bibinfo{person}{Shruti Sannon}, \bibinfo{person}{Natalya~N Bazarova}, {and} \bibinfo{person}{Samuel~Hardman Taylor}.} \bibinfo{year}{2017}\natexlab{}.
\newblock \showarticletitle{" Alexa is my new BFF" social roles, user satisfaction, and personification of the Amazon Echo}. In \bibinfo{booktitle}{\emph{Proceedings of the 2017 CHI conference extended abstracts on human factors in computing systems}}. \bibinfo{pages}{2853--2859}.
\newblock


\bibitem[Ravi et~al\mbox{.}(2024)]%
        {SAM2}
\bibfield{author}{\bibinfo{person}{Nikhila Ravi}, \bibinfo{person}{Valentin Gabeur}, \bibinfo{person}{Yuan-Ting Hu}, \bibinfo{person}{Ronghang Hu}, \bibinfo{person}{Chaitanya Ryali}, \bibinfo{person}{Tengyu Ma}, \bibinfo{person}{Haitham Khedr}, \bibinfo{person}{Roman R{\"a}dle}, \bibinfo{person}{Chloe Rolland}, \bibinfo{person}{Laura Gustafson}, {et~al\mbox{.}}} \bibinfo{year}{2024}\natexlab{}.
\newblock \showarticletitle{Sam 2: Segment anything in images and videos}.
\newblock \bibinfo{journal}{\emph{arXiv preprint arXiv:2408.00714}} (\bibinfo{year}{2024}).
\newblock


\bibitem[Redmon et~al\mbox{.}(2016)]%
        {yolo}
\bibfield{author}{\bibinfo{person}{Joseph Redmon}, \bibinfo{person}{Santosh Divvala}, \bibinfo{person}{Ross Girshick}, {and} \bibinfo{person}{Ali Farhadi}.} \bibinfo{year}{2016}\natexlab{}.
\newblock \showarticletitle{You only look once: Unified, real-time object detection}. In \bibinfo{booktitle}{\emph{Proceedings of the IEEE conference on computer vision and pattern recognition}}. \bibinfo{pages}{779--788}.
\newblock


\bibitem[Reeves and Nass(1996)]%
        {Reeves1996}
\bibfield{author}{\bibinfo{person}{Byron Reeves} {and} \bibinfo{person}{Clifford Nass}.} \bibinfo{year}{1996}\natexlab{}.
\newblock \showarticletitle{The media equation: How people treat computers, television, and new media like real people}.
\newblock \bibinfo{journal}{\emph{Cambridge, UK}} \bibinfo{volume}{10}, \bibinfo{number}{10} (\bibinfo{year}{1996}), \bibinfo{pages}{19--36}.
\newblock


\bibitem[Rook(1987)]%
        {rook1987}
\bibfield{author}{\bibinfo{person}{Karen~S Rook}.} \bibinfo{year}{1987}\natexlab{}.
\newblock \showarticletitle{Social support versus companionship: effects on life stress, loneliness, and evaluations by others.}
\newblock \bibinfo{journal}{\emph{Journal of personality and social psychology}} \bibinfo{volume}{52}, \bibinfo{number}{6} (\bibinfo{year}{1987}), \bibinfo{pages}{1132}.
\newblock


\bibitem[Roshanaei et~al\mbox{.}(2024)]%
        {Roshanaei2024}
\bibfield{author}{\bibinfo{person}{Mahnaz Roshanaei}, \bibinfo{person}{Rezvaneh Rezapour}, {and} \bibinfo{person}{Magy~Seif El-Nasr}.} \bibinfo{year}{2024}\natexlab{}.
\newblock \showarticletitle{Talk, Listen, Connect: Navigating Empathy in Human-AI Interactions}.
\newblock \bibinfo{journal}{\emph{arXiv preprint arXiv:2409.15550}} (\bibinfo{year}{2024}).
\newblock


\bibitem[Rostami and Navabinejad(2023)]%
        {rostami2023}
\bibfield{author}{\bibinfo{person}{Mehdi Rostami} {and} \bibinfo{person}{Shokouh Navabinejad}.} \bibinfo{year}{2023}\natexlab{}.
\newblock \showarticletitle{Artificial empathy: User experiences with emotionally intelligent chatbots}.
\newblock \bibinfo{journal}{\emph{AI and Tech in Behavioral and Social Sciences}} \bibinfo{volume}{1}, \bibinfo{number}{3} (\bibinfo{year}{2023}), \bibinfo{pages}{19--27}.
\newblock


\bibitem[Sharma et~al\mbox{.}(2023)]%
        {Shama2023}
\bibfield{author}{\bibinfo{person}{Ashish Sharma}, \bibinfo{person}{Inna~W. Lin}, \bibinfo{person}{Adam~S. Miner}, \bibinfo{person}{David~C. Atkins}, {and} \bibinfo{person}{Tim Althoff}.} \bibinfo{year}{2023}\natexlab{}.
\newblock \showarticletitle{Human–{AI} collaboration enables more empathic conversations in text-based peer-to-peer mental health support}.
\newblock \bibinfo{journal}{\emph{Nature Machine Intelligence}} \bibinfo{volume}{5}, \bibinfo{number}{1} (\bibinfo{date}{Jan.} \bibinfo{year}{2023}), \bibinfo{pages}{46--57}.
\newblock
\showISSN{2522-5839}
\href{https://doi.org/10.1038/s42256-022-00593-2}{doi:\nolinkurl{10.1038/s42256-022-00593-2}}


\bibitem[Sorkin et~al\mbox{.}(2002)]%
        {sorkin2002}
\bibfield{author}{\bibinfo{person}{Dara Sorkin}, \bibinfo{person}{Karen~S Rook}, {and} \bibinfo{person}{John~L Lu}.} \bibinfo{year}{2002}\natexlab{}.
\newblock \showarticletitle{Loneliness, lack of emotional support, lack of companionship, and the likelihood of having a heart condition in an elderly sample}.
\newblock \bibinfo{journal}{\emph{Annals of Behavioral Medicine}} \bibinfo{volume}{24}, \bibinfo{number}{4} (\bibinfo{year}{2002}), \bibinfo{pages}{290--298}.
\newblock


\bibitem[Stoner et~al\mbox{.}(2018)]%
        {Stoner2018}
\bibfield{author}{\bibinfo{person}{Jennifer~L Stoner}, \bibinfo{person}{Barbara Loken}, {and} \bibinfo{person}{Ashley Stadler~Blank}.} \bibinfo{year}{2018}\natexlab{}.
\newblock \showarticletitle{The name game: How naming products increases psychological ownership and subsequent consumer evaluations}.
\newblock \bibinfo{journal}{\emph{Journal of Consumer Psychology}} \bibinfo{volume}{28}, \bibinfo{number}{1} (\bibinfo{year}{2018}), \bibinfo{pages}{130--137}.
\newblock


\bibitem[Su and Bao(2024)]%
        {Su2024}
\bibfield{author}{\bibinfo{person}{Megan Su} {and} \bibinfo{person}{Yuwei Bao}.} \bibinfo{year}{2024}\natexlab{}.
\newblock \showarticletitle{User Modeling Challenges in Interactive AI Assistant Systems}. In \bibinfo{booktitle}{\emph{The Twelfth International Conference on Learning Representations: Tiny Papers Track}}.
\newblock


\bibitem[Tejwani(2020)]%
        {tejwani2020migratabledefense}
\bibfield{author}{\bibinfo{person}{Ravi Tejwani}.} \bibinfo{year}{2020}\natexlab{}.
\newblock \emph{\bibinfo{title}{Migratable AI}}.
\newblock \bibinfo{thesistype}{Ph.\,D. Dissertation}. \bibinfo{school}{Massachusetts Institute of Technology}.
\newblock


\bibitem[Tejwani et~al\mbox{.}(2020)]%
        {tejwani2020migratable}
\bibfield{author}{\bibinfo{person}{Ravi Tejwani}, \bibinfo{person}{Felipe Moreno}, \bibinfo{person}{Sooyeon Jeong}, \bibinfo{person}{Hae~Won Park}, {and} \bibinfo{person}{Cynthia Breazeal}.} \bibinfo{year}{2020}\natexlab{}.
\newblock \showarticletitle{Migratable AI: Effect of identity and information migration on users' perception of conversational AI agents}. In \bibinfo{booktitle}{\emph{2020 29th IEEE International Conference on Robot and Human Interactive Communication (RO-MAN)}}. IEEE, \bibinfo{pages}{877--884}.
\newblock


\bibitem[Ten~Bruggencate et~al\mbox{.}(2018)]%
        {ten2018}
\bibfield{author}{\bibinfo{person}{TINA Ten~Bruggencate}, \bibinfo{person}{Katrien~G Luijkx}, {and} \bibinfo{person}{Janienke Sturm}.} \bibinfo{year}{2018}\natexlab{}.
\newblock \showarticletitle{Social needs of older people: A systematic literature review}.
\newblock \bibinfo{journal}{\emph{Ageing \& Society}} \bibinfo{volume}{38}, \bibinfo{number}{9} (\bibinfo{year}{2018}), \bibinfo{pages}{1745--1770}.
\newblock


\bibitem[Vatavu(2025)]%
        {nonnatural2025Vatavu}
\bibfield{author}{\bibinfo{person}{Radu-Daniel Vatavu}.} \bibinfo{year}{2025}\natexlab{}.
\newblock \showarticletitle{Non-Natural Interaction Design}. In \bibinfo{booktitle}{\emph{Proceedings of the 2025 CHI Conference on Human Factors in Computing Systems}} \emph{(\bibinfo{series}{CHI '25})}. \bibinfo{publisher}{Association for Computing Machinery}, \bibinfo{address}{New York, NY, USA}, Article \bibinfo{articleno}{435}, \bibinfo{numpages}{16}~pages.
\newblock
\showISBNx{9798400713941}
\href{https://doi.org/10.1145/3706598.3713459}{doi:\nolinkurl{10.1145/3706598.3713459}}


\bibitem[Wan and Chen(2021)]%
        {Wan2021}
\bibfield{author}{\bibinfo{person}{Echo~Wen Wan} {and} \bibinfo{person}{Rocky~Peng Chen}.} \bibinfo{year}{2021}\natexlab{}.
\newblock \showarticletitle{Anthropomorphism and object attachment}.
\newblock \bibinfo{journal}{\emph{Current Opinion in Psychology}}  \bibinfo{volume}{39} (\bibinfo{year}{2021}), \bibinfo{pages}{88--93}.
\newblock
\showISSN{2352-250X}
\href{https://doi.org/10.1016/j.copsyc.2020.08.009}{doi:\nolinkurl{10.1016/j.copsyc.2020.08.009}}
\newblock
\shownote{Object Attachment}.


\bibitem[Wrigley et~al\mbox{.}(2021)]%
        {Wrigley2021}
\bibfield{author}{\bibinfo{person}{Cara Wrigley}, \bibinfo{person}{Sean Peel}, \bibinfo{person}{Kimmi Keum~Hee Ko}, {and} \bibinfo{person}{Karla Straker}.} \bibinfo{year}{2021}\natexlab{}.
\newblock \showarticletitle{Patient names for mechanical circulatory support devices: Developing emotional insights}.
\newblock \bibinfo{journal}{\emph{Heart \& Lung}} \bibinfo{volume}{50}, \bibinfo{number}{6} (\bibinfo{year}{2021}), \bibinfo{pages}{953--967}.
\newblock


\bibitem[Yang and Lee(2019)]%
        {Yang2019}
\bibfield{author}{\bibinfo{person}{Heetae Yang} {and} \bibinfo{person}{Hwansoo Lee}.} \bibinfo{year}{2019}\natexlab{}.
\newblock \showarticletitle{Understanding user behavior of virtual personal assistant devices}.
\newblock \bibinfo{journal}{\emph{Information Systems and e-Business Management}}  \bibinfo{volume}{17} (\bibinfo{year}{2019}), \bibinfo{pages}{65--87}.
\newblock


\bibitem[Yang et~al\mbox{.}(2024)]%
        {10.1145/3659625}
\bibfield{author}{\bibinfo{person}{Ziqi Yang}, \bibinfo{person}{Xuhai Xu}, \bibinfo{person}{Bingsheng Yao}, \bibinfo{person}{Ethan Rogers}, \bibinfo{person}{Shao Zhang}, \bibinfo{person}{Stephen Intille}, \bibinfo{person}{Nawar Shara}, \bibinfo{person}{Guodong~Gordon Gao}, {and} \bibinfo{person}{Dakuo Wang}.} \bibinfo{year}{2024}\natexlab{}.
\newblock \showarticletitle{Talk2Care: An LLM-based Voice Assistant for Communication between Healthcare Providers and Older Adults}.
\newblock \bibinfo{journal}{\emph{Proc. ACM Interact. Mob. Wearable Ubiquitous Technol.}} \bibinfo{volume}{8}, \bibinfo{number}{2}, Article \bibinfo{articleno}{73} (\bibinfo{date}{May} \bibinfo{year}{2024}), \bibinfo{numpages}{35}~pages.
\newblock
\href{https://doi.org/10.1145/3659625}{doi:\nolinkurl{10.1145/3659625}}


\bibitem[Zhan et~al\mbox{.}(2019)]%
        {Zhan2019}
\bibfield{author}{\bibinfo{person}{Yibing Zhan}, \bibinfo{person}{Jun Yu}, \bibinfo{person}{Ting Yu}, {and} \bibinfo{person}{Dacheng Tao}.} \bibinfo{year}{2019}\natexlab{}.
\newblock \showarticletitle{On exploring undetermined relationships for visual relationship detection}. In \bibinfo{booktitle}{\emph{Proceedings of the IEEE/CVF Conference on Computer Vision and Pattern Recognition}}. \bibinfo{pages}{5128--5137}.
\newblock


\bibitem[Zhang and Li(2025)]%
        {Zhang2025}
\bibfield{author}{\bibinfo{person}{Shuning Zhang} {and} \bibinfo{person}{Shixuan Li}.} \bibinfo{year}{2025}\natexlab{}.
\newblock \showarticletitle{The Real Her? Exploring Whether Young Adults Accept Human-AI Love}.
\newblock \bibinfo{journal}{\emph{arXiv preprint arXiv:2503.03067}} (\bibinfo{year}{2025}).
\newblock


\end{thebibliography}

\appendix
\end{document}